\documentclass[12pt,reqno]{article}
\usepackage{amsmath,amssymb,amsthm,graphics,color,a4wide,hyperref,verbatim}
\usepackage{graphicx}
\usepackage{psfrag} 
\usepackage{marvosym}

\usepackage[table]{xcolor}

\usepackage{longtable}
\usepackage{multicol}

\usepackage{mathrsfs}
\input xy
\xyoption{all}

\usepackage[margin=2.5cm,includefoot]{geometry}
\usepackage{pdflscape}
\usepackage[symbol*]{footmisc}

\def\be{\begin{equation}}
\def\ee{\end{equation}}
\def\ba{\begin{eqnarray}}
\def\ea{\end{eqnarray}}

\newcommand{\C}{{\mathcal C}}

\renewcommand{\H}{{\mathcal H}}

\renewcommand{\L}{{\mathcal L}}
\newcommand{\M}{{\mathcal M}}
\newcommand{\N}{{\mathcal N}}

\renewcommand{\P}{{\mathcal P}}
\newcommand{\cC}{\mathcal{C}}
\newcommand{\T}{\mathcal{T}}
\newcommand{\U}{\mathcal{U}}

\newcommand{\RR}{\mathbb{R}}

\newcommand{\ZZ}{\mathbb{Z}}
\newcommand{\QQ}{\mathbb{Q}}
\newcommand{\CC}{\mathbb{C}}

\newcommand{\HH}{\mathbb{H}}

\newcommand{\NN}{\mathbb{N}}

\newcommand{\II}{\mathbb{I}}

\newcommand{\nn}{\nonumber}

\DeclareMathOperator{\Mat}{Mat}
\DeclareMathOperator{\Tr}{Tr}

\newlength{\picwidth} 
\newlength{\miniwidth} 
\newlength{\nanowidth} 
\newlength{\melowidth} 
\newlength{\leftminiwidth} 
\newlength{\rightminiwidth} 
\newlength{\minipagewidth} 
\newtheorem{theorem}{Theorem}
\newtheorem{conjecture}[theorem]{Conjecture}

\numberwithin{theorem}{section}
\numberwithin{equation}{section}

\begin{document}

\thispagestyle{empty}
\begin{flushright}AEI-2013-269\end{flushright}

\vskip 3cm

%\begin{center}
%{\bf \Large{Second Quantized Mathieu Moonshine}}
%\end{center}
\begin{center}
{\LARGE{S}\Large{ECOND} \LARGE{Q}\Large{UANTIZED} \LARGE{M}\Large{ATHIEU} \LARGE{M}\Large{OONSHINE}}
\end{center}

\bigskip

\begin{center}
{\bf Daniel Persson}
\\ Fundamental Physics, Chalmers University of Technology,\\
  412 96, Gothenburg, Sweden\\
{\tt daniel.persson@chalmers.se}\\[8pt]
{\bf Roberto Volpato}\\
Max-Planck-Institut f\"ur Gravitationsphysik,\\
Am M\"uhlenberg 1, 14476 Golm, Germany\\
{\tt roberto.volpato@aei.mpg.de}
\end{center}

\vspace{1cm}

\begin{abstract}

\noindent We study the second quantized version of the twisted twining genera of generalized  Mathieu moonshine, and prove that they give rise to Siegel modular forms with infinite product representations. Most of these forms are expected to have an interpretation as twisted partition functions counting 1/4 BPS dyons in type II superstring theory on K3$\times T^2$  or in heterotic CHL-models. We show that all these Siegel modular forms, independently of their possible physical interpretation, satisfy an ``S-duality'' transformation and a ``wall-crossing formula''. The latter reproduces all the eta-products of an older version  of generalized Mathieu moonshine proposed by Mason in the '90s. Surprisingly, some of the Siegel modular forms we find coincide with the multiplicative (Borcherds) lifts of Jacobi forms in umbral moonshine.

\end{abstract}
\newpage
\tableofcontents
%\newpage

\section{Introduction and Summary}

\noindent Mathieu moonshine \cite{Eguchi:2010ej,Cheng:2010pq,Gaberdiel:2010ch,Gaberdiel:2010ca,Eguchi:2010fg,Taormina:2010pf,Cheng:2011ay,Gaberdiel:2011fg,Taormina:2011rr,Volpato:2012qe,Gaberdiel:2012um,GannonMathieu,Taormina:2013jza,Taormina:2013mda,Gaberdiel:2013psa,Creutzig:2013mqa}  and its generalized version \cite{Gaberdiel:2012gf,Gaberdiel:2013nya} pertain to the association of a class of weak Jacobi forms $\phi_{g,h}(\tau, z)$, called twisted twining genera, to each commuting pair  of elements $(g,h)$ in $M_{24}$. It has been verified that these functions satisfy all the requirements of Norton's generalized moonshine conjectures \cite{Norton}; in particular they decompose into (projective) graded characters of the centralizer of $g$ in $M_{24}$. Many of these Jacobi forms arise as supersymmetric indices in certain non-linear sigma models with target space K3; in particular, $\phi_{e,e}$ is the K3 elliptic genus. The obvious idea that $M_{24}$ might be the symmetry group of some of these non-linear sigma models has been ruled out quite quickly \cite{Gaberdiel:2011fg, Volpato:2012qe, Gaberdiel:2012um} and the most recent works on the subject \cite{Cheng:2013kpa,Harrison:2013bya,Harvey:2013mda} seek for an explanation of Mathieu moonshine in the framework of K3 compactifications of full-fledged superstring theory rather than within conformal field theory.

In this paper we adopt a similar viewpoint, and construct the second-quantized\footnote{The terminology ``second-quantized'' originates in the work of Dijkgraaf, Moore, Verlinde, Verlinde \cite{Dijkgraaf:1996xw}.} version $\Psi_{g,h}$  of the twisted twining genera. From a mathematical perspective, the functions $\Psi_{g,h}$ are obtained from the Jacobi forms $\phi_{g,h}$ via a certain twisted equivariant version of Borcherds multiplicative lift. Physically, many of these functions can be interpreted as twisted supersymmetric indices counting 1/4 BPS states in type II superstring theory compactified on K3$\times T^2$ and in CHL models \cite{Sen:2009md,Sen:2010ts,Govindarajan:2010fu,Govindarajan:2011em}. A similar construction for the twining genera of the ordinary Mathieu moonshine was first proposed by Cheng  \cite{Cheng:2010pq}. 

We show that these multiplicative lifts  are Siegel modular forms, generalizing results for the case $(g,h)=(e,h)$ \cite{Cheng:2010pq,Cheng:2012uy,Raum12}. In some cases we can identify $\Psi_{g,h}$ with known Siegel modular forms which have previously appeared in the context of umbral moonshine \cite{Cheng:2012tq}.  Furthermore, they satisfy a ``wall-crossing formula'', which reproduces all of Mason's generalized eta-products $\eta_{g,h}$ for $M_{24}$ \cite{MasonM24,MasonEtas,MasonGen,MasonGelliptic}. This therefore establishes the link between the two existing versions of generalized moonshine for $M_{24}$. 

\subsection{Background}

The first hint of Mathieu moonshine was found by Eguchi, Ooguri and Tachikawa \cite{Eguchi:2010ej}, who noticed a connection between the the elliptic genus of K3 and the largest Mathieu group $M_{24}$. The elliptic genus of K3 $\phi_{K3}(\tau,z)$ is the unique (up to normalization) weak Jacobi form of weight $0$ and index $1$ and it can be defined as a supersymmetric index of non-linear sigma models with target space K3. It therefore is natural to consider a decomposition of $\phi_{K3}(\tau, z)$ into characters of the $\mathcal{N}=4$ superconformal algebra. The authors of \cite{Eguchi:2010ej} noticed that  the coefficients in this decomposition are sums of dimensions of irreducible representations of $M_{24}$. This observation led to the Mathieu moonshine conjecture: for each conjugacy class $[g]$ of $M_{24}$ there should exist a weak Jacobi form $\phi_{g} : \mathbb{H}\times \mathbb{C} \to \mathbb{C}$  such that the Fourier coefficients of  $\phi_g$ are the characters of these $M_{24}$-representations evaluated at $g$. In particular, for the identity element $e$ of $M_{24}$, $\phi_e(\tau, z)$ coincides with the elliptic genus of K3.   In subsequent work \cite{Cheng:2010pq,Gaberdiel:2010ch,Gaberdiel:2010ca,Eguchi:2010fg}, all the functions $\phi_g$, dubbed ``twining elliptic genera'', were found and substantial evidence was given for the validity of Mathieu moonshine, namely the existence of a graded $M_{24}$-module such that the $\phi_g$'s are its  graded characters. This conjecture has now been proven rigorously by Gannon \cite{GannonMathieu}.

The story outlined above is of course in close analogy with Monstrous moonshine \cite{ConwayNorton,FLM,Borcherds} which assigns modular functions $T_g : \mathbb{H}\to \mathbb{C}$ (McKay-Thompson series) with each conjugacy class $[g]$ of the Fischer-Griess Monster group $\mathbb{M}$, the largest of the finite sporadic simple groups. After the initial conjecture \cite{ConwayNorton}, Norton proposed his generalized  Monstrous moonshine \cite{Norton}, which assigns modular functions $f(g,h;\tau)$ to each commuting pair $g,h\in \mathbb{M}$. For fixed $g\in \mathbb{M}$, these generalized  moonshine functions should then have (possibly rational) Fourier coefficients that correspond to projective characters of the centralizer $C_\mathbb{M}(g)=\{k\in \mathbb{M}\, |\,  gk=kg\}$. Dixon-Ginsparg-Harvey subsequently suggested that Norton's functions $f(g,h;\tau)$ naturally arise in string theory as the path integral on a torus $\mathbb{C}/(\mathbb{Z}+\tau\mathbb{Z})$ with boundary conditions twisted by $(g, h)$ along the $(a,b)$-cycles. Although the full generalized  Monstrous moonshine conjecture is still open, considerable progress has been made toward proving it \cite{Tuite:1994ni,Dong:1997ea,Hohn,CarnahanI,CarnahanII,CarnahanIII,CarnahanIV}.

In earlier work \cite{Gaberdiel:2012gf,Gaberdiel:2013nya}, we gave substantial evidence that Norton's generalization also holds, with small modifications, for Mathieu moonshine. We found that for each commuting pair $g,h\in M_{24}$ there exists a weak Jacobi form $\phi_{g,h}(\tau, z)$, dubbed \emph{twisted twining genus}, whose Fourier coefficients are characters of a projective representation of $C_{M_{24}}(g)$. Inspired by orbifolds of holomorphic CFT's  \cite{Dijkgraaf:1989pz,Roche:1990hs,Bantay:1990yr,Coste:2000tq}, we further showed that the modular properties of the functions $\phi_{g,h}$ are controlled by a cohomology class $[\alpha]$ in the third cohomology group $H^3(M_{24}, U(1))$, as was anticipated in \cite{Gannontalk} (see also \cite{GannonMathieu}).

A different kind of generalized  moonshine for $M_{24}$ had in fact already been established in old work by Mason \cite{MasonM24,MasonGen,MasonGelliptic}. Mason associated to each commuting pair $g,h$ in $M_{24}$ a so called multiplicative eta product $\eta_{g,h}(\tau)$, based on the action of $M_{24}$ on $24$ chiral free bosons. This leads to the natural question: \emph{Is there a relation between the recently discovered Mathieu moonshine, pertaining to weak Jacobi forms, and Mason's $M_{24}$-moonshine involving eta-products?}

For the special case of commuting pairs $(g,h)=(e,h)$, where $e$ is the identity element of $M_{24}$, this was given an affirmative answer by Cheng in \cite{Cheng:2010pq}. Cheng's idea was to generalize the known fact that the elliptic genus of K3, $\phi_{e}(\tau, z)$, exhibits an exponential Borcherds lift to the unique weight 10 Siegel modular form for $Sp(4;\mathbb{Z})$,  referred to as the ``Igusa cusp form'' and commonly denoted by $\Phi_{10}$. She proposed that to all twining genera $\phi_h(\tau, z)$ one should have a similar lift of the form
\be 
\Phi_h(\sigma,\tau,  z)=pqy\prod_{(n,m,\ell)>0} \exp\left[-\sum_{k=1}^{\infty}\frac{c_{h^k}(4mn-\ell^2)}{k} (p^mq^ny^\ell)^k\right],
\label{twininglift}
\ee
where $q=e^{2\pi i \tau}, y=e^{2\pi i z}, p=e^{2\pi i \sigma}$, and $c_{g}$ denotes the Fourier coefficients of $\phi_g(\tau, z)$:
\be
\phi_g(\tau, z)=\sum_{m\geq 0, \ell\in \mathbb{Z}} c_g(m,\ell) q^m y^{\ell}.
\ee
It was conjectured in \cite{Cheng:2012uy} that $\Phi_h$  is automorphic with respect to a subgroup $\Gamma_h^{(2)}$ of $Sp(4;\mathbb{Z})$. This  was then proven by Raum \cite{Raum12} for most of the conjugacy classes $[h]\subset M_{24}$.

A central point for us is that $\Phi_h(\sigma,\tau,  z)$ has a double pole at $z=0$ and in the limit one 
finds
\be 
\displaystyle \lim_{z \rightarrow 0}\frac{\Phi_h(\sigma,\tau,  z)}{(2\pi i z)^2}=\eta_h(\tau)\, \eta_h(\sigma),
\label{zlimit}
\ee
where $\eta_h=\eta_{e,h}$ are the eta-products of Mason's $M_{24}$-moonshine \cite{MasonM24}. Hence, the process of taking the multiplicative lift of $\phi_h$ followed by studying the limiting behavior as $z\rightarrow 0$ provides a link between the two moonshines. 

\subsection{Summary of results}

In this paper we answer the question above in the general case of commuting pairs $(g,h)$ in $M_{24}$. In other words, we establish a link between the generalized Mathieu moonshine proposed in \cite{Gaberdiel:2012gf,Gaberdiel:2013nya} and the generalized eta-products of Mason. To describe our results, we recall the notion of ``second quantized elliptic genus" as defined in \cite{Dijkgraaf:1996xw}. Suppose $\phi_X(\tau, z)$ is the elliptic genus of some Calabi-Yau manifold $X$. Then one defines the second quantized genus $\Psi_X$ as the generating function of the elliptic genus $\phi_{S^n X}$ for the $n$:th symmetric product $S^nX$: $\Psi_X=\sum_{n\geq 0}p^n \phi_{S^n X}$. In \cite{Dijkgraaf:1996xw}, the following remarkable formula is proved
\be
\Psi_X(\sigma,\tau,  z)=\exp\Big[\sum_{n=1}^{\infty} p^n T_n \phi_X(\tau, z)\Big], 
\label{DMVVformula}
\ee
where $p=e^{2\pi i \sigma}$, and $T_n$  is the standard Hecke operator acting on Jacobi forms. Subsequently, Gritsenko showed \cite{Gritsenko:1999fk} that by multiplying the inverse $\Psi_X^{-1} $ by a certain factor $A_X(\sigma,\tau,  z)$ (the ``Hodge anomaly'') one obtains a Siegel modular form 
\be 
\Phi_X(\sigma,\tau, z)=\frac{A_X(\sigma,\tau, z)}{\Psi_X(\sigma,\tau, z)},
\ee
of weight $c_X(0,0)/2$ for a subgroup $\Gamma^{(2)}_X\subset Sp(4;\mathbb{Z})$, where $c_X(m,n)$ are the Fourier coefficients of $\phi_X$. 

Inspired by these results, we define for each commuting pair $(g,h)$ in $M_{24}$ the associated \emph{second quantized twisted twining genus} $\Psi_{g,h}$ via a generalization of the formula (\ref{DMVVformula}):
\be
\Psi_{g,h}(\sigma,\tau, z) = \exp\Big[\sum_{n=1}^{\infty} p^n \mathcal{T}^{\alpha}_n \phi_{g,h}(\tau, z)\Big],
\ee
where $\mathcal{T}^{\alpha}_n$  is now a certain twisted equivariant Hecke operator  (for the precise definition see section (\ref{s:tHecke})) which reduces to the $T_n$ in (\ref{DMVVformula}) in the special case $(g,h)=(e,e)$. Notice that, when $g=e$, the presence of  a non trivial 3-cocycle $\alpha$ governing the modular properties of the twisted twining genera  can be safely ignored (see \cite{Gaberdiel:2012gf} and section \ref{sec_twist}). This leads to the simpler definition of the multiplicative lift adopted in \cite{Cheng:2012uy}. However, for the  general case considered in this paper, the technical subtleties associated with non-trivial 3-cocycles cannot be avoided.

Multiplying the second quantized twisted twining genera by a correction $A_{g,h}(\sigma,\tau, z)$ (see eq. (\ref{SiegelAutoCorr}) for the explicit form of $A_{g,h}$) we obtain a new class of functions
\be
\Phi_{g,h}(\sigma,\tau, z)=\frac{A_{g,h}(\sigma,\tau, z)}{\Psi_{g,h}(\sigma, \tau, z)},
\label{DefSieg}
\ee
and the main result of this paper is:

\vspace{.2cm} 

\noindent {\bf Theorem.} {\it The functions $\Phi_{g,h}(\sigma,\tau, z)$ defined by (\ref{DefSieg}) are meromorphic Siegel modular forms for certain discrete subgroups $\Gamma^{(2)}_{g,h}\subset Sp(4;\mathbb{R})$, such that $\Gamma_{g,h}^{(2)}\cap Sp(4,\ZZ)$ is a congruence subgroup of $Sp(4,\ZZ)$. %Furthermore, $\Gamma_{g,h}$ containing the invariance subgroups $\Gamma_{g,h}\subset SL(2,\mathbb{Z})$ of the twisted twining genera $\phi_{g,h}$.
}
\vspace{.2cm}Ê

\noindent For a more precise formulation, see Theorem 5.3 in section \ref{sec_modular}. 
\vspace{.2cm}Ê

\noindent The most noticeable new automorphic property is the `S-duality' transformation
\be\label{Sduality} \Phi_{g,h}(\sigma,\tau,z)=\Phi_{g,h'}(\frac{\tau}{N\lambda},N\lambda\sigma,z)\ ,
\ee where $N$ is the order of $g$ and $\lambda$ is the length of the shortest cycle of $g$ in the $24$-dimensional permutation representation. Notice that $h,h'\in C_{M_{24}}(g)$ are not necessarily in the same conjugacy class. This is related with the `relabeling' phenomenon described in \cite{Gaberdiel:2012gf}. In particular, as described in section \ref{s:orbirelabel}, when $g$ is an element of an $M_{23}$ subgroup, then $h'$ is the symmetry induced by $h$ in the $g$-orbifold theory. We proved \eqref{Sduality} for at least one commuting pair of generators $(g,h)$ in each group $\langle g,h\rangle$, but we conjecture that similar relations hold for all commuting pairs.

\begin{figure}[h!]
\begin{center}
\includegraphics[width=13cm]{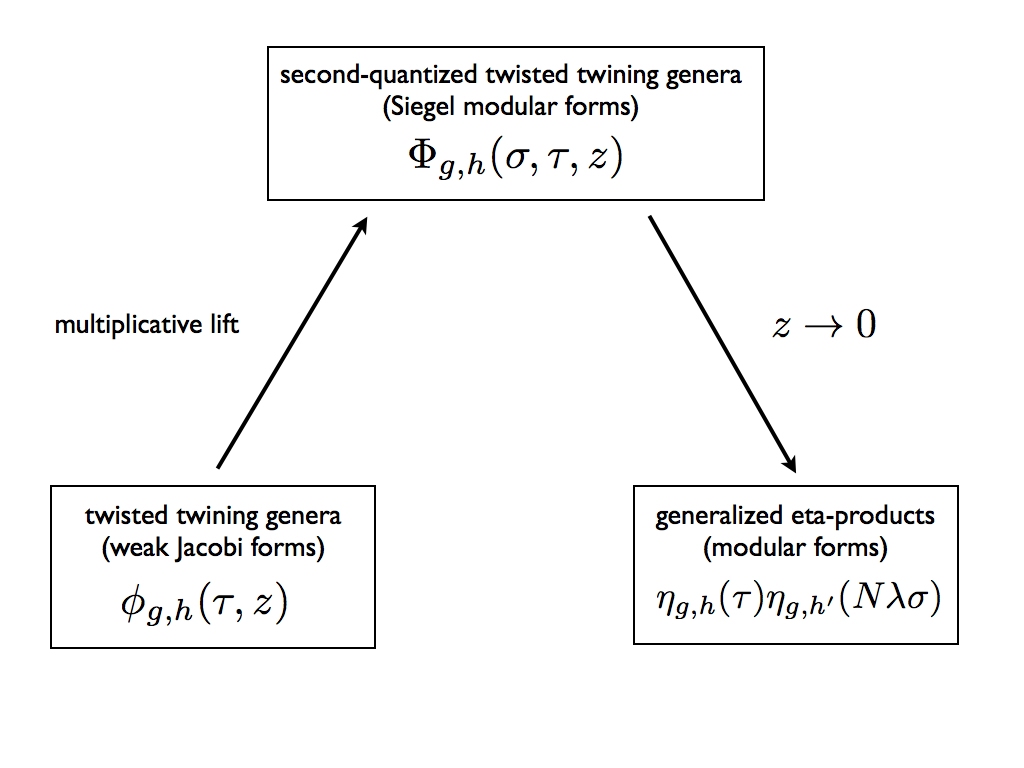}
\label{triangle}
\caption{Pictorial description of the relation between generalized Mathieu moonshine, involving the twisted twining genera $\phi_{g,h}$, and Mason's generalized eta-products $\eta_{g,h}$. Starting from $\phi_{g,h}$ one constructs a second-quantized twisted twining genus $\Psi_{g,h}$, whose reciprocal becomes a Siegel modular form $\Phi_{g,h}$ after multiplying by a factor $A_{g,h}$. This is the content of the multiplicative (automorphic) lift indicated in the leftmost arrow. At the level of the Siegel modular form $\Phi_{g,h}$ one then takes the limit $z\to 0$ which produces the generalized eta-products $\eta_{g, h}(\tau)\eta_{g, h'}(N\lambda\sigma)$. Here, $h'$ is possibly in a different class from $h$, as a consequence of the `relabeling phenomenon' described in \cite{Gaberdiel:2012gf}. See section \ref{sec_connecting} for more details. }
\end{center}
\end{figure}

In section \ref{sec_MasonConnection} we prove the following ``wall-crossing formula''
\be \lim_{z\to 0}\frac{\Phi_{g,h}(\sigma,\tau,z)}{(2\pi iz)^2}=\eta_{g,h}(\tau)\eta_{g,h'}(N\lambda\sigma),
\ee
which generalizes (\ref{zlimit}). This shows that the limit $z\to 0$ of $\Phi_{g,h}(\sigma, \tau, z)$ reproduces all of Mason's generalized eta-products $\eta_{g,h}$  \cite{MasonGen,MasonEtas}, thus providing the desired link between the two moonshines. See figure \ref{triangle} for a pictorial overview of the relation between the various modular objects. 

When $(g,h)$ lies in the conjugacy classes $(2A, 2A), (3A, 3A)$ and $(4B, 4B)$ we can identify $\Phi_{g,h}$ with known Siegel modular forms:
\ba
(2A, 2A) &:& \Phi_{g,h} =(\Delta_2)^2\, \, \, \, =\Phi^{(3)},
\nn \\ 
(3A, 3A) &:& \Phi_{g,h}= (\Delta_1)^2\, \, \, \, = \Phi^{(4)},  
\nn \\ 
(4B, 4B) &:& \Phi_{g,h}= (\Delta_{1/2})^2=\Phi^{(5)}, 
\label{Siegel}
\ea
where the first equality identifies them with the squares of the weight $k$ Siegel modular forms $\Delta_k$ obtained by Gritsenko and Nikulin \cite{GritsenkoNikulin2}, and the second equality with the  umbral Siegel modular forms $\Phi^{(\ell)}$ of lambency $\ell=3,4,5$ \cite{Cheng:2012tq}. Thus we obtain a surprising connection between generalized Mathieu moonshine and umbral moonshine, which clearly deserves further investigation.\footnote{We thank Miranda Cheng for a discussion that pointed us in this  direction.}

Finally, we would like to comment further on the possible physical interpretation of the functions $\Phi_{g,h}$. For $\Phi_{e,e}=\Phi_{10}$, it is of course well-known that the reciprocal $1/\Phi_{10}$ is the generating function of 1/4-BPS dyons in heterotic string theory on $T^6$ \cite{Dijkgraaf:1996it}, and in this case the decomposition (\ref{zlimit}) becomes $1/ \Phi_{10}\sim \eta(\tau)^{-24}\eta(\rho)^{-24}$, where each factor on the right hand side is identified with the partition function of 1/2 BPS-states. More generally, $1/\Phi_{e,h}$  should correspond to the partition function of ``twisted dyons'', and the limit (\ref{zlimit}) reflects the wall-crossing phenomenon when a 1/4 BPS-state splits into two 1/2 BPS-states. Similarly we expect that in general $1/\Phi_{g,h}$ should be the generating function of certain ``twisted dyons'' in CHL orbifolds $T^6/\mathbb{Z}_N$ of heterotic string theory \cite{Sen:2009md,Sen:2010ts,Govindarajan:2010fu,Govindarajan:2011em}. In particular, this should be true whenever $g,h$ can be interpreted as a pair of commuting symmetries of some K3 surface and the orbifold by $g$ is consistent. Given the results of \cite{Gaberdiel:2011fg}, the first condition amounts to the group $\langle g,h\rangle$ having at least four orbits in the standard $24$-dimensional permutation representation of $M_{24}$, while the second condition simply requires $g$ to belong to some $M_{23}$ subgroup of $M_{24}$. From the tables in appendix \ref{app:tables}, it is easy to check that the first condition is satisfied by all $34$ groups except the groups numbered 22, 30, 31, 32. As for the remaining groups, the second condition is satisfied by at least one generator $g$, except for groups 4, 5, 6, 13, 14.

\subsection{Outline}

Our paper is organized as follows. In section \ref{sec_ttgenera} we recall some relevant facts about holomorphic orbifolds, focussing on properties of the associated twisted twinining partition functions. We explain the important role played by the third cohomology group $H^3(G, U(1))$, where $G$ is a finite automorphism group of the holomorphic CFT. In particular, this cohomology group controls the modular properties of the twisted twining partition functions. After discussing these properties in a general context we restrict to the relevant case of $G=M_{24}$ for which we recall the main results of \cite{Gaberdiel:2012gf,Gaberdiel:2013nya}. Section \ref{sec_SymProd} introduces some aspects of symmetric products of holomorphic CFT's and the connection with Hecke operators. In section \ref{s:tHecke} we then introduce the key notion of twisted equivariant Hecke operators that will play a crucial role in what follows. In section \ref{sec_SecQuant} we use the twisted equivariant Hecke operators to define second-quantized twisted twining genera $\Psi_{g,h}$ and we compute their infinite product expansions. The connection with Mason's eta-products is then made in section \ref{sec_MasonConnection} by evaluating $\Psi_{g,h}$ in the limit $z\to 0$. The modular properties of the second-quantized twisted twining genera are analyzed in section \ref{sec_modular}. Finally, in the concluding section \ref{sec_conclusions} we summarize our results and give some ideas for future research. Various background material, technical calculations and tables have been relegated to the appendices.

\section{Twisted twining partition functions}\label{sec_ttgenera}
\noindent In this section we review the main properties of the twisted twining partition functions associated with the orbifold of a conformal field theory (CFT)  $\mathcal{C}$  by a finite group $G$  of symmetries \cite{Dixon:1986jc,Dolan:1989vr,Dong:1997ea}. When $\mathcal{C}$ is a bosonic CFT, these are modular functions $Z_{g, h}(\tau)$ on the upper-half plane $\mathbb{H}$, associated with each commuting pair $g,h\in G$. For superconformal field theories the analogous functions include the twisted twining genera $\phi_{g,h}(\tau, z)$, which are Jacobi forms on $\mathbb{H} \times \mathbb{C}$. Our main interest is the case when $\mathcal{C}$ is a putative  $\mathcal{N}=4$ super-vertex algebra at central charge $c=6$ with automorphism group $G=M_{24}$, but we shall keep the discussion general whenever possible. We emphasize that the existence of such a super-vertex algebra is not established and its definition seems to be problematic. Despite these issues, the properties satisfied by the twisted-twining genera are analogous to the ones expected for a non-linear sigma model with target space K3, although no such model has a group of symmetries containing $M_{24}$.
Finally, we will provide an abstract formulation of Mathieu moonshine \cite{Eguchi:2010ej,Cheng:2010pq,Gaberdiel:2010ch,Gaberdiel:2010ca,Eguchi:2010fg,GannonMathieu} and of its generalized version \cite{Gaberdiel:2012gf,Gaberdiel:2013nya}.

  \subsection{Definition and basic properties}

Consider a two-dimensional conformal field theory $\C$ and let $G$ be its symmetry group, i.e. the group of linear automorphisms of its space of states $\H$ that preserves the OPE of the corresponding fields and fixes (at least) the left- and right-moving Virasoro algebra. Given such data, one can construct a new conformal field theory by considering the orbifold of $\C$ by (a subgroup of) $G$. The first step in the orbifold construction consists in introducing, for each $g\in G$, the $g$-twisted fields $\Xi_g$, generating a $g$-twisted space of states $\H_g$. The defining property of the $g$-twisted fields is that every field of $\C$ has a non-trivial monodromy $g$ when it is moved along a (sufficiently small) closed path encircling the twisted field $\Xi_g$. The orbifold theory is defined by including all twisted sectors in the spectrum  and then restricting to the $G$-invariant part. The orbifold theory is a consistent CFT provided certain conditions (in particular, the level-matching condition), assuring the locality of the OPE of twisted fields, are satisfied.

\bigskip

Even if the consistency conditions for the orbifold theory are not satisfied, it makes sense to consider the $g$-twisted sector $\H_g$ as a module over the Virasoro algebra (or more generally, over the $G$-invariant part of the chiral algebra of $\C$). Let us consider the case where $\C$ is a holomorphic CFT, so that it coincides with its chiral algebra, which is a self-dual vertex operator algebra. Then, under suitable assumptions, there exists a unique irreducible $g$-twisted sector $\mathcal{H}_g$ for each $g\in G$ \cite{Dong:1997ea}.
The action of a generic element $h\in G$ on $\C$ induces a linear map
\be \rho_g(h):\H_g\to \H_{h^{-1}gh}\ , 
\ee from the $g$-twisted to the $h^{-1}gh$-twisted sector. In particular, when $h$ commutes with $g$, it defines an endomorphism of $\H_g$, so that $\H_g$ carries a representation $\rho_g$ of the centralizer \be C_G(g):=\{h\in G: gh=hg\}\ ,\ee
of $g$ in $G$.  
It should be noticed that the representation $\rho_g$ of $C_G(g)$ is in general only projective. This fact will play a crucial role in what follows. 
 
Let us define a $g$-twisted $h$-twining partition function on the upper half-plane $\HH_+$ by
\be\label{Zgh} Z_{g,h}(\tau)=\Tr_{\H_g}(\rho_g(h) q^{L_0-\frac{c}{24}})\ , \qquad q:=e^{2\pi i\tau},\quad g,h\in G,\ gh=hg\ .
\ee When $g$ and $h$ are the identity, this reduces to the usual torus partition function for $\C$. 
On general physical grounds, one expects the twisted twining partition function $Z_{g,h}$ to be given by a path-integral on an elliptic curve $E_\tau\cong\CC/(\ZZ+\tau\ZZ)$ with modular parameter $\tau$, where the fields have monodromies $g$ and $h$ around the generators $-1,\tau$ of the first homology group $H_1(E_\tau,\ZZ)\cong \ZZ+\tau\ZZ$.
Furthermore, each twisted twining partition function $Z_{g,h}$ is a holomorphic function of the modular parameter $\tau$ and is expected to be a modular function under the subgroup $\Gamma_{g,h}\subset SL(2,\ZZ)$, that preserves the monodromies of the fields around the non-trivial cycles of the torus.

\bigskip

Analogous properties are expected for the twisted twining genera in superconformal field theories with an extended (at least $\N=(2,2)$) superconformal algebra and a group $G$ of symmetries preserving such an algebra. The twisted twining genera are defined as traces
\be\label{twtwgendef} \phi_{g,h}(\tau,z)= \Tr_{\H^{RR}_g}(h q^{L_0-\frac{c}{24}}\bar{q}^{\tilde{L}_0-\frac{\tilde{c}}{24}}(-1)^{F+\tilde{F}}y^{J_0})\ ,\qquad q:=e^{2\pi i\tau},\ y:=e^{2\pi iz},
\ee over the Ramond-Ramond (RR) $g$-twisted sector $\H^{RR}_g$. Here $J_0$ is the zero mode of a $\mathfrak{u}(1)$ current algebra contained in the left $\N=2$ superconformal algebra (normalized so that the charges are integral) and $(-1)^{F+\tilde{F}}$ is the total fermion number. This trace is computed by a path-integral with the same $g$- and $h$-twisted periodicity conditions for both bosonic and fermionic fields, with the insertion of an operator $y^{J_0}$. Although these conformal field  theories are not chiral, the twisted twining partition functions are holomorphic with respect to both $\tau$ and   $z$ and are expected to be Jacobi forms of weight $0$ and index $\frac{c}{6}$ under the same subgroup $\Gamma_{g,h}\subset SL(2,\ZZ)$ as above.

\subsection{Modular properties and the $\alpha$-twist}
\label{sec_twist}
We mentioned above that for a holomorphic CFT $\mathcal{C}$ with automorphism group $G$, the twisted sectors $\mathcal{H}_g$, carry projective representations $\rho_g$ of the centralizer $C_G(g)$. This implies that $\rho_g$ satisfies
\be\label{grouplaw}
\rho_g(h)\rho_g(h')=c_g(h, h') \rho_g(hh'), \qquad h,h'\in C_G(g)\ ,
\ee
where $c_g(h, h')$ is a phase, or, more precisely, a 2-cocycle representing a class $[c_g]$ in $H^2(C_G(g), U(1))$. As a consequence, the twisted twining partition function $Z_{g, h}$ is not an honest class function on $G$, but rather satisfies 
\be
Z_{g,h}(\tau)=\frac{c_g(h,k)}{c_g(k, k^{-1}h k)}Z_{g , k^{-1} hk}(\tau),\qquad k\in C_G(g),
\label{conjrestr}
\ee
which puts strong constraints on $Z_{g, h}$ (see appendix \ref{app:modtt} and \cite{Gaberdiel:2013nya}). The same formula also applies to the twisted twining elliptic genera $\phi_{g,h}$. 

The role of the cohomology group $H^2(C_G(g), U(1))$ can in fact be derived from a more fundamental property of $\mathcal{C}$, namely that consistent orbifolds of 
a holomorphic CFT are classified by the third cohomology $H^3(G, U(1))$. Concretely, this implies there exists a $U(1)$-valued 3-cocycle $\alpha(g,h,k)$, with $(g,h,k)\in G\times G\times G$, 
representing a class $[\alpha]$ in $H^3(G, U(1))$, which in turn determines the class $[c_g]\in H^2(C_G(g), U(1))$ via the formula \cite{Bantay:1990yr,Coste:2000tq}
\be\label{ciggi} c_g(x,y):=\frac{\alpha(g,x,y)\alpha(x,y,(xy)^{-1}g (xy))}{\alpha(x,x^{-1} g x,y)}\ ,\qquad g,x,y\in G\ .
\ee
Thus, $\alpha$ characterizes the projective representations $\rho_g$. The 3-cocycle $\alpha$ is normalized, i.e. $\alpha(g,h,k)=1$ whenever at least one of its arguments $g,h,k$ is the identity. The 
cohomology group $H^3(G, U(1))$ also controls the modular properties of the twisted twining partition functions $Z_{g,h}$, in the sense that under a modular transformation $(\begin{smallmatrix} a & b \\ c & d  \end{smallmatrix})\in SL(2,\mathbb{Z})$ they satisfy 
\be
\epsilon_{g,h}(\begin{smallmatrix} a & b \\ c & d  \end{smallmatrix})Z_{g^dh^{-c},g^{-b}h^a}\left(\frac{a\tau+b}{c\tau+d}\right)= \,
Z_{g, h}(\tau),
\label{modtrans}
\ee
where $\epsilon_{g,h}(\begin{smallmatrix} a & b \\ c & d  \end{smallmatrix})$ is a $U(1)$-valued multiplier system\footnote{This was called $\chi_{g,h}$ in \cite{Gaberdiel:2013nya}.} for $SL(2,\mathbb{Z})$ that can be explicitly constructed from the 3-cocycle $\alpha(g,h,k)$ (see appendix \ref{app:modtt}). Furthermore, \eqref{conjrestr} generalizes to
\be
Z_{g,h}(\tau)=\frac{c_g(h,k)}{c_g(k, k^{-1}h k)}Z_{k^{-1}gk , k^{-1} hk}(\tau),\qquad k\in G,
\label{conj}
\ee where conjugation is taken within the whole group $G$.
In \cite{Gaberdiel:2013nya,Gaberdiel:2012gf} these properties were used to constrain and determine all of the twisted twining elliptic genera $\phi_{g,h}$ for $g,h$ a commuting pair in $ M_{24}$. 

Given the rather complicated modular action in (\ref{modtrans}) it turns out to be very convenient to introduce a certain slash operator which combines the action of $SL(2,\mathbb{Z})$ on $(\tau; g,h)$ with the conjugation by $G$ in (\ref{conj}). In addition this must incorporate the multiplier induced by the cocycle twist. To define the slash operator we must first define the following twisted modular action on commuting pairs of elements $(g,h)$ in $G$:
\be (\gamma,k): (g,h)\mapsto(kgk^{-1},khk^{-1})\gamma^{-1}=(kg^{d}h^{-c}k^{-1},kg^{-b}h^ak^{-1})\ .
\ee Here, $(\gamma,k)=\left(\big(\begin{smallmatrix} a & b \\ c & d \end{smallmatrix}\big), k\right)\in SL(2,\ZZ)\times G$.

When acting on twisted twining partition functions $Z_{g,h}$ the $\alpha$-twisted equivariant slash operator is defined as follows
\be\label{slashalphaZgh} Z_{g,h}(\tau)\rvert_\alpha{(\gamma,k)}:=\epsilon_{g,h}(\gamma,k) Z_{(\gamma,k)\cdot(g,h)}(\gamma\cdot\tau)\ ,\qquad (\gamma,k)\in SL(2,\ZZ)\times G\ ,
\ee 
where $\epsilon_{g,h}(\gamma,k)\in U(1)$ is a phase which depends on the choice of  3-cocycle $\alpha$ representing a class $[\alpha]\in H^3(G,U(1))$ (see appendix \ref{sec_genperm} for more details). When $k=e$ this phase reduces to the phase $\epsilon_{g,h}(\gamma, e)=\epsilon_{g,h}(\gamma)$ in (\ref{modtrans}), while for $\gamma =1$ we have that $\epsilon_{g,h}(1,k)$ is identified with the phase in (\ref{conj}).\footnote{This phase was called $\xi_{g,h}$ in \cite{Gaberdiel:2013nya}.}

The class $[\alpha]\in H^3(G, U(1))$ is part of the defining data of a holomorphic CFT $\mathcal{C}$ with finite automorphism group $G$. In general the twisted twining partition functions $Z_{g,h}(\tau)$ associated with $\mathcal{C}$ are invariant under the $\alpha$-twisted slash operator
\be\label{slashalphainv} Z_{g,h}(\tau)\rvert_\alpha{(\gamma,k)}=Z_{g,h}(\tau).\ee 
Different choices of representative $\alpha$ in the class $[\alpha]$  are related to  each other by rescalings of $Z_{g,h}(\tau)$ by overall $(g,h)$-dependent phases. More precisely, if two normalized $3$-cocycles $\alpha$ and $\alpha'$ differ by a $3$-coboundary $\partial\beta$, i.e.
\be\label{alphashift} \alpha(g_1, g_2, g_3)=\alpha'(g_1, g_2, g_3) \,
\frac{\beta(g_1, g_2g_3)\beta(g_2, g_3)}{\beta(g_1 g_2, g_3)\beta(g_1, g_2)}\ ,\ee
for some $\beta:G\times G\to U(1)$, with $\beta(e,g)=\beta(g,e)=1$ for all $g\in G$, then the corresponding twisted twining  partition functions $Z_{g,h}$ and $Z'_{g,h}$ are related by
\be\label{Zghredef} Z'_{g,h}(\tau,z)= \frac{\beta(g,h)}{\beta(h,g)} Z_{g,h}(\tau,z)\ .
\ee %Note that equation \eqref{slashalphainv} reduce to \eqref{slashinv} when $\alpha$ is trivial. 

We also define the $\alpha$-twisted  slash operators on the twisted twining genera $\phi_{g,h}$ by the natural generalization of (\ref{slashalphaZgh}) to Jacobi forms of weight 0 and index 1:
\be\label{JacSlashAlpha} \phi_{g,h}(\tau,z)\rvert_\alpha\bigl(\gamma ,k\bigr):= \epsilon_{g,h}(\gamma ,k)e^{-\frac{2\pi i m cz^2}{c\tau+d}}\phi_{ k\,g^dh^{-c}\,k^{-1},\;k\,g^{-b}h^a\,k^{-1}}(\gamma\cdot \tau,\frac{z}{c\tau+d}).\ee

\subsection{(Generalized) Mathieu moonshine}\label{sec_ttprop}

\noindent The elliptic genus of K3 is defined in physics as a refined partition function \be \phi_{K3}(\tau,z)= \Tr_{\H}(q^{L_0-\frac{c}{24}}\bar{q}^{\tilde{L}_0-\frac{\tilde{c}}{24}}(-1)^{F+\tilde{F}}y^{J_0})\ ,\qquad q:=e^{2\pi i\tau},\ y:=e^{2\pi iz},
\ee of certain conformal field theories with $\N=(4,4)$ superconformal algebra at central charge $c=6$, namely non-linear sigma models with target space K3. The only states contributing to this partition function are the Ramond-Ramond right-moving ground states, i.e. with $\bar L_0-\frac{\bar c}{24}=0$. 
%More generally, the twisted twining genera, defined as in eq.\eqref{twtwgendef}, are obtained by `twisting' and `twining' this refined partition function by some symmetries $g,h$ of the K3 model that preserve the $\N=(4,4)$ algebra. 

In \cite{Eguchi:2010ej}, Eguchi, Ooguri and Tachikawa conjectured a relationship (subsequently dubbed Mathieu moonshine) between the elliptic genus of K3 and the largest Mathieu group $M_{24}$.
It is useful to formulate the Mathieu moonshine conjecture (now a theorem, thanks to \cite{GannonMathieu}) in terms of an abstract representation $\H$ of the (holomorphic) $\N=4$ superconformal algebra at central charge $c=6$, which is also a module for the Mathieu group $M_{24}$. Heuristically, $\H$ can be interpreted as the spectrum of R-R right-moving ground states in a generic K3 model, although it is known that no such model has $M_{24}$ as its symmetry group \cite{Gaberdiel:2011fg,Gaberdiel:2012um}. The module $\H$ also admits a $\ZZ_2$-grading by a `right-moving fermion number' $(-1)^{\bar F}$, which is preserved both by the $\N=4$ superconformal algebra and by the action of $M_{24}$.
The precise statement of the conjecture is as follows.

\medskip

\noindent {\bf Theorem (Mathieu moonshine).} {\sl There exists unitary Ramond representation $\H$ of the $\N=4$ superconformal algebra at central charge $c=6$, $\ZZ_2$-graded by the `right-moving fermion number' $(-1)^{\bar F}$, that carries a non-trivial action of the Mathieu $M_{24}$ commuting with both the $\N=4$ algebra and the $\ZZ_2$-grading. Furthermore, for each $g\in M_{24}$, the twining genus
\be\label{twingen1} \phi_g(\tau,z):=\Tr_{\H}(g\,q^{L_0-\frac{c}{24}}y^{J_0^3}(-1)^{F+\bar F})\ ,\qquad g\in M_{24},
\ee ($q:=e^{2\pi i\tau}$, $y:=e^{2\pi iz}$) is a weak Jacobi form of weight $0$ and index $1$ (possibly with multiplier) under $\Gamma_0(N)$, where $N$ is the order of $g$. In particular, $\phi_e$ is the elliptic genus of K3.}

\medskip

A list of Jacobi forms $\phi_g$, $g\in M_{24}$, satisfying the expected modular properties was proposed in \cite{Cheng:2010pq}--\cite{Eguchi:2010fg} (see table \ref{t:cycl} in appendix \ref{app:tables} for the explicit expressions) and the existence of the corresponding module $\H$ was proved in \cite{GannonMathieu}. The module $\H$ is not uniquely determined by this description, since one can always adjoin to $\H$ a pair of isomorphic representations of the $\N=4$ algebra and $M_{24}$ with opposite right-moving fermion number, so that they do not contribute to any twisted twining genus. However, there is a `minimal' module that contains no such pair of representations and this is uniquely determined by the twining genera $\phi_g$. Remarkably, in this minimal module, the only $\N=4$ representations with negative $(-1)^{\bar F}$ are the BPS ones. It is believed that the module $\H$ and the Jacobi forms $\phi_g$ of \cite{Cheng:2010pq}--\cite{Eguchi:2010fg} are the unique solutions of the Mathieu moonshine conjecture, although this has not been proved yet.

\medskip

In the same spirit, generalized Mathieu moonshine can be expressed in terms of the existence of twisted modules $\H_g$ and twisted twining genera $\phi_{g,h}$.

\medskip

\noindent {\bf Conjecture (Generalized Mathieu moonshine).} {\sl For each $g\in M_{24}$, there exists a $\ZZ_2$-graded unitary Ramond representation $\H_g$ of the $\N=4$ algebra at central charge $c=6$, that carries a projective representation $\rho_g\colon C_{M_{24}}(g)\to GL(\H_g)$ of the centralizer of $g$ in $M_{24}$, with group law \be \rho_{g}(h)\rho_{g}(k)=c_{g}(h,k)\rho_{g}(hk)\ ,\qquad h,k\in C_{M_{24}}(g)\ ,
\ee where $c_{g}(h,k)$ is defined by \eqref{ciggi} in terms of a normalized 3-cocyle $\alpha$ representing a primitive class in $H^3(M_{24},U(1))\cong\ZZ_{12}$. The operators $\rho_{g}(h)$ commute with the $\N=4$ algebra and with the right-moving fermion number $(-1)^{\bar F}$, and we have
\be \rho_{g}(g)=e^{2\pi i (L_0-\frac{c}{24})}\ .
\ee
Furthermore, for each pair of commuting elements $g,h\in M_{24}$, the function
\be\label{twisttwingen} \phi_{g,h}(\tau,z):=\Tr_{\H_g}(\rho_g(h)q^{L_0-\frac{c}{24}}y^{J_0^3}(-1)^{F+\bar F})\ ,
\ee satisfies the following properties:
\begin{enumerate}
\item[(A)] For $g=e$, where $e$ is the identity element of $M_{24}$, the functions $\phi_{e,h}$ coincide with the 
twining genera $\phi_h$ constructed in \cite{Cheng:2010pq}--\cite{Eguchi:2010fg}. In particular, $\phi_{e,e}$ is the elliptic genus of $K3$. 
\item[(B)] Elliptic properties:
\begin{align} 
&\phi_{g,h}(\tau,z+ \ell \tau + \ell') = e^{-2 \pi i(\ell^2 \tau+ 2 \ell z)}\, \phi_{g,h}(\tau,z) &
\ell,\ell'\in \ZZ\ .%\\
%&\phi_{g,h}\Bigl(\frac{a \tau + b}{c \tau + d} , \frac{z}{c \tau + d}\Bigr) =
%\chi_{g,h}(\begin{smallmatrix} a & b \\ c & d  \end{smallmatrix})\,
%e^{ 2 \pi i m \frac{c z^2}{c \tau + d} } \, \phi_{g^ah^c,g^bh^d}(\tau,z) \ ,
%& \hspace*{-0.9cm} (\begin{smallmatrix} a & b \\ c & d  \end{smallmatrix} )\in SL(2,\ZZ) 
%\label{modcond}
\end{align} 
\item[(C)] Invariance under the $\alpha$-twisted slash operator (see eq.\eqref{JacSlashAlpha})
\be \label{JacSlashInv} 
\phi_{g,h}(\tau,z)\rvert_\alpha (\gamma,k)=\phi_{g,h}(\tau,z)\ ,\qquad
(\gamma,k)\in SL(2,\ZZ)\times M_{24}\ . 
\ee
In particular, each $\phi_{g,h}$ is an even weak Jacobi form of weight $0$ and index $1$ under a group $\Gamma_{g,h}\subseteq PSL(2,\ZZ)$, with multiplier $\chi_{g,h}$ depending on $\alpha$.
\end{enumerate}}

\medskip
 
These conditions were presented in  \cite{Gaberdiel:2012gf} where all the functions $\phi_{g,h}$ were explicitly found, and strong numerical evidence was given that they decompose into projective characters of $C_{M_{24}}(g)$ with respect to the $\mathcal{N}=4$ algebra. It was also proven in  \cite{Gaberdiel:2012gf} that there exists a unique cohomology class $[\alpha]\in H^3(M_{24},U(1))$ for which  (A)-(C) can be satisfied and that, for each choice of a normalized representative $\alpha$, the functions $\phi_{g,h}$ satisfying  (A)-(C) are uniquely determined. If two normalized $3$-cocycles $\alpha$ and $\alpha'$ differ by a $3$-coboundary $\partial\beta$, as in \eqref{alphashift}, then the corresponding twisted twining genera $\phi_{g,h}$ and $\phi'_{g,h}$ are related as in \eqref{Zghredef}
\be \phi'_{g,h}(\tau,z)= \frac{\beta(g,h)}{\beta(h,g)} \phi_{g,h}(\tau,z)\ .
\ee The explicit expressions for $\phi_{g,h}$ derived in \cite{Gaberdiel:2012gf} are collected in table \ref{t:noncycl} in appendix \ref{app:tables}. Although the existence of the modules $\H_g$ matching \eqref{twisttwingen} is not proven yet for $g$ different from the identity,  strong evidence in this direction  has been given \cite{Gaberdiel:2012gf}. In the rest of the paper we will assume that the conjecture holds.
As in the ordinary moonshine case, one can choose minimal modules $\H_g$, such that the only states with negative $(-1)^{\bar F}$ are contained in irreducible BPS representations of $\N=4$. In fact, the results of \cite{Gaberdiel:2012gf} suggest that such states only appear in the untwisted sector $\H_{g=e}$.

\subsection{Geometric perspective}
\label{sec_geometric}
Before we close this section we shall offer a more geometric perspective on the twisted twining genera which will be useful later on. 
In \cite{Ganter}, Ganter showed that the natural home for the Norton series $f(g,h;\tau)$ \cite{Norton} is the equivariant elliptic cohomology developed in \cite{GKV}.
We shall here give a short review of Ganter's perspective, adapted to the case of Jacobi forms  on $\mathbb{H}_+\times \mathbb{C}$, as opposed to 
modular forms on $\mathbb{H}_+$.

Let $\P$ be the set of pairs of commuting  elements of $M_{24}$
\be \P:=\{(g,h)\in M_{24}\mid gh=hg\}\ ,
\ee
and $\bar\P$ the set of conjugacy classes of such pairs, i.e. the quotient of $\P$ with respect to $(g,h)\sim (k^{-1}gk,k^{-1}hk)$, for any $k\in M_{24}$.
The set $\bar \P$ can be identified with the set  
of isomorphism classes of principal $M_{24}$-bundles over the elliptic curve 
$E_\tau=\mathbb{C}/(\mathbb{Z}+\tau\mathbb{Z})$. One can then consider the associated moduli 
space\footnote{This should really be a moduli \emph{stack} but we ignore this technical point; 
see \cite{Ganter} for a more precise description.} of principal $M_{24}$-bundles on $E_\tau$
\be
\M=\P \times (\mathbb{H}_+\times \mathbb{C})\, / \, M_{24}\times (SL(2,\mathbb{Z})\ltimes \mathbb{Z}^2)\ . 
\ee
For each  pair $(g,h)\in \P$ there exists an $(SL(2,\mathbb{Z})\times \mathbb{Z}^2)$-equivariant line bundle
\be
\begin{array}{c}
  \L_{g,h}^{\alpha}\\
   \downarrow \\
   \M \\
\end{array}
\ee
which is twisted by the 3-cocycle $\alpha\in H^{3}(M_{24}, U(1))$  \cite{Ganter}. Thus, for fixed $(g,h)$ we can think of the twisted 
twining genus $\phi_{g,h}$ as a section of $\L_{g,h}^{\alpha}$.
The twist by $\alpha$  in the bundle $\L_{g,h}^{\alpha}$ accounts for the $\alpha$-dependent multiplier phases 
occurring in the modular transformations of the twisted twining genera. In the cases when $\alpha$ describes a trivial class in 
$H^{3}(M_{24}, U(1))$, the bundle $\L_{g,h}^{\alpha}$ is canonically trivialized and the associated 
$\phi_{g,h}$ exhibits no multiplier phase. The interpretation of the twisted twining genera as sections of 
$\L_{g,h}^{\alpha}$ will in particular play a role in section \ref{sec_SecQuant} when we discuss twisted equivariant Hecke operators.

\section{Second quantization of Mathieu moonshine}
\label{sec_SymProd}

\noindent In this section we construct the second quantized twisted twining genera $\Psi_{g,h}$ 
as the exponentiated generating function of certain twisted equivariant Hecke operators acting on $\phi_{g,h}$. The forms $\Psi_{g,h}$  have a natural interpretation in terms of generating functions of the twisted twining genera in the symmetric orbifold CFTs and we therefore begin by reviewing some relevant facts about symmetric orbifold theory. Having defined the functions $\Psi_{g,h}$ we also show that they have infinite 
product expansions.

\subsection{Symmetric Orbifolds and Hecke Operators}

%\subsubsection{Symmetric Orbifolds}

Given a holomorphic CFT, or self-dual Vertex Operator Algebra (VOA), $\cC$ with a group $G\subseteq {\rm Aut}(\cC)$ of automorphisms, let us consider the theory $\cC ^{\otimes L}$ obtained by taking the tensor product of $L$ copies of $\cC$, for some $L\ge 1$. The group of automorphisms of $\cC ^{\otimes L}$ contains the direct product of the symmetric group $S_L$ and the group $G$ acting diagonally on all copies of $\cC$.\footnote{Actually, ${\rm Aut}(\cC^{\otimes L})$ contains the wreath product $S_L\wr G:= S_L\rtimes (G\times \ldots \times G)$, i.e. the semidirect product of $S_L$ and the direct product $G^L:=G\times \ldots \times G$ of $L$ copies of $G$, with $S_L$ acting on $G^L$ by the obvious permutation.}  The twisted twining partition function $Z_{g,h}^{\cC ^{\otimes L}}$ of the theory $\cC ^{\otimes L}$, associated to a pair of commuting elements $(g,h)$ of the diagonal group $G$, is simply the $L$:th-power of the twisted twining partition function $Z_{g,h}$ in the original theory $\cC$:
\be Z_{g,h}^{\cC ^{\otimes L}}(\tau)=Z_{g,h}(\tau)^L\ .
\ee In other words, the path integral on the torus with boundary conditions $g,h$ in the product theory $\cC ^{\otimes L}$ is simply the product of $L$ copies of the path integral in the original theory with the same boundary conditions. In particular, if $Z_{g,h}$ is invariant under the $\alpha$-twisted slash operator for some $3$-cocycle $\alpha$, then $Z_{g,h}^{ \cC^{\otimes L}}$ is invariant under the $\alpha^L$-twisted slash operator.

The $L$:th symmetric orbifold $S^L\cC$ of $\cC$ is defined as the orbifold of $\cC^{\otimes L}$ by $S_L$, and its group of automorphisms contains the diagonal $G$.
Therefore, one can define the twisted twining partition functions 
\be
Z_{g,h}^{(L)}(\tau) \equiv Z_{g,h}^{S^L \cC}(\tau)
\ee
 in each symmetric product. These partition functions can be computed using the usual formulae for orbifold CFTs.

The orbifold formula for $Z^{(L)}_{g,h}$ can be nicely understood in the geometric setting of section \ref{sec_geometric}. 
Recall that the twisted twining partition function $Z_{g,h}(\tau)$ of a VOA $\cC$ with Aut$(\cC)=G$ is a section of a ($\alpha$-twisted) line bundle $\L^\alpha_{g,h}$ over the moduli space $\M_G $ of $G$-bundles over the  elliptic curve $E=\CC/\Lambda$ with $\Lambda =\ZZ+\tau\ZZ$. In this picture $g$ and $h$ correspond to the monodromies around the cycles $-1$ and $\tau$ of the elliptic curve.

As shown in \cite{Dijkgraaf:1996xw}, $Z^{(L)}_{g,h}$ can be expressed as a sum of contributions from all isomorphism classes of unramified $L$-fold coverings $\Upsilon:E'\to E$, namely
\be Z^{(L)}_{g,h}(\tau)\sim \sum_{\substack{\text{isomorphism classes of}\\L\text{-fold coverings}\\ \Upsilon:E' \to E}} \Upsilon^* Z_{g,h}(\tau)\ ,
\ee up to a suitable normalization. Here, $\Upsilon^* Z_{g,h}$ is simply the partition function associated with the pull-back $\Upsilon^*\L^\alpha_{g,h}\to E'$ of the line bundle $\L^\alpha_{g,h}$.

Up to isomorphisms, we can consider coverings that preserve the base-point of the elliptic curve. Any \emph{connected} unramified base-point preserving $L$-fold cover of $E=\CC/\Lambda$ is given by $E'\cong \CC/\Lambda'\to E$, $z\mapsto z$, where $\Lambda'$ is a sublattice of index $L$ in $\Lambda$. In turn, the sublattices of index $L$ are given by $\Lambda'=M\Lambda$ for any $M$ in the set 
\be \Mat_L(\ZZ) :=\{\begin{pmatrix}
a & b\\ c &d
\end{pmatrix}\mid a,b,c,d\in\ZZ, ad-bc=L\}
\label{MatN}
\ee 
of integral $2\times 2$ matrices $M$ of determinant $L$. In particular, when $L=1$, the covers $\Upsilon:E'\to E$ are actually isomorphisms and are classified by $\Mat_1(\ZZ)\equiv SL(2,\ZZ)$. Let $\Upsilon_\gamma$ denote the isomorphism associated with a specific choice of $\gamma\in SL(2,\mathbb{Z})$. The pull-back $\Upsilon_\gamma^*Z_{g,h}$ then coincides with the slash operator
\be \Upsilon_\gamma^* Z_{g,h}(\tau)=Z_{g,h}(\tau)\rvert_\alpha (\gamma,e)\ ,\qquad \gamma\in SL(2,\ZZ)\ .
\ee

 For generic $L>0$, the isomorphism classes of coverings are in one to one correspondence with the cosets
\be SL(2,\ZZ)\backslash \Mat_L(\ZZ)\ ,
\ee and a set of coset representatives is given by
\be\label{cosetrep} \begin{pmatrix}
a & b\\ 0 & d
\end{pmatrix}\ ,\qquad ad=L,\ a,d>0,\ 0\le b<d\ .
\ee

Therefore, the contribution to the partition function $Z^{(L)}_{g,h}(\tau)$ from the \emph{connected} $L$-fold coverings is given by
\be \T^\alpha_LZ_{g,h}(\tau):=\frac{1}{L}\sum_{M\in SL_2(\ZZ)\backslash \Mat_L(\ZZ)} \Upsilon_M^* Z_{g,h}(\tau)\equiv \frac{1}{L}\sum_{M\in SL_2(\ZZ)\backslash \Mat_L(\ZZ)} Z_{g,h}(\tau)\rvert_\alpha (M,e)\ ,
\label{connected}
\ee where $\rvert_\alpha (M,e)$ is a suitable generalization of the slash operator to $\Mat_L(\ZZ)$, that will be discussed in the following sections and in appendix \ref{app:twistHecke}. 

The full twisted twining partition function $Z^{(L)}_{g,h}$ is given by the sum over connected and disconnected $L$-coverings. The result is most easily expressed in terms of a generating function, which is given by the plethystic exponential of the connected contribution
\be\label{plethystic} \sum_{L=0}^\infty p^L Z^{(L)}_{g,h}(\tau)=\exp \Bigl(\sum_{L=1}^\infty p^L \T^\alpha_L Z_{g,h}(\tau)\Bigr)\ .\ee
Explicit formulae for the ($\alpha$-twisted) equivariant Hecke operators $\T^\alpha_L$ defined by (\ref{connected}) will be discussed in more detail in section \ref{s:tHecke} and appendix \ref{app:twistHecke}.

\subsection{Twisted equivariant Hecke operators}\label{s:tHecke}

The exponential lifts $\Psi_{g,h}$ of the twisted twining genera $\phi_{g,h}$ are the analogue of the generating functions \eqref{plethystic} for superconformal field theories. In particular, they can be defined in terms of some set of Hecke operators that are compatible with 
the properties (A)-(C) in section \ref{sec_ttprop}. This means that the Hecke action should be compatible, i.e. equivariant, with respect to the $M_{24}$-action on the pair 
$(g,h)$, and it should incorporate the $\alpha$-twist in the modular transformation. Equivariant versions of Hecke operators acting on modular forms was proposed by Ganter \cite{Ganter} in the context of generalized  Monstrous moonshine, and we shall see that this result also applies here after some minor modifications. 

\subsubsection{Generalities on Hecke operators}
Hecke operators are linear operators acting on modular forms. Specifically, for each $L\geq 1$ there is an operator $T_L$ that acts on a weight $k$ modular form $f(\tau)$ and 
produces another modular form $T_L f(\tau)$ of the same weight. On modular functions, i.e. weight $k=0$ modular forms, the Hecke operator is defined by
\be
T_L f(\tau):= \frac{1}{L} \sum_{\left(\begin{smallmatrix} a & b \\ c & d \\ \end{smallmatrix}\right)\in SL(2,\mathbb{Z})\backslash \Mat_L(\mathbb{Z})}  f\left(\frac{a\tau+b}{c\tau+d}\right).
\ee
These  operators have a natural generalization to Jacobi forms given by \cite{EichlerZagier} (restricting to the weight $0$, index $m$ case)
\be
T_L\phi(\tau, z):=\frac{1}{L} \sum_{\left(\begin{smallmatrix} a & b \\ c & d \\ \end{smallmatrix}\right)\in SL(2,\mathbb{Z})\backslash \Mat_L(\mathbb{Z})} e^{\frac{-mLcz^2}{c\tau+d}} \phi\left(\frac{a\tau+b}{c\tau+d}, \frac{Lz}{c\tau+d}\right).
\label{HeckeJacobi}
\ee
In fact, on Jacobi forms there exists an additional Hecke operator $U_L$ with the simple action
\be U_L \phi(\tau,z):=\phi(\tau,Lz).
\ee
These operators map Jacobi forms of weight $0$ and index $m$ to Jacobi forms of weight $0$ and index $Lm$ and $L^2m$, respectively. They also form a Hecke algebra, with the relations
\begin{align}\label{hecke1}
&T_m\cdot T_n =T_{mn}\ ,& \text{for }\gcd(m,n)=1\ ,\\
\label{hecke1a} &U_m\cdot U_n=U_{mn}\ ,\\
\label{hecke2}&T_m\cdot U_n=U_n\cdot T_m\ , \\
\label{hecke3}&T_p\cdot T_{p^m} = T_{p^{m+1}}+\frac{1}{p} T_{p^{m-1}}\cdot U_p\ , &\text{for } p\text{ prime}\ .
\end{align}

It is easy to see that a set of representatives for the quotient $SL(2,\ZZ)\backslash \Mat_L(\ZZ)$ can always be chosen of the form \eqref{cosetrep}.
This allows us to write the action of $T_L$ on $\phi(\tau, z)$ as follows
\be\label{Heckestand} T_L\phi(\tau,z)%=\frac{1}{l}\sum_{u \in SL_2(\ZZ)\backslash M_l(\ZZ)} \phi(u\cdot(\tau,z))
=\frac{1}{L}\sum_{a,d>0, ad=L}\sum_{b=0}^{d-1} \phi(\frac{a\tau+b}{d},az)\ .
\ee 

\subsubsection{Twisted equivariant Hecke operators: main definition}
The standard Hecke-operators $T_L$ map weak Jacobi forms of weight 0, index 1 to weak Jacobi forms of weight 0, index $L$. In the generalized  setting the analogous statement implies that we should have equivariant Hecke operators $\T_L$ that map sections of $\L_{g,h}^{\alpha}$ to sections of the product bundle $(\L_{g,h}^{\alpha})^{\otimes L}$, where $\L_{g,h}^{\alpha}\rightarrow \M$ is the line bundle discussed in section \ref{sec_geometric}. In fact, because of the cocycle-dependent multiplier phases arising in the modular transformations of the $\phi_{g,h}$ we need to introduce a certain $\alpha$-twisted version $\T^{\alpha}_L$ of the equivariant Hecke operator. 
More explicitly, if the sections of $\L_{g,h}^{\alpha}$ correspond to holomorphic functions $\phi_{g,h}$ on the covering space $\P\times \HH_+\times\CC$ of $\M$,  with an automorphy factor determined by a 3-cocycle $\alpha$, then $\T^{\alpha}_L\phi_{g,h}$ should correspond to holomorphic functions whose automorphy factor derives from the $L$:th power $\alpha^L$ of the 3-cocycle. 

Since the detailed analysis is rather technical, we will postpone it to appendix \ref{app:twistHecke} and just quote the explicit formula for the action of the Hecke operators on a twisted twining character $\phi_{g,h}$
\begin{align}\label{twistedHecke2} \T^\alpha_L\phi_{g,h}(\tau,z)&:=\frac{1}{L}\sum_{\substack{a,d>0,\\ ad=L}}\sum_{b=0}^{d-1} \epsilon_{g,h}\left(\begin{smallmatrix}
a & b\\ 0 & d
\end{smallmatrix}\right) \phi_{g^d, g^{-b}h^a}\bigl(\frac{a\tau+b}{d},az\bigr)\ ,\\
\U^\alpha_L\phi_{g,h}(\tau,z)&:=\epsilon_{g,h}\left(\begin{smallmatrix}
L & 0\\ 0 & L
\end{smallmatrix}\right)\phi_{g^L,h^L}(\tau,Lz)\ ,
\end{align}
where 
\be \epsilon_{g,h}\bigl(\begin{smallmatrix}
 a & b\\ 0 & d
\end{smallmatrix}\bigr):= \frac{\prod_{i=1}^{a-1}c_g(h,h^i)^d}{\prod_{j=1}^{d-1}c_{g^{-b}h^a}(g,g^j)\prod_{k=1}^bc_g(g,g^{-k}h^a)^d}\ ,
\ee and $c_g(x,y)$, $g,x,y\in G$, depends on the cocycle $\alpha$ that determines the automorphy factor of $\phi_{g,h}$  via \eqref{ciggi}.

As discussed in appendix \ref{app:twistHecke}, these Hecke operators satisfy the Hecke algebra \eqref{hecke1}--\eqref{hecke3}. The explicit expressions for $\T^\alpha_L\phi_{g,h}$, for $L=1,2,3,4$, in the case of Mathieu moonshine are collected in appendix \ref{app:tables}.

Notice that  the phases $\epsilon_{g,h}\left(\begin{smallmatrix}
a & b\\ c & d
\end{smallmatrix}\right)$ are trivial whenever $g$ is the identity or when the restriction of the  cocycle $\alpha$ to the group $\langle g,h\rangle$ is trivial. An example where non-trivial phases appear is given by $\T^\alpha_2\phi_{g,g}$, where $g$ is in class $2B$ of $M_{24}$. These phases can be understood by first considering  \be \T^\alpha_2\phi_{e,g}(\tau,z)=\frac{1}{2}\bigl(\phi_{e,e}(2\tau,2z)+\phi_{e,g}(\frac{\tau}{2},z)+\phi_{e,g}(\frac{\tau+1}{2},z)\bigr)\ ,
\ee
and then imposing the relations
\be \T^\alpha_2\phi_{g,e}(\tau,z)=e^{\frac{-4\pi iz^2}{\tau}}\T^\alpha_2\phi_{e,g}(-\frac{1}{\tau},\frac{z}{\tau})\ ,\qquad \T^\alpha_2\phi_{g,g}(\tau,z)=\T^\alpha_2\phi_{g,e}(\tau+1,z)\ .
\ee Using the modular properties of $\phi_{e,g}$, we obtain
\begin{align} \T^\alpha_2\phi_{g,e}(\tau,z)&=\frac{1}{2}\bigl(\phi_{g,e}(2\tau,2z)+\phi_{e,e}(\frac{\tau}{2},z)-\phi_{e,g}(\frac{\tau+1}{2},z)\bigr)\ ,\\
 \T^\alpha_2\phi_{g,g}(\tau,z)&=\frac{1}{2}\bigl(-\phi_{g,e}(2\tau,2z)-\phi_{e,g}(\frac{\tau}{2},z)+\phi_{e,e}(\frac{\tau+1}{2},z)\bigr)\ ,
\end{align} where the minus signs in the last two equations are due to the non-trivial multiplier system of the twining genus $\phi_{e,g}$. 

\subsubsection{Central extensions}\label{s:centrext}

Rather than considering the projective representation $\rho_g$ of the centralizer $C_{M_{24}}(g)$, it is often useful to work with a linear representation of a central extension
\be 1\to  U(1) \to C^\alpha_{M_{24}}(g) \to C_{M_{24}}(g) \to 1\ .
\ee The group $C^\alpha_{M_{24}}(g)$ can be constructed explicitly in terms of generators and relations. Let $q(x):=e^{2\pi ix}$, $x\in \RR/\ZZ$, be the generator of the central $U(1)$ factor and consider one generator $h_\alpha$ for each $h\in C_{M_{24}}(g)$. Then, $C^\alpha_{M_{24}}(g)$ is generated by all such elements subject to the relations
\be h_\alpha k_\alpha =q\bigl(\mu_g(h,k)\bigr)\, (hk)_\alpha \ ,\qquad q(x)h_\alpha=h_\alpha q(x)
\ee for all $h,k\in C_{M_{24}}(g)$. Here,  $\mu_g(h,k)\in\RR/\ZZ$ is defined by \be e^{2\pi i\mu_g(h,k)}=c_g(h,k)\ ,\ee where $c_g(h,k)$ is the $2$-cocycle determining the projective representation $\rho_g$ as in \eqref{grouplaw}. This definition provides also a canonical lift 
\be h\mapsto h_\alpha\ ,
\ee from $C_{M_{24}}(g)$ to its central extension $C^\alpha_{M_{24}}(g)$.
By construction, the $g$-twisted sector carries a genuine representation $\tilde\rho_g$ of $C^\alpha_{M_{24}}(g)$, with
\be \tilde\rho_g(q(x))=e^{2\pi ix}\ ,\qquad \tilde\rho_g(h_\alpha)=\rho_g(h)\ ,
\ee for all $h\in C_{M_{24}}(g)$, $x\in\RR/\ZZ$. A less trivial observation is that also the $g^r$-twisted sectors, for all $r\in \ZZ_{\ge 0}$, carry a genuine representation $\tilde\rho_{g,r}$ of $C^\alpha_{M_{24}}(g)$, defined by\footnote{We denote by the same symbol $\tilde \rho_{g,r}$ also the projective representation of $C_{M_{24}}(g)$ on $\H_{g^r}$ with $\tilde\rho_{g,r}(h):=\rho_{g^r}(h)/f_{g,r}(h)$}
\be \tilde\rho_{g,r}(q(x)):=e^{2\pi i rx}\ ,\qquad \tilde\rho_{g,r}(h_\alpha)\equiv \tilde\rho_{g,r}(h):=\frac{\rho_{g^r}(h)}{f_{g,r}(h)}\ ,
\ee for all $h\in C_{M_{24}}(g)\subseteq C_{M_{24}}(g^r)$, $x\in\RR/\ZZ$, $r\in \ZZ_{\ge 0}$, where
\be\label{fgrh} f_{g,r}(h)=\prod_{i=1}^{r-1}c_h(g,g^i)\ .
\ee
The fact that $\tilde\rho_{g,r}$ are well defined representations of $C^\alpha_{M_{24}}(g)$ is an immediate consequence of the identity
\be\label{cigierre} c_{g^r}(h,k)=\frac{f_{g,r}(h)f_{g,r}(k)}{f_{g,r}(hk)} \, c_g(h,k)^r\ ,\qquad h,k\in C(g)\ ,
\ee which follows from the definition of $c_g$ in terms of $\alpha$ and repeated applications of the cocycle condition for $\alpha$. Notice that $f_{g,1}(h)=1$ for all $h\in C_{M_{24}}(g)$, so that $\tilde\rho_g\equiv \tilde\rho_{g,1}$.

If $\alpha$ and $\alpha'$ are different cocycle representatives of the same cohomology class $[\alpha]$, then the central extensions are isomorphic $C^\alpha_{M_{24}}(g)\cong C^{\alpha'}_{M_{24}}(g)$, although the corresponding lifts $h_\alpha$ and $h_{\alpha'}$ are different. In particular, for special choices of the representative cocycle $\alpha$, it is sufficient to consider a central extension of $C_{M_{24}}(g)$ by a finite subgroup of $U(1)$. With slight abuse of notation, we will denote also these finite central extensions by $C^\alpha_{M_{24}}(g)$. See appendix \ref{a:centrext} for more details on these special choices for $\alpha$.

\subsubsection{Twisted equivariant Hecke operators: alternative definition}

The $\alpha$-twisted equivariant Hecke operators are more easily defined in terms of the central extension $C^\alpha_{M_{24}}(g)$ described in section \ref{s:centrext}. Set
\be \phi_{g_\alpha^r,h_\alpha}(\tau,z):=\Tr_{\H_{g^r}}\bigl(\tilde\rho_{g,r}(h_\alpha)q^{L_0-\frac{c}{24}}y^{J_0^3}(-1)^{F+\bar F}\bigr)\ .\ee
so that
\be\begin{split} 
\phi_{g_\alpha^d,g_\alpha^{-b}h_\alpha^a}(\tau,z)&=\Tr_{\H_{g^d}}\bigl(\tilde\rho_{g,d}( g)^{-b}\tilde\rho_{g,d}(h)^aq^{L_0-\frac{c}{24}}y^{J_0^3}(-1)^{F+\bar F}\bigr)\\
&=\frac{\prod_{i=1}^{a-1}c_g(h,h^i)^d}{\prod_{k=1}^{b}c_g(g,g^{-k}h^a)^d}\Tr_{\H_{g^d}}\bigl(\tilde\rho_{g,d}(g^{-b}h^a)q^{L_0-\frac{c}{24}}y^{J_0^3}(-1)^{F+\bar F}\bigr)\\
&=\frac{\prod_{i=1}^{a-1}c_g(h,h^i)^d}{\prod_{j=1}^{d-1}c_{g^{-b}h^a}(g,g^j)\prod_{k=1}^{b}c_g(g,g^{-k}h^a)^d}\Tr_{\H_{g^d}}\bigl(\rho_{g^d}(g^{-b}h^a)q^{L_0-\frac{c}{24}}y^{J_0^3}(-1)^{F+\bar F}\bigr)\\
&=
\epsilon_{g,h}\left(\begin{smallmatrix}
a & b\\ 0 & d
\end{smallmatrix}\right)\phi_{g^d,g^{-b}h^a}(\tau,z)\ .
\end{split}\ee
Therefore, we can reinterpret the $\alpha$-twisted Hecke operators $\T^\alpha_L$ acting on the twisted twining genera $\phi_{g,h}$ as (untwisted) equivariant operators acting on   the twisted twining genera $\phi_{g_\alpha ,h_\alpha}$ 
\be \T^\alpha_L\phi_{g,h}(\tau,z)=\frac{1}{L}\sum_{ad=L}\sum_{b=0}^{d-1}\phi_{g_\alpha^d,g_\alpha^{-b}h_\alpha^a}\bigl(\frac{a\tau+b}{d},az\bigr)=:\T_L\phi_{g_\alpha,h_\alpha}(\tau,z)\ .
\ee 
This form of the Hecke operators will turn out to be very useful for some of the calculations in sections \ref{sec_SecQuant} and \ref{sec_MasonConnection}.

\subsection{Second-Quantized Twisted Twining Genera}
\label{sec_SecQuant}
\noindent
\subsubsection{Definition}

For any Calabi-Yau manifold $X$, one can  define its \emph{second-quantized} elliptic genus $\Psi_X$ as the exponentiated generating function of the orbifold elliptic genus $\phi_{S^LX}(\tau, z)$ of the symmetric products $S^{L}X$ \cite{Dijkgraaf:1996xw}. In fact, ref. \cite{Dijkgraaf:1996xw} gave three equivalent expressions for $\Psi_X$:
\be
\Psi_X(\sigma,\tau, z)= \sum_{L=0}^{\infty} p^L \phi_{S^L X}(\tau, z)=\text{exp} \big[\sum_{L=1}^{\infty}p^L (T_L\phi_X)(\tau, z)\big]=\prod_{\substack{n>0, m\geq 0,\\  \ell\in \mathbb{Z}}}\big(1-p^n q^m y^{\ell}\big)^{-c(mn,\ell)},
\label{SQEG}
\ee
where we have set 
\be
q=e^{2\pi i \tau}, \qquad y=e^{2\pi i z},\qquad p=e^{2\pi i \rho},
\ee
and $c(mn,\ell)$ are the Fourier coefficients of the elliptic genus of $X$:
\be
\phi_X(\tau,z)=\sum_{m\geq 0, \ell\in \mathbb{Z}}c(m,\ell) q^{m} y^{\ell}.
\ee  Here, $T_L$ are standard Hecke operators defined in (\ref{Heckestand}).

It was shown in \cite{Gritsenko:1999fk} that $\Phi\equiv A_X/\Psi_X$, where $A_X$ is a simple correction factor (``Hodge anomaly'') that is determined by the Hodge numbers of $X$, transforms as a Siegel modular form for some congruence subgroup of $Sp(4;\mathbb{Z})$. 

Taking the formula (\ref{SQEG}) as a starting point, we now wish to define the \emph{second quantized twisted twining genera} as follows
\be
\Psi_{g,h}(\sigma,\tau,z):=\text{exp}\big[\sum_{L=1}^{\infty} p^L (\T^\alpha_L\phi_{g,h})(\tau, z)\big],
\label{SQTTG}
\ee
where $\T^\alpha_L$ is the twisted equivariant Hecke operator defined in \eqref{twistedHecke2}. We will later show that after including  a correction factor $A_{g,h}$ the functions $\Psi_{g,h}$ transform as Siegel modular forms for some discrete subgroup $\Gamma_{g,h}^{(2)}\subset Sp(4;\mathbb{R})$ that contains the invariance group $\Gamma_{g,h}\subset SL(2,\mathbb{Z})$ of the twisted twining genera $\phi_{g,h}$.

Notice that the definition \eqref{SQTTG} depends on the choice of a normalized $3$-cocycle $\alpha$. If $\alpha$ and $\alpha'$ differ by a $3$-coboundary $\partial\beta$ as in\eqref{alphagauge}, the corresponding twisted twining genera are related as
\be\label{rescale}  \T^{\alpha'}_L\phi'_{g,h}(\tau,z)=e^{2\pi i\nu_{g,h} L}\T^{\alpha}_L\phi_{g,h}(\tau,z)\ ,
\ee where $\nu_{g,h}\in \RR/\ZZ$ is defined as
\be e^{2\pi i\nu_{g,h}}=\frac{\beta(g,h)}{\beta(h,g)}\ ,
\ee so that
\begin{align}
\Psi'_{g,h}(\sigma,\tau,z)=&\exp\Bigl(\sum_{L=1}^\infty p^L\T^{\alpha'}_L\phi'_{g,h}(\tau,z)\Bigr)\\
=& \exp\Bigl(\sum_{L=1}^\infty (e^{2\pi i\nu_{g,h}}p)^L\,\T^{\alpha}_L\phi_{g,h}(\tau,z)\Bigr)
= \Psi_{g,h}(\sigma+\nu_{g,h},\tau,z)\ .
\end{align} 
Therefore, a different choice for the cocycle representative $\alpha$ simply amounts to a redefinition $\sigma \to \sigma+\nu_{g,h}$ of the variable $\sigma$.

\subsubsection{Infinite product representation}

We shall now derive infinite product representations for the second quantized twisted twining genera $\Psi_{g,h}$ in terms of the Fourier coefficients of $\phi_{g,h}$.

Let us consider a generic commuting pair of elements $g,h\in M_{24}$, with the cocycle $\alpha$ inducing a possibly non-trivial multiplier for $\phi_{g,h}$. Let $N=o(g)$ be the order of $g$ and $\lambda$ the length of the shortest cycle of $g$ in the $24$-dimensional permutation representation (see Table \ref{t:cycl} in appendix \ref{app:tables}). 
Then,  the lift $g_\alpha$ of $g$ to the central extension $C^\alpha_{M_{24}}(g)$, as defined in section \ref{s:centrext}, has order $N\lambda$ (see appendix \ref{a:centrext}). For any $h\in C_{M_{24}}(g)$ the associated twisted twining genus $\phi_{g,h}$ has a Fourier expansion of the form
\be \phi_{g,h}(\tau,z)=\sum_{n=0}^\infty\sum_{\ell\in\ZZ} c_{g,h}\bigl(\frac{n}{N\lambda},\ell\bigr)q^{\frac{n}{N\lambda}}y^\ell\ ,
\ee where, in particular, $c_{g,h}\bigl(\frac{n}{N\lambda},\ell\bigr)=0$ unless $n\equiv -1\mod \lambda$.
Let $M=o(h_\alpha)$ be the order of the lift $h_\alpha$ of $h$ to the central extension $C^\alpha_{M_{24}}(g)$.\footnote{We assume that $\alpha$ is chosen in such a way that the order is finite.} The logarithm of the second quantized twisted twining genus of $\phi_{g,h}$ is
\be\begin{split}&\log\Psi_{g,h}(\sigma,\tau,z)=\sum_{L=1}^\infty p^L\T^\alpha_L\phi_{g,h}(\tau,z)\\
&=\sum_{a,d=1}^\infty\frac{1}{ad} \sum_{b=0}^{d-1}\sum_{\ell\in\ZZ}\sum_{n=0}^\infty \epsilon_{g,h}\left(\begin{smallmatrix}
a & b\\ 0 & d
\end{smallmatrix}\right)c_{g^d,g^{-b}h^a}(\frac{n}{N\lambda},\ell)e^{\frac{2\pi i bn}{N\lambda d}}q^{\frac{an}{N\lambda d}}y^{a\ell}p^{ad}\\
&=\sum_{a,d=1}^\infty\frac{1}{ad} \sum_{\ell\in\ZZ}\sum_{n=0}^\infty \sum_{b=0}^{d-1}c_{g_\alpha^d,g_\alpha^{-b}h_\alpha^a}(\frac{n}{N\lambda},\ell)e^{\frac{2\pi i bn}{N\lambda d}}q^{\frac{an}{N\lambda d}}y^{a\ell}p^{ad}\label{secquant1}\\
&=\sum_{a,d=1}^\infty\frac{1}{a} \sum_{t,k=0}^{M-1}\frac{e^{\frac{2\pi it(a-k)}{M}}}{M}\sum_{\ell\in\ZZ}\sum_{n=0}^\infty \frac{1}{d}\sum_{b=0}^{d-1}c_{g_\alpha^d,g_\alpha^{-b}h_\alpha^k}(\frac{n}{N\lambda},\ell)e^{\frac{2\pi i bn}{N\lambda d}}q^{\frac{an}{N\lambda d}}y^{a\ell}p^{ad}\\
&=\sum_{a,d=1}^\infty\frac{1}{a} \sum_{t,k=0}^{M-1}\frac{e^{\frac{2\pi it(a-k)}{M}}}{M}\sum_{\ell\in\ZZ} F_{g,h}(a,d,k,\ell)y^{a\ell}p^{ad}\ ,\end{split}\ee
where
\be F_{g,h}(a,d,k,\ell):=\sum_{n=0}^\infty \frac{1}{d}\sum_{b=0}^{d-1} e^{\frac{2\pi i bn}{N\lambda d}}\epsilon_{g,h}\left(\begin{smallmatrix}
k & b\\ 0 & d
\end{smallmatrix}\right)c_{g^d,g^{-b}h^k}(\frac{n}{N\lambda},\ell) q^{\frac{an}{dN\lambda}}\ .\ee
In appendix \ref{s:prova} we show that this sum can be rewritten as follows
\be\label{identica2}
F_{g,h}(a,d,k,\ell)=\sum_{m=0}^\infty \frac{1}{N\lambda}\sum_{b=0}^{\lambda N-1}  e^{\frac{2\pi i bm}{N\lambda}}\epsilon_{g,h}\left(\begin{smallmatrix}
k & b\\ 0 & d
\end{smallmatrix}\right)c_{g^{d},g^{-b}h^k}(\frac{md}{N\lambda},\ell) q^{\frac{am}{N\lambda}}\ .\ee
By plugging this expression into \eqref{secquant1} we obtain
\be\begin{split} \log\Psi_{g,h}(\sigma,\tau,z)=&\sum_{d=1}^\infty\sum_{m=0}^\infty\sum_{\ell\in\ZZ} \sum_{t=0}^{M-1} \hat c_{g,h}(d,m,\ell ,t) \sum_{a=1}^\infty\frac{1}{a} \Bigl(e^{\frac{2\pi it}{M}} q^{\frac{m}{N\lambda}}y^{\ell}p^{d}\Bigr)^a\\
&=-\sum_{d=1}^\infty\sum_{m=0}^\infty\sum_{\ell\in\ZZ} \sum_{t=0}^{M-1} \hat c_{g,h}(d,m,\ell ,t)\log\bigl(1-e^{\frac{2\pi it}{M}}q^{\frac{m}{N\lambda}}y^\ell p^d\bigr),
\end{split}\ee
where
\begin{align} \hat c_{g,h}(d,m,\ell ,t):=&\sum_{k=0}^{M-1}\sum_{b=0}^{\lambda N-1}  \frac{e^{-\frac{2\pi itk}{M}}}{M}\frac{e^{\frac{2\pi i bm}{\lambda N}}}{\lambda N}\epsilon_{g,h}\left(\begin{smallmatrix}
k & b\\ 0 & d
\end{smallmatrix}\right)c_{g^{d},g^{-b}h^k}(\frac{md}{N\lambda},\ell)\nonumber\\
=&
\sum_{k=0}^{M-1}\sum_{b=0}^{\lambda N-1}  \frac{e^{-\frac{2\pi itk}{M}}}{M}\frac{e^{\frac{2\pi i bm}{\lambda N}}}{\lambda N}\Tr_{\H_{g^d}(\frac{md}{N\lambda},\ell)}\bigl(\tilde\rho_{g,d}(g)^{-b}\tilde\rho_{g,d}(h)^k(-1)^{F+\bar F}\bigr)\ .\label{chatgh}
\end{align} Eq.\eqref{chatgh} implies that $\hat c_{g,h}(d,m,\ell ,t)$ admit an interpretation as a $\ZZ_2$-graded dimension (with the grading given by $(-1)^{F+\bar F}$) of the simultaneous eigenspace for $\tilde\rho_{g,d}(g)$ and $\tilde\rho_{g,d}(h)$, restricted to $\H_{g^d}(\frac{md}{N\lambda},\ell)$, relative to the eigenvalues $e^{\frac{2\pi i m}{\lambda N}}$ and $e^{\frac{2\pi i t}{M}}$, respectively. In particular, $\hat c_{g,h}(d,m,\ell ,t)$ is always an integer.

Thus, for the inverse of the second quantized twisted twining genus we obtain the infinite product expression
\be \frac{1}{\Psi_{g,h}(\sigma,\tau,z)}=\exp\left(-\sum_{L=1}^\infty p^L\T^\alpha_L\phi_{g,h}(\tau,z)\right)=\prod_{d=1}^\infty\prod_{m=0}^\infty\prod_{\ell\in\ZZ}\prod_{t=0}^{M-1}(1-e^{\frac{2\pi it}{M}}q^{\frac{m}{N\lambda}}y^\ell p^d)^{\hat c_{g,h}(d,m,\ell ,t)}\ .
\ee
  Note that \eqref{chatgh} makes sense also for $d=0$, so that  the infinite product in $d$ and $m$ can be symmetrized to obtain
\begin{align}\label{infprod} \Phi_{g,h}(\sigma,\tau,z):= p q^{\frac{1}{N\lambda}} y\prod_{(d,m,\ell)>0}\prod_{t=0}^{M-1}(1-e^{\frac{2\pi it}{M}}q^{\frac{m}{N\lambda}}y^\ell p^d)^{\hat c_{g,h}(d,m,\ell ,t)},
\end{align} where the first product runs over
\be d,m\in\ZZ_{\ge 0}\quad \text{and} \quad \begin{cases} \ell\in\ZZ,\ \ell<0, & \text{if }m=0=d\ ,\\
 \ell\in\ZZ, & \text{otherwise.}
\end{cases}
\ee The prefactor  $p q^{\frac{1}{N\lambda}} y$ in (\ref{infprod}) has been chosen in such a way that $\Phi_{g,h}$ satisfies suitable automorphic properties, that will be described in section \ref{sec_modular}.

One can rewrite $\Phi_{g,h}$ in terms of the original $\Psi_{g,h}$ as
\be
 \Phi_{g,h}(\sigma, \tau, z)=\frac{p\psi_{g,h}(\tau,z)}{\Psi_{g,h}(\sigma,\tau,z)},
 \label{SiegelAutoCorr}
 \ee
 where $\psi_{g,h}$ includes the $d=0$ factors in \eqref{infprod} and takes the form
\be \psi_{g,h}(\tau,z)=q^{\frac{1}{N\lambda}}y\prod_{t=0}^{M-1}\Bigl(\prod_{\ell<0} (1-e^{\frac{2\pi it}{M}}y^\ell)^{\hat c_{g,h}(0,0,\ell ,t)}\Bigr)\Bigl(\prod_{\ell\in\ZZ}\prod_{m=1}^\infty (1-e^{\frac{2\pi it}{M}}q^{\frac{m}{N\lambda}}y^\ell )^{\hat c_{g,h}(0,m,\ell ,t)}\Bigr)\ .
\label{psigh}
\ee
The numerator in (\ref{SiegelAutoCorr}) is the analogue of the ``Hodge anomaly'' in \cite{Gritsenko:1999fk}. 

Infinite products of the form \eqref{infprod} were studied in \cite{Borcherds}, where it was shown that they converge for  $\left(\begin{smallmatrix}
\tau & z\\ z & \sigma
\end{smallmatrix}\right)$ in a suitable domain of the Siegel upper half-space of $2\times 2$ complex symmetric matrices with positive definite imaginary part
\be  \mathbb{H}_2:=\Bigl\{ Z \in \Mat_2(\CC)\mid Z=Z^t,\  {\rm Im}Z>0 \Bigr\}\ .
\ee Furthermore, these products can be analytically continued to meromorphic functions on the whole $\mathbb{H}_2$, with zeroes and (possibly) poles along the rational quadratic divisor \cite{Borcherds2} (see section \ref{sec_connecting}). As will be discussed in section \ref{sec_modular}, the $\Phi_{g,h}$   are Siegel modular forms under certain discrete subgroups of $Sp(4,\RR)$.

\medskip

The forms $\Phi_{g,h}$ depend on the choice of the cocycle representative $\alpha$. If $\alpha$ and $\alpha'$ differ by a $3$-coboundary $\partial\beta$, then the corresponding forms $\Phi_{g,h}$ and $\Phi'_{g,h}$ are related by a shift in $\sigma$ and an overall phase
\be\label{newPhi} \Phi'_{g,h}(\sigma,\tau,z)=e^{-2\pi i\nu_{g,h}}\Phi_{g,h}(\sigma+\nu_{g,h},\tau,z)\ ,
\ee where $\nu_{g,h}$ satisfies $e^{2\pi i\nu_{g,h}}=\frac{\beta(g,h)}{\beta(h,g)}$.

\subsubsection{Multiplicative versus additive lift}\label{sec_addmult}

The construction of a Siegel modular form $\Phi(\sigma,\tau, z)$ from a (weak) Jacobi form $\phi(\tau, z)$ via an infinite product representation, as exemplified by (\ref{infprod}), is generally referred to as a \emph{multiplicative (automorphic) lift}. The Jacobi form $\phi$ is said to be the \emph{multiplicative seed} of the lift (see, for instance, \cite{Borcherds2,Gritsenko:1999fk,GritsenkoNikulin2,Dabholkar:2006xa}), and we write 
\be
\Phi = \text{Mult}[\phi].
\ee
It should be stressed that the definition of $\Phi_{g,h}$ in this paper is based on the ($\alpha$-twisted) equivariant Hecke operators $\T^\alpha_L$ rather than the ordinary $T_L$, as in the standard multiplicative lifts. As a consequence, the infinite product expression \eqref{infprod} involves the Fourier coefficients of many distinct Jacobi forms $\phi_{g^d,g^{-b}h^a}$, rather than a unique $\phi$ as, for example, in \cite{Borcherds2}.

\bigskip

In some cases one can also obtain the Siegel modular form using a different procedure, known as the \emph{additive lift}\footnote{This generalizes the \emph{Saito-Kurokawa-Maass lift} defined in \cite{EichlerZagier}.}. In this case the  Siegel modular form is constructed as a certain generating function (without exponentiation) of Hecke operators acting on a different Jacobi form $\psi$, which is then called the \emph{additive seed}. For $\psi(\tau, z)$ a Jacobi form of weight $k$ we write $\tilde{\psi}(\sigma, \tau, z)=p \psi(\tau, z)$ and define the additive lift as 
\be
\Phi=\text{Add}[\psi]:= \sum_{m\geq 1} m^{2-k} (T_m^{-}\tilde{\psi})(\sigma, \tau, z),
\label{addlift}
\ee
where $T_m^{-}$ is a certain Hecke operator; see, e.g. \cite{GritsenkoNikulinSimplest} for the precise definition and properties of the right hand side. 

As an example, consider the case of the Igusa cusp form $\Phi_{10}$ (corresponding to $\Phi_{e,e}$). The multiplicative seed is the K3 elliptic genus $\phi_{e,e}=\phi_{0,1}$ which is the unique weak Jacobi form of weight 0 and index 1. The multiplicative lift  yields an infinite product formula for $\Phi_{10}$ \cite{GritsenkoNikulin}:
\be
\Phi_{10}=\text{Mult}[\phi_{0,1}]=pqy \prod_{(d,m,\ell)> 0} \left(1-q^m y^{\ell} p^d\right)^{c(d,m,\ell)}
\ee
where $c(d,m,\ell)$ are the Fourier coefficients of $\phi_{0,1}$. This is indeed obtained from (\ref{infprod}) by restricting to $(g,h)=(e,e)$. As explained in \cite{GritsenkoNikulin2} one can also obtain this Siegel modular form via an additive lift from the seed $\phi_{10,1}$ (the unique weak Jacobi form of weight 10 and index 1):
\be
\Phi_{10}=\text{Add}[\phi_{10,1}].
\label{addliftPhi10}
\ee

We now observe that the additive seed $\phi_{10,1}$ can be expressed as $\vartheta(\tau, z)^{2} \eta(\tau)^{18}$ which is precisely the symmetrization factor $\psi_{g,h}$ in (\ref{psigh}) when restricting to $(g,h)=(e,e)$. This is in fact a general feature that holds whenever a Siegel modular form can be realized both as an additive lift as well as a multiplicative lift. It is a consequence of the fact that the additive seed $\psi$ Êis the first Fourier-Jacobi coefficient in the expansion of $\Phi$ and it is precisely this coefficient which appears as the prefactor in the symmetrization of the infinite product. For purposes we thus expect that for all $(g,h)$ for which an additive lift exists we should have the following equalities
\be
\Phi_{g,h}=\text{Mult}[\phi_{g,h}]=\text{Add}[\psi_{g,h}]. 
\ee
We stress that one can of course define an additive lift of a Jacobi form via (\ref{addlift}). However it is not guaranteed that when applying this procedure to $\psi_{g,h}$ one will always reproduce the functions $\Phi_{g,h}$. Moreover, not all automorphic infinite products have associated additive lifts. In general we expect that whenever $\Phi_{g,h}$ can be represented through an additive lift, the seed will be $\psi_{g,h}$. In section \ref{s:examples} we indeed verify this for a number of examples. For $(g,h)=(e,h)$ modified versions of these additive lifts have also been considered in \cite{Eguchi:2011aj}.

\section{Wall-crossing and Mason's generalized moonshine for $M_{24}$}
\label{sec_MasonConnection}

\noindent Already in 1990, Mason proposed an $M_{24}$ version of Norton's generalized  moonshine conjecture \cite{MasonEtas,MasonGen}. In this section, we will  establish the connection between Mason's generalized moonshine for $M_{24}$ and the recent 
Mathieu moonshine involving the $K3$ elliptic genus. This involves taking the multiplicative lift $\phi_{g,h}\to \Phi_{g,h}$, defined in section \ref{sec_SymProd},  after which the limit $z\to 0$  reproduces the generalized  eta-products $\eta_{g,h}$ 
constructed by Mason  \cite{MasonGen}. Thus, the 
exponential lift $\Phi_{g,h}$ links Mason's generalized  $M_{24}$-moonshine to the $M_{24}/{\rm K3}$-moonshine considered here. 
This connection was first suggested in \cite{Cheng:2010pq} for the special case of the twining genera $\phi_{1,h}=\phi_h$.
A pictorial overview of this relation is given in figure \ref{triangle}. 
We begin by reviewing Mason's construction of the generalized eta-products $\eta_{g,h}$ and explain their interpretation in terms of 
partition functions of twisted chiral bosons. Physically, the Siegel modular forms $\Phi_{g,h}$ have interpretations as generating 
functions of twisted dyons in (CHL) orbifolds of $\mathcal{N}=4$ string theory, and the eta-products emerge as a result 
of a wall-crossing formula, generalizing the one written down in \cite{Cheng:2010pq} for $\Phi_{e,h}$. 

\subsection{Mason's generalized $M_{24}$-moonshine}
 For each commuting pair $(g,h)\in M_{24}$, Mason associates a modular function on the upper half-plane satisfying the requirements posed by Norton in \cite{Norton}. The starting point of Mason was the 24-dimensional permutation representation of $M_{24}$ in which each element can be associated to a \emph{cycle shape}, which describes the element as a product of 
cycles of permutations. For instance, in this representation the identity element is represented by the cycle shape $1^{24}$, corresponding to the product of 24 identity permutations, while the elements in class $2A$ are represented by $1^8 2^8$, corresponding to the product of 8 identity permutations followed by 8 consecutive order 2 permutations. 

Mason's functions were all given in terms of so called \emph{eta-products}, namely products of Dedekind eta-functions.  Suppose an order $M$ element $h\in M_{24}$ has cycle shape 
\be\prod_{\ell |M} \ell^{i(\ell)}\ ,
\ee
on the $24$-dimensional representation of $M_{24}$, for some integers $i(\ell)$. Then one can  associate an eta-product to $h$ by
\be
\eta_h(\tau):=\prod_{\ell |M}\eta( \ell \tau)^{i(\ell)}.
\label{etaprod}
\ee
Mason found a generalization of such eta-products  associated to ``generalized  cycle shapes'' labelled by commuting pairs $(g,h)$ in $M_{24}$. We shall denote these generalized  moonshine functions by $\eta_{g,h}(\tau)$.\footnote{In \cite{MasonGen} these functions were denoted by $f(g,h;\tau)$.}
 Let $g,h\in M_{24}$ be a pair of commuting elements of order $o(g)=N$ and $o(h)=M$ and consider their action in the  standard $24$-dimensional representation $V$ of $M_{24}$. Let $v_1,\ldots, v_{24}$ be a basis of simultaneous eigenvectors for $g$ and $h$, relative to the eigenvalues $(e^{\frac{2\pi i r_i}{N}},e^{\frac{2\pi i t_i}{M}})$, $i=1,\ldots,24$, with $1\le r_i\le N$ and $1\le t_i\le M$. Then, the eta-product $\eta_{g,h}$ is defined as
 \be \eta_{g,h}(\tau)=q^{\frac{1}{N\lambda}}\prod_{i=1}^{24}\prod_{n=0}^\infty (1-e^{\frac{2\pi it_i}{M}}q^{\frac{r_i}{N}+n})\ ,
 \ee where $\lambda$ is the length of the shortest cycle of $g$.
 In particular, the eta-products $\eta_h$ of equation \eqref{etaprod} correspond to $(g,h)=(e,h)$. The products $\eta_{g,h}$ (or rather their inverse) can be interpreted as $h$-twining partition functions for $24$ $g$-twisted chiral free bosons.
 
As proven in \cite{MasonEtas}, the eta products $\eta_{g,h}$ satisfy the modular transformations
\be\label{etamodular} \upsilon_{g,h}\left(\begin{smallmatrix}
a & b\\ c & d
\end{smallmatrix}\right)(c\tau+d)^{-w}\eta_{g^dh^{-c},g^{-b}h^a}\bigl(\frac{a\tau+b}{c\tau+d}\bigr)=\eta_{g,h}(\tau)\ ,
\ee for some $\upsilon_{g,h}\left(\begin{smallmatrix}
a & b\\ c & d
\end{smallmatrix}\right)\in U(1)$. The full list of Mason's generalized  eta-products and the corresponding weights $w$ can be found in Table \ref{TableMason2}. Furthermore, by definition, $\eta_{g,h}$ are invariant under conjugation in $M_{24}$
\be\label{etaconj} \eta_{kgk^{-1},khk^{-1}}(\tau)=\eta_{g,h}(\tau)\ .
\ee

\subsection{Connecting the two moonshines via wall-crossing} 
\label{sec_connecting}

The connection between the $K3/M_{24}$-moonshine for twining genera $\phi_h(\tau, z)$ and the $M_{24}$-moonshine associated with the functions $\eta_h(\tau)$ was pointed out in \cite{Cheng:2010pq}. We now want to use our results above to establish this link also between the two different \emph{generalized} moonshines for $M_{24}$. 

As already mentioned, infinite products of the form \eqref{infprod} have been considered by Borcherds \cite{Borcherds2}. Although in general $\Phi_{g,h}$  are not expected to be Borcherds products in a strict sense \cite{Raum12}, the same argument as in  \cite{Borcherds2} shows  that the infinite products \eqref{infprod} converge in some region of the Siegel upper half-space $\mathbb{H}_2$ and they can be analytically continued to meromorphic functions on the whole of $\mathbb{H}_2$.
Furthermore, in the domain of absolute convergence of the infinite products \eqref{infprod}, the zeroes and poles of $\Phi_{g,h}$ can only be located at rational quadratic divisors of the form:
\be H_{d,m,\ell,v}=\left\{\left(\begin{smallmatrix}
\tau & z\\ z & \sigma
\end{smallmatrix}\right)\mid \tfrac{m}{N\lambda}\tau+\ell z+d\sigma + \frac{v}{M} =0\right\}\ ,
\label{RQD}
\ee
 where $d,m, \ell,v \in \ZZ$ are coprime integers ($\gcd(d,m, \ell,v)=1$) that  satisfy a positive discriminant condition
\be \ell^2-\frac{4md}{N\lambda}>0\ .
\ee This property is an immediate consequence of the infinite product representation \eqref{infprod} of $\Phi_{g,h}$; the divisors of Borcherds products admit a similar description \cite{Borcherds,Borcherds2,GritsenkoClery}. 
Using the fact that every point in the Siegel upper half-space can be brought into a domain of convergence using an $Sp(4;\mathbb{Z})$-transformation, and that the family of functions $\Phi_{g,h}$ is preserved under this action (see section \ref{s:autom}), it follows that all divisor components can be brought to the form (\ref{RQD}), for suitable $N\lambda, M$, by some $Sp(4;\mathbb{Z})$-transformation.\footnote{Note that $Sp(4,\ZZ)$ acts non-linearly on the variables $\sigma,\tau,z$; therefore, even though the divisor \eqref{RQD} is defined by a linear equation in $\sigma,\tau,z$, this is in general not true for its modular images.} This formula can be viewed as a generalisation of the rational quadratic divisor found in \cite{Sen:2010ts} in the context of twisted dyon counting in CHL-models.

The multiplicity of the zero or pole of $\Phi_{g,h}$ at the divisor  $H_{d,m,\ell,v}$  is given by $\hat c_{g,h}(d,m,\ell,v)$. In particular, $\Phi_{g,h}$ is holomorphic if and only if $\hat c_{g,h}(d,m,\ell,v)$ is non-negative at every rational quadratic divisor.  As will be shown in section \ref{s:autom}, $\Phi_{g,h}$ is an automorphic form for a discrete subgroup $\Gamma_{g,h}^{(2)}\subset Sp(4,\RR)$. Therefore, one only needs to consider the distinct orbits of rational quadratic divisors under the action of $\Gamma_{g,h}^{(2)}$ to determine the full divisor of $\Phi_{g,h}$. The precise calculation of the divisor of each $\Phi_{g,h}$ requires knowledge of the modular group $\Gamma^{(2)}_{g,h}$ for each commuting pair $(g,h)$ and will be left to future work.

A special role is played by (the modular orbit of) the divisor $H_{0,0,-1,0}$, corresponding to the locus $z=0$ in $\mathbb{H}_2$. Since $c_{e,g}(0,-1)=2$ for all $g\in M_{24}$, we have
\be \hat c_{g,h}(0,0,-1,0)=\frac{1}{MN\lambda}\sum_{k=0}^{M-1}\sum_{b=0}^{\lambda N-1}  c_{e,g^{-b}h^k}(0,-1)=2\ .
\ee
It follows that every $\Phi_{g,h}$ has a double zero at this divisor. Using the relation
\be \sum_{\ell\in\ZZ}c_{g,h}(r,\ell)=0\ ,\qquad \text{for }r>0\ ,
\ee together with the relations 
\begin{align} &\Tr_{\H_{e}(0,\pm 1)}( g^{-b} h^k(-1)^{F+\bar F})=2\ ,\label{trace1}\\
 & \Tr_{\H_{e}(0,\pm \ell)}(g^{-b} h^k(-1)^{F+\bar F})=0\qquad \text{for }\ell>1\ ,\label{trace2}\\
&\sum_{\ell\in\ZZ}\Tr_{\H_{e}(0,\ell)}( g^{-b} h^k(-1)^{F+\bar F})=\Tr_{\bf 24}(g^{-b}h^k)\ ,\label{trace3}
\end{align}
 we obtain
\begin{align} \lim_{z\to 0}\frac{\Phi_{g,h}(\sigma,\tau,z)}{(2\pi iz)^2}=\lim_{z\to 0} &\frac{y(1-y^{-1})^2}{(2\pi iz)^2}\Bigl(q^{\frac{1}{N\lambda}}\prod_{t=1}^M\prod_{m=1}^\infty (1-e^{\frac{2\pi it}{M}}q^{\frac{m}{N\lambda}})^{\sum_{\ell\in\ZZ}\hat c_{g,h}(0,m,\ell,t)}\Bigr)\nn \\
&\times\Bigl(p\prod_{t=1}^M\prod_{d=1}^\infty (1-e^{\frac{2\pi it}{M}}p^{d})^{\sum_{\ell\in\ZZ}\hat c_{g,h}(d,0,\ell,t)}\Bigr)\ .
\label{Philim}
\end{align} 

Next we use the relations (\ref{trace1})-(\ref{trace3}) to find an  expression for $\psi_{g,h}$ in (\ref{psigh}) in terms of $\eta_{g,h}$ and known modular objects. Using the aforementioned equations we obtain
\begin{align} \psi_{g,h}(\tau,z)&=q^{\frac{1}{N\lambda}}y\prod_{t=0}^{M-1}\Bigl((1-e^{\frac{2\pi it}{M}}y^{-1})^{\hat c_{g,h}(0,0,-1 ,t)}\prod_{m=1}^\infty\prod_{\ell\in\ZZ}(1-e^{\frac{2\pi it}{M}}q^{\frac{m}{N\lambda}}y^\ell )^{\hat c_{g,h}(0,m,\ell ,t)}\Bigr)\nn \\
&=q^{\frac{1}{N\lambda}}y(1-y^{-1})^{2}\prod_{n=1}^\infty \frac{(1-q^ny)^2
(1-q^ny^{-1})^2}{(1-q^n)^4}\prod_{t=0}^{M-1}(1-e^{\frac{2\pi it}{M}}q^{\frac{m}{N\lambda}})^{\sum_{\ell\in\ZZ}\hat c_{g,h}(0,m,\ell ,t)}.
\label{psirewrite}
\end{align}
To proceed we note the identity
\begin{align} &q^{\frac{1}{N\lambda}}\prod_{t=0}^{M-1}\prod_{m=1}^\infty(1-e^{\frac{2\pi it}{M}}q^{\frac{m}{N\lambda}})^{\sum_{k=0}^{M-1}\sum_{b=0}^{N\lambda-1}  \frac{e^{-\frac{2\pi itk}{M}}}{M}\frac{e^{\frac{2\pi i bm}{N\lambda}}}{N\lambda}\Tr_{\bf 24}(g^{-b}h^k)}\nn \\
&=q^{\frac{1}{N\lambda}}\prod_{t=0}^{M-1}\prod_{m=1}^\infty(1-e^{\frac{2\pi it}{M}}q^{\frac{m}{N\lambda}})^{\sum_{i=1}^{24}\frac{1}{M}\sum_{k=0}^{M-1}\frac{1}{N\lambda}\sum_{b=0}^{N\lambda-1}  e^{-\frac{2\pi i(t-t_i)k}{M}}e^{\frac{2\pi i b(m-\lambda r_i)}{N\lambda}}}\nn \\
&=q^{\frac{1}{N\lambda}}\prod_{i=1}^{24}\prod_{n=0}^\infty(1-e^{\frac{2\pi it_i}{M}}
q^{n+\frac{r_i}{N}})=\eta_{g,h}(\tau)\ ,\end{align}
where $1\le r_i\le N$ and $1\le t_i\le M$ are such  that $(e^{\frac{2\pi i r_i}{N}},e^{\frac{2\pi i t_i}{M}})$, $i=1,\ldots,24$, are the $(g,h)$-eigenvalues of a basis of simultaneous eigenvectors for $g$ and $h$ in the $24$-dimensional representation of $M_{24}$. Using this relation in (\ref{psirewrite}) we  obtain
\begin{align} \psi_{g,h}(\tau, z) &=-\frac{\vartheta_1(\tau,z)^2}{\eta(\tau)^6}\eta_{g,h}(\tau) =\phi_{-2,1}(\tau,z)\eta_{g,h}(\tau)\ ,
\label{psighjacobi}
\end{align}
where $\phi_{-2,1}$ is the standard weak Jacobi forms of weight $-2$ and index $1$ for $SL(2,\ZZ)$ (see appendix \ref{app:moduJaco}).

To complete the analysis of the limit in (\ref{Philim}) we note that 
\be \lim_{z\to 0}\frac{y(1-y^{-1})^2}{(2\pi iz)^2}\Bigl(q^{\frac{1}{N\lambda}}\prod_{t=1}^M\prod_{m=1}^\infty (1-e^{\frac{2\pi it}{M}}q^{\frac{m}{N\lambda}})^{\sum_{\ell\in\ZZ}\hat c_{g,h}(m,0,\ell,t)}\Bigr)=\lim_{z\to 0}\frac{\psi_{g,h}(\tau,z)}{(2\pi iz)^2}=\eta_{g,h}(\tau)\ .
\ee For the pair $(g,h)=(e,e)$, the function $\Phi_{e,e}(\sigma,\tau,z)$ is invariant under the exchange $\sigma\leftrightarrow \tau$. From a physical viewpoint, this is a consequence of S-duality of type II superstring compactified on K3$\times T^2$.  In section \ref{sec_modular}, we will prove that $\Phi_{g,h}$ satisfy analogous transformations
\be \Phi_{g,h}(\sigma,\tau,z)=\Phi_{g,h'}(\frac{\tau}{N\lambda},N\lambda\sigma,z)\ ,
\ee where $h'\in C_{M_{24}}(g)$ is not necessarily in the same conjugacy class as $h$. Using this identity, we conclude that
\be \lim_{z\to 0}\frac{\Phi_{g,h}(\sigma,\tau,z)}{(2\pi iz)^2}=\eta_{g,h}(\tau)\eta_{g,h'}(N\lambda\sigma)\ .
\label{wcf}
\ee

We conjecture that this equation has a physical interpretation as a wall-crossing formula whenever $1/\Phi_{g,h}$ is the generating function of BPS-states in string theory. Indeed, for some pairs $(g,h)$ we know that $\Phi_{g,h}^{-1}$ corresponds to the generating function for the $h$-twisted degeneracies of 1/4 BPS states in a CHL model. More precisely,  these degeneracies are the Fourier coefficients of the automorphic form $\Phi_{g,h}^{-1}$ and the region where the Fourier expansion is performed depends on the moduli. These multiplicities jump as one crosses the pole at $z=0$; the physical interpretation is that some 1/4 BPS dyon corresponding to a bound state of 1/2 BPS configurations becomes unstable in a certain region of the moduli space and thus disappears from the spectrum. The mismatch $\eta_{g,h}(\tau)^{-1}\eta_{g,h'}(N\lambda\sigma)^{-1}$ between the Fourier coefficients at the two sides of the pole represents the degeneracy of a bound state of two 1/2 BPS states.

\section{Automorphic Properties}\label{s:autom}
\noindent In this final section we analyze the modular transformation properties of $\Phi_{g,h}(\sigma, \tau, z)$ with respect to discrete subgroups of $Sp(4,\mathbb{R})$. We show that they are Siegel modular forms and in some cases we are able to identify them with previously known objects. Our proof of modularity is done in two steps. The crucial first step is to determine the transformation properties of $\Phi_{g,h}(\sigma, \tau, z)$ with respect to the interchange $\sigma \leftrightarrow \tau$. We refer to this as ``S-duality'' since in the cases when $\Phi_{g, h}$ has an interpretation as a partition function in an $\mathcal{N}=4$ string theory this corresponds precisely to the S-duality that exchanges electric and magnetic charges (see, e.g., \cite{Dijkgraaf:1996it,David:2006ji,David:2006ud,Sen:2009md,Sen:2010ts}). It turns out that this transformation involves the subtle concept of ``relabeling'' of the group elements $(g,h)$ introduced in \cite{Gaberdiel:2012gf}. By a careful analysis of this relabeling phenomenon we establish the S-duality symmetry in section \ref{sec:Sduality}. To complete the analysis of the automorphic properties of $\Phi_{g,h}(\sigma, \tau, z)$ we must also verify the transformation properties with respect to the remaining generators of the relevant modular subgroups of $Sp(4,\mathbb{R})$. This is rather straightforward since it essentially follows from the modularity of the seed functions $\phi_{g,h}(\tau, z)$ and $\psi_{g,h}(\tau, z)$. We present this analysis in section \ref{sec_modular}, and in section \ref{s:examples} we also investigate some examples in detail.  

\subsection{Relabeling and S-duality}
\label{sec:Sduality}
\noindent In this section, we will show that the functions $\Phi_{g,h}(\sigma,\tau,z)$ defined by the infinite product \eqref{infprod} satisfy some `S-duality' identities that exchange $\sigma$ and $\tau$. This property, together with the modular properties of the twisted twining genera $\phi_{g,h}$ proved in \cite{Gaberdiel:2012gf}, will be sufficient to prove that all $\Phi_{g,h}$ are automorphic functions under some subgroup of $Sp(4,\RR)$ (see section \ref{sec_modular}).

\subsubsection{Orbifolds and relabeling}\label{s:orbirelab}

In the case where $g$ belongs to some $M_{23}$ subgroup of $M_{24}$, the main step in the derivation of S-duality for $\Phi_{g,h}$ is the relabeling phenomenon \cite{Gaberdiel:2012gf}. This can be understood by considering the example of a holomorphic conformal field theory by a cyclic group. In this subsection, we review some of the standard properties of these orbifolds. See, for example, \cite{Dixon:1986jc,Dolan:1989vr,Dong:1997ea} for more details.

Let $\C$ be a holomorphic conformal field theory with automorphism group $G$. 
Given an element $g\in G$ of order $N$, let $\C'$ be the orbifold\footnote{Here and in the following, we use the word `orbifold' in the physicists' sense, i.e. as a full-fledged two dimensional conformal field theory with modular invariant partition function.} of $\C$ by $\langle g\rangle$. The spectrum $\H'$ of $\C'$ is constructed by considering the direct sum of the $g^r$-twisted representations $\H_{g^r}$ of $\C$, for all $r=0,\ldots, N-1$ and then restricting to the $g$-invariant sector, i.e.
\be \H' = \oplus_{r=0}^{N-1} \Bigl(\H_{g^r}^{\langle g\rangle}\Bigr)\ ,
\ee where $\H_{g^r}^{\langle g\rangle}$ denotes the $g$-invariant part of the $g^r$-twisted sector. The orbifold theory is a well-defined CFT, provided that one can define a consistent (in particularly, local) OPE between the $g$-invariant twisted fields. A necessary and sufficient consistency condition is that the cohomology class $[\alpha]\in H^3(G,U(1))$ is trivial when restricted to $H^3(\langle g\rangle,U(1))$ (level-matching). In what follows, we will assume that this condition is satisfied.

Each twisted sector carries a representation  of some central extension $C^\alpha_G(g)$ of the centralizer $C_G(g)$ of $g$ in $G$, that is compatible with the structure of twisted $\C$-module. As explained in section \ref{s:centrext}, there is an ambiguity in the definition of this representation. If the orbifold is consistent, the action of $C^\alpha_G(g)$ can be always chosen to be compatible also with the OPE of $g$-invariant twisted fields, so that it defines a group of automorphisms of the orbifold theory. In particular, for any $r,s\in\ZZ$, the representation of $C^\alpha_G(g)$ corresponding to the $g^{r+s}$-twisted sector must be contained in the tensor product of the representations corresponding to the  $g^r$- and $g^s$-twisted sectors. A central extension satisfying this constraint can be chosen of the form
\be 1 \to \langle Q\rangle \cong \ZZ_N \to C^\alpha_G(g)\to C_G(g)\to 1\ ,
\ee where the central element $Q$ (quantum symmetry) acts by $e^{\frac{2\pi i r}{N}}$ on the $g^r$-twisted sector.

 Since the spectrum $\H'$ of $\C'$ consists of $g$-invariant states, the group acting faithfully on $\H'$ is $C^\alpha_G(g)/\langle g\rangle$. We conclude that $C^\alpha_G(g)/\langle g\rangle$ must be a subgroup of the group $G'$ of automorphisms of the orbifold CFT $\C'$. 
For each $h\in C^\alpha_G(g)$, one can define the $h$-twining genera $Z'_{e,h}$ as in \eqref{Zgh}, such that
\be\label{Zprime} Z'_{e,h}(\tau)=\frac{1}{N}\sum_{r,s\in \ZZ/N\ZZ} Z_{g^r,g^s h}(\tau)\ .
\ee Note that $Z'_{e,h}(\tau)$ only depends on the image of $h$ in the projection $C^\alpha_G(g)\to C^\alpha_G(g)/\langle g\rangle\subseteq G'$.

\medskip

The group $G'$ of automorphisms of $\C'$ contains, in particular, an element $g'$ corresponding to the quantum symmetry $g'\equiv Q$. It is well-known that by taking the orbifold of $\C'$ by the cyclic group $\langle g'\rangle$ one re-obtains the original CFT $\C$. More precisely, the ${g'}^n$-twisted sector $\H'_{{g'}^n}$ can be identified with the direct sum
\be\label{twistidentif} \H'_{{g'}^n} = \oplus_{r=0}^{N-1} \H_{g^r,n}\ ,
\ee
where $\H_{g^r,n}$ is the $g$-eigenspace with eigenvalue $e^{\frac{2\pi in}{N}}$ in the $g^r$-twisted sector $\H_{g^r}$ of the CFT $\C$. 
Under the identification \eqref{twistidentif}, the $g'$-invariant sector of the direct sum $\oplus_{n=0}^{N-1} \H'_{{g'}^n}$ corresponds indeed to the spectrum $\H$ of the original theory $\C$ and the quantum symmetry $Q'$ corresponds to $g$. The identification \eqref{twistidentif} can be refined as
\be\label{Hidentif} \H'_{{g'}^n,r}\cong \H_{g^r,n}\ ,
\ee where $\H'_{{g'}^n,r}$ is the $g'$-eigenspace with eigenvalue $e^{\frac{2\pi im}{N}}$ in the twisted sector $\H'_{{g'}^n}$. Eq.\eqref{Hidentif} is an isomorphism between (untwisted) modules over the common subalgebra $\C^{\langle g\rangle}={\C'}^{\langle g'\rangle}$ of $\C$ and $\C'$.

Each eigenspace $\H_{g^r,n}$ is a representation of $C_G^\alpha(g)$. Analogously, each twisted sector $\H'_{{g'}^n}$ (and, in fact, each eigenspace $\H'_{{g'}^n,r}$) carries an action of  a suitable central extension 
\be\label{cextprime} 1\to \langle Q'\rangle \to C^{\alpha'}_{G'}(g') \to C_{G'}(g')\to 1\ ,\ee of $C_{G'}(g')$ for some $3$-cocycle $\alpha'$ representing a class in $H^3(G',U(1))$. In fact, there is an isomorphism $\varphi:C_G^\alpha(g)\stackrel{\cong}{\rightarrow}  C^{\alpha'}_{G'}(g')$ such that $\varphi(Q)=g'$, $\varphi(g)=Q'$ and the identification \eqref{Hidentif} is equivariant with respect to the action of the corresponding groups. As a consequence, eq.\eqref{Zprime} can be generalized to obtain the formulae for the twisted twining partition functions in the theory $\C'$
\be\label{Zrelabel1} Z'_{{g'}^n,\varphi(h)}(\tau)=\frac{1}{N}\sum_{r,s\in \ZZ/N\ZZ} e^{-\frac{2\pi i ns}{N}}
Z_{g^r,g^s h}(\tau)\ , \qquad h\in C^\alpha_G(g)\ .
\ee

A special case occurs when the two CFTs $\C'$ and $\C$ are isomorphic, so that $G'\cong G$. If, in addition, $g'$ and $g$ are in the same conjugacy class of $G$, the isomorphism $\varphi$ can be chosen to be an outer automorphism of the group $C^\alpha_G(g)$ that exchanges $g$ and $Q$, i.e. $\varphi(g)=Q$ and $\varphi(Q)=g$ and such that
\be\label{Zrelabel} \sum_{b=0}^{N-1} \frac{e^{\frac{2\pi i rb}{N}}}{N} Z_{g^m,g^{-b}\varphi(h)}(\tau)=
\sum_{s=0}^{N-1} \frac{e^{\frac{2\pi i ms}{N}}}{N} Z_{g^r,g^{-s}h}(\tau)\ .
\ee 
%which corresponds to the identification
%\be\label{Hidentif} \H_{Q^n,r}\cong \H_{g^r,n}\ .
%\ee 
Since $\varphi$ is outer, it might happen that $\varphi(h)$ and $h$ belong to different conjugacy classes of $C^\alpha_G(g)$. If this is the case, we say that the class of $h$ has been relabeled. Roughly speaking, a similar phenomenon occurs for the Mathieu moonshine. In particular, we will see that S-duality relates a function  $\Phi_{g,h}$ to a function $\Phi_{g,h'}$, where $h'$ belongs to the `relabeled' class.

\subsubsection{Relabeling for Mathieu Moonshine}
\label{s:orbirelabel}

The construction described in the previous section in the case of a holomorphic conformal field theory can be repeated with minor modifications for the abstract modules $\H_g$ underlying the generalized Mathieu Moonshine. It should be stressed, however, that since no consistent CFT  with automorphism group $M_{24}$ and spectrum $\H$ exists, these properties do not follow directly  from the general theory of holomorphic CFTs and need to be proved independently.

Consider an element $g\in G=M_{24}$ of order $N$ and let $\H_g$ be the $g$-twisted sector in the generalized Mathieu Moonshine conjecture. Let $\H'$ be the $g$-invariant subspace of the direct sum of all twisted sectors
\be\label{identif} \H' = \oplus_{r=0}^{N-1} \Bigl(\H_{g^r}^{\langle \tilde\rho_{g,r}(g)\rangle}\Bigr)\ ,
\ee where $\H_{g^r}^{\langle \tilde\rho_{g,r}(g)\rangle}$ denotes the $\tilde\rho_{g,r}(g)$-invariant part of the $g^r$-twisted sector. The level-matching condition, i.e. the requirement that the restriction of the cohomology class $[\alpha]$ is trivial in $H^3(\langle g\rangle,U(1))$, is satisfied if and only if $g$ is contained in some $M_{23}$ subgroup of $M_{24}$, so we will consider only this case.
 If $\H$ can be interpreted as the spectrum of R-R right-moving ground states in a non-linear sigma model on K3 with a symmetry $g$, then $\H'$ corresponds to the spectrum of right-moving ground states in the $g$-orbifold theory.  The orbifold is a consistent $\N=(4,4)$ SCFT  with central charge $6$, which turns out to be again a non-linear sigma model on K3, since its elliptic genus is
\begin{align} \Tr_{\H'}\bigl(q^{L_0-\frac{c}{24}}{\bar q}^{\bar L_0-\frac{\bar c}{24}}y^{J_0^3}(-1)^{F+\bar F}\bigr)&=\sum_{r,s=0}^{N-1}\frac{1}{N}\Tr_{\H_{g^r}}\bigl(\tilde\rho_{g,r}(g)^sq^{L_0-\frac{c}{24}}{\bar q}^{\bar L_0-\frac{\bar c}{24}}y^{J_0^3}(-1)^{F+\bar F}\bigr)\notag\\
&=\frac{1}{N}\sum_{r,s=1}\phi_{g^r,g^s}(\tau,z)=\phi_{e,e}(\tau,z)\ .\label{primaident}
\end{align} Here, the twisted twining genera $\phi_{g^r,g^s}$ are relative to a cocycle $\alpha$ satisfying the conditions \eqref{easycoc} and \eqref{triviac}, in particular with trivial restriction to $\langle g\rangle$. As shown in appendix \ref{a:centrext}, in this case the associated central extension $C^\alpha_{M_{24}}(g)$ is finite
\be\label{cextagain} 1\to \langle Q\rangle\cong \ZZ_N \to C^{\alpha}_{M_{24}}(g) \to C_{M_{24}}(g)\to 1\ .\ee
In the theory of orbifold CFTs, the conditions \eqref{easycoc} and \eqref{triviac} ensure that the action of $C^\alpha_{M_{24}}(g)$ is compatible with the OPE of the twisted fields. The identity \eqref{primaident} can be verified case by case for all $g\in M_{24}$ satisfying the level-matching condition, and holds independently of the existence of a physical interpretation in terms of orbifolds of K3 sigma models.

By \eqref{primaident}, the space $\H'$ defined by \eqref{identif} is isomorphic to $\H$ as a module over the $\N=4$ superconformal algebra. As a consequence, one can define a representation of $G'=M_{24}$ over $\H'$ satisfying the properties of Mathieu moonshine. Furthermore, $\H'$ also carries a representation $\rho'$ of $ C^\alpha_{M_{24}}(g)/\langle g_\alpha\rangle$ given by the restriction of $\oplus_{r=0}^{N-1}\tilde\rho_{g,r}$ to $\H'$. It can be proved that, as an abstract group, $C^\alpha_{M_{24}}(g)/\langle g_\alpha\rangle$ is isomorphic to a subgroup of $M_{24}$ \cite{Gaberdiel:2012gf}. More precisely, $Q$ is identified with an element $g'\in G'\cong M_{24}$ in the same conjugacy class as $g$ and the image of $C^\alpha_{M_{24}}(g)/\langle g\rangle$ is the centralizer $C_{M_{24}}(g')$ of $g'$. Therefore, the `orbifold' construction for the Mathieu moonshine modules is perfectly analogous to the case, described in the subsection \ref{s:orbirelab}, where a holomorphic CFT $\C$ and its orbifold $\C'$ are isomorphic. 

In analogy with the subsection \ref{s:orbirelab}, we can introduce some ${g'}^n$-twisted sectors $\H'_{g'^n}$ satisfying \eqref{twistidentif} and \eqref{Hidentif}, now considered as isomorphisms of $\N=4$ modules. Each twisted sector $\H'_{g'^n}$ carries a representation of a central extension $C^{\alpha'}_{M_{24}}(g')$ defined as in eq.\eqref{cextprime}. It is natural to conjecture that there is an isomorphism $\varphi:C^\alpha_{M_{24}}(g) \stackrel{\cong}{\rightarrow} C^{\alpha'}_{M_{24}}(g')$ such that the identifications \eqref{Hidentif} are equivariant with respect to the action of the corresponding groups. Since $g\in G$ and $g'\in G'$ belong to the same conjugacy class of $G\cong G'\cong M_{24}$, they can be identified through a suitable choice of the isomorphism $G\cong G'$. With this identification, $\varphi$ defines an outer automorphism of $C^\alpha_{M_{24}}(g)$ exchanging $g$ and $Q$. Furthermore, we expect the analogue of \eqref{Zrelabel} to hold. More precisely, we can state the following:

\begin{conjecture}\label{conj:relab} For all $g\in M_{23}\subset M_{24}$ and all choices of cocycle $\alpha$ satisfying \eqref{easycoc} and \eqref{triviac} and the corresponding central extension $C^\alpha_{M_{24}}(g)$ as in \eqref{cextagain}, there is an automorphism $\varphi:C^\alpha_{M_{24}}(g)\to C^{\alpha}_{M_{24}}(g)$ such that $\varphi(g)=Q$, $\varphi(Q)=g$, and for all $k\in C^\alpha_{M_{24}}(g)$, $m,r\in\ZZ$,
%\be\label{conja} \Tr_{ \H'_{{g'}^m,r}}(\tilde\rho'_{g',m}(\varphi(k))q^{L_0-\frac{c}{24}}y^{J_0^3}(-1)^{F+\bar F})=\Tr_{ \H_{g^r,m}}(\tilde\rho_{g,r}(k)q^{L_0-\frac{c}{24}}y^{J_0^3}(-1)^{F+\bar F})\ .
%\ee
\be\label{conja}  \sum_{b=0}^{N-1}\frac{e^{\frac{2\pi i rb}{N}}}{N}\phi_{{g_{\alpha}}^m,{g_{\alpha}}^{-b}\varphi(k)}(\tau,z)
=
\sum_{s=0}^{N-1}\frac{e^{\frac{2\pi i ms}{N}}}{N}\phi_{g^r_\alpha,g^{-s}_\alpha k}(\tau,z)\ .
\ee
\end{conjecture}

%\medskip

%Notice that if such an isomorphism exists for a certain choice of $\alpha$ and $\alpha'$, then for any other choices it can be obtained by composing $\varphi$ with \eqref{isocobou}.  Obviously, since $g$ and $g'$ are elements in the same conjugacy classes of $M_{24}$, there exist also isomorphisms $C^\alpha_{M_{24}}(g)\cong C^{\alpha'}_{M_{24}}(g')$ that map $g$ to $g'$ and $Q$ to $Q'$. In particular, since $g'$ is defined up to conjugation in $M_{24}$, we can simply choose $g'=g$ so that there is a natural identification of the centralizers $C_{M_{24}}(g)=C_{M_{24}}(g')$ and, upon choosing the same cocycles $\alpha=\alpha'$, also of their central extensions $C^\alpha_{M_{24}}(g)=C^{\alpha'}_{M_{24}}(g')$. With these conventions, the conjecture is equivalent to the existence of an outer automorphism of $C^\alpha_{M_{24}}(g)$ that exchanges $g$ and $Q$ and the representations $\H_{g^m,r}$ and $\H_{g^r,m}$.

The existence of automorphisms that exchange $g$ and $Q$  is easy to check for the elements $g$ of order higher than $4$ (thus, excluding the classes 2A and 4B of $M_{24}$). Indeed, in all these cases the central extension of  $C_{M_{24}}(g)$ has the form $C^\alpha_{M_{24}}(g)\cong \langle Q\rangle \times (\langle Qg\rangle.G ) $. This group clearly admits an automorphism that fixes the factor $\langle Qg\rangle.G$ and exchanges $Q$ and $g$. 
For the classes $2A$ and $4B$, the existence of such automorphisms can be verified  with the aid of the software GAP (see also the ancillary files in the arXiv version of \cite{Gaberdiel:2012gf}). 

In order to prove the conjecture, one needs to show that, for each $g$, one of these automorphisms $\varphi$ satisfies \eqref{conja}. Heuristically, by standard CFT arguments, a conjecture of the form \ref{conj:relab} and the corresponding identities \eqref{conja} are expected to hold for any group $H\subseteq C_{M_{24}}(g)$ admitting an interpretation as a group of symmetries of a K3 sigma model. According to the analysis in \cite{Gaberdiel:2011fg}, such an interpretation exists whenever $H\subset M_{24}$ has at least four orbits in the $24$-dimensional permutation representation; this condition is satisfied by most of the non-cyclic abelian groups $\langle g,h\rangle\subset M_{24}$  (see Table \ref{t:noncycl} in appendix \ref{app:tables}).

A more rigorous and general proof can be obtained through a direct computation. To this aim,  it is useful to rephrase eq.\eqref{conja}  in terms of identities among the twisted twining genera $\phi_{g,h}$ relative to the centralizer $C_{M_{24}}(g)$ rather than the genera $\phi_{g_\alpha,h_\alpha}$ relative to its central extension $C^\alpha_{M_{24}}(g)$. Recall that any element $k\in C^\alpha_{M_{24}}(g)$ can be written as $k=Q^x h_\alpha$, for some $x\in\ZZ/N\ZZ$, $h\in  C_{M_{24}}(g)$. Therefore, one needs to show that, for each $h\in C_{M_{24}}(g)$, the image $\varphi(h_\alpha)=Q^{x'}h'_{\alpha}$ of the lift $h_\alpha \in C^\alpha_{M_{24}}(g)$ satisfies
 \begin{align}\label{conja2} e^{\frac{2\pi i mx'}{N}}\sum_{b=0}^{N-1}&\frac{e^{\frac{2\pi i rb}{N}}}{N}\epsilon_{g,h'}\left(\begin{smallmatrix}
1 & b\\0 & m 
 \end{smallmatrix}\right)\phi_{{g}^m,{g}^{-b}h'}(\tau,z)
=
\sum_{s=0}^{N-1}\frac{e^{\frac{2\pi i ms}{N}}}{N}\epsilon_{g,h}\left(\begin{smallmatrix}
1 & s\\0 & r 
 \end{smallmatrix}\right)\phi_{g^r,g^{-s}h}(\tau,z)\ ,
 \end{align} for all $m,r\in \ZZ$.  Given the explicit knowledge of the twisted twining genera $\phi_{g,h}$, the proof of the conjecture then amounts to verifying a long series of identities between Jacobi forms. 
 
We have verified that, for each conjugacy class of $g$, there is essentially only one choice for the conjugacy class $h'$ and $x'\in \ZZ/N\ZZ$ for which all these identities can be possibly satisfied. In particular, if $g$ is not in one of the classes $2A$ or $4B$, then the automorphism $\varphi$ necessarily satisfies
\be h'=h\ ,\qquad x'=0\ ,
\ee for all $h\in C_{M_{24}}(g)\setminus \langle g\rangle$. For $g$ in one of the classes 2A or 4B, the conjugacy class of $h'$ in $C_{M_{24}}(g)$ for each $C_{M_{24}}(g)$-conjugacy class of $h$ is reported in Tables \ref{t:relabeling} and \ref{t:relabeling2}.  Notice that, in general, $h$ and $h'$ might belong to different conjugacy classes of $M_{24}$; this phenomenon has been dubbed `relabeling' in \cite{Gaberdiel:2012gf}.
When $h$ and $h'$ are in different conjugacy classes of $C_{M_{24}}(g)$, we can choose the cocycle $\alpha$ in such a way that $x'=0$. When $h$ and $h'$ are conjugated in $C_{M_{24}}(g)$, then $x'$ can be set to $0$ if and only if $\varphi(h_\alpha)$ is conjugated with $h_\alpha$ in the central extension $C^\alpha_{M_{24}}(g)$. This is always the case, except for $g$ in $M_{24}$-class $4B$ and $h$ in one of the classes  2B$_3$ or 4B$_1$ of $C_{M_{24}}(g)$. In both these cases, the pair $g,h$ generates group 20 in the list of appendix \ref{app:tables} and $\varphi(h_\alpha)$ is conjugated with $Q^2h_\alpha$, i.e. $x'=2$.

We have  proved a subset of the identities \eqref{conja2}, required for the analysis in section \ref{sec:Sdualityinvariance}. This leads to the following theorem:

\medskip

\begin{theorem}\label{thm:relabel}
Conjecture \ref{conj:relab} holds for all $g$ in the classes $1A$, $2A$, $3A$, $4B$ and $8A$ of $M_{24}$.
% For all $g\in M_{23}\subset M_{24}$ and all choices of cocycle $\alpha$ and central extension $C^\alpha_{M_{24}}(g)$ satisfying \eqref{easycoc} and \eqref{triviac}, there is an automorphism $\varphi:C^\alpha_{M_{24}}(g)\to C^{\alpha}_{M_{24}}(g)$ such that $\varphi(g)=Q$ and $\varphi(Q)=g$. Furthermore, in the following cases:
%\begin{enumerate}
%\item $g$ in class $2A$, $3A$ or $4B$ and $h$ is any element of $C^\alpha_{M_{24}}(g)$, or
%%\item $g$ in class $4B$ and $h$ in one of the classes $2B_3$, $4B_1$ (group 20), $2B_1$, $4B_5$ (group 21), $4A_3$ (group 23), $4A_4$ (group 24), $4B_4$ (group 25), $4B_7$ (group 26) of $C_{M_{24}}(g)$, or
%\item $g$ in class 8A and $h$ in class $2B_{1,2}$  (group 27) of $C_{M_{24}}(g)$
%\end{enumerate}
%the identities \be\label{conja3}  \sum_{b=0}^{N-1}\frac{e^{\frac{2\pi i rb}{N}}}{N}\phi_{g_{\alpha}^m,g_{\alpha}^{-b}\varphi(h^k_\alpha)}(\tau,z)
%=
%\sum_{s=0}^{N-1}\frac{e^{\frac{2\pi i ms}{N}}}{N}\phi_{g^r_\alpha,g^{-s}_\alpha h_\alpha^k}(\tau,z)\ ,
%\ee
% are satisfied for all $m,r,k\in \ZZ$.
\end{theorem}

\medskip

The identities \eqref{conja2} can be easily verified by considering the modular properties of both sides and comparing a sufficient number of Fourier coefficients. This is a tedious but totally straightforward calculation, so we omit the details. We leave a systematic analysis of these identities and the complete proof of the conjecture to future work.

Notice that whenever a generator $g$ of a non-cyclic abelian subgroup $\langle g,h\rangle\subset M_{24}$ is also an element of $M_{23}\subset M_{24}$, then it is necessarily in one of the classes 2A, 3A, 4B or 8A and theorem \ref{thm:relabel} applies (see appendix \ref{app:tables} for a list of all such groups and their generators).  As will be explained in the subsection \ref{sec:Sdualityinvariance}, the identities \eqref{conja2} are sufficient to establish the `S-duality invariance'  of the corresponding automorphic forms $\Phi_{g,h}$. The conjecture \ref{conj:relab} is trivially true in the degenerate case where $g$ is the identity, so the automorphic forms $\Phi_{e,h}$ associated with cyclic groups satisfy an analogous property \cite{Cheng:2012uy,Raum12}.  For the groups $\langle g,h\rangle$ that have no generator in $M_{23}$, the modular properties of $\Phi_{g,h}$ will be determined in a different way in section \ref{sec:Sdualityinvariance}.

\medskip

\subsubsection{$S$-duality invariance}
\label{sec:Sdualityinvariance}

In this section, we will show that the functions $\Phi_{g,h}(\sigma,\tau,z)$ defined in terms of the infinite product \eqref{infprod}, satisfy identities of the form
\be\label{Sdual} \Phi_{g,h}(\sigma,\tau,z)\sim \Phi_{g,h'}(\frac{\tau}{N\lambda}+x,N\lambda \sigma,z)\ ,
\ee for some suitable $h'\in C_{M_{24}}(g)$ and real $x$ that depends on the cocycle $\alpha$. Here, $\sim$ denotes equality up to a phase, that depends on the cocycle $\alpha$. More precisely, for each abelian subgroup $\langle g,h\rangle\subset M_{24}$, we will prove identities of the form \eqref{Sdual} for at least one pair of generators $(g,h)$. Furthermore, in all such cases the cocycle $\alpha$ can be chosen in such a way that $x=0$. For the remaining pairs $(g,h)\in \P$, the identities \eqref{Sdual} can be derived through a complete analysis of the automorphic properties of $\Phi_{g,h}$, as discussed in section \ref{sec_modular}.

When the function $\Phi_{g,h}$ admits a physical interpretation as the generating functions for twisted multiplicities of 1/4 BPS states in a CHL model, the identity \eqref{Sdual} corresponds to S-duality exchanging electric and magnetic charges in the low energy effective action.

\medskip
\noindent {\bf Case 1:} {\it $g$ in one of the classes $1A$, $2A$, $3A$, $4B$ or $8A$}

\vspace{.2cm}

\noindent Let $g$ be an element of $M_{23}\subset M_{24}$ of order $N$ in one of the classes considered in theorem \ref{thm:relabel}. The restriction of $[\alpha]$ to $H^3(\langle g\rangle,U(1))$ is trivial, so that $\lambda=1$ and we can choose a representative $\alpha$ satisfying \eqref{easycoc} and \eqref{triviac}. Consider an element $h\in C_{M_{24}}(g)$ and let $M$ be the order of the lift $h_\alpha \in C^\alpha_{M_{24}}(g)$.
%Let us identify $g=g'$ and the centralizers $C_{M_{24}}(g)= C_{M_{24}}(g')$ and consider the same cocycle $\alpha=\alpha'$. 
Then
eq.\eqref{conja2} implies that for any $d,m,k\in \ZZ_{\ge 0}$, there are $h'\in C_{M_{24}}(g)$ and $x'\in \ZZ/N\ZZ$ such that
 \begin{align} \sum_{s=0}^{N-1}&\frac{e^{\frac{2\pi i ds}{N}}}{N}\epsilon_{g,h}\left(\begin{smallmatrix}
k & s\\0 & m 
 \end{smallmatrix}\right)\phi_{{g}^m,{g}^{-s}{h'}^k}(\tau,z)
=
\sum_{b=0}^{N-1}\frac{e^{\frac{2\pi i m(b-kx')}{N}}}{N}\epsilon_{g,h}\left(\begin{smallmatrix}
k & b\\0 & d 
 \end{smallmatrix}\right)\phi_{g^d,g^{-b}h^k}(\tau,z)\ .
 \end{align}
 Thus,
 \begin{align} \hat c_{g,h}(d,m,\ell ,t)=&\sum_{k=0}^{M-1} \frac{e^{-\frac{2\pi itk}{M}}}{M}\sum_{b=0}^{N-1} \frac{e^{\frac{2\pi i bm}{N}}}{ N}\epsilon_{g,h}\left(\begin{smallmatrix}
k & b\\ 0 & d
\end{smallmatrix}\right)c_{g^{d},g^{-b}h^k}(\frac{md}{N},\ell)\\
=&\sum_{k=0}^{M-1} e^{\frac{2\pi ik mx'}{N}}\frac{e^{-\frac{2\pi itk}{M}}}{M}\sum_{s=0}^{N-1}\frac{e^{\frac{2\pi i ds}{N}}}{N}\epsilon_{g,h'}\left(\begin{smallmatrix}
k & s\\0 & m 
 \end{smallmatrix}\right)c_{{g}^{m},{g}^{-s}{h'}^k}(\frac{md}{N},\ell)\\
 &=\hat c_{g,h'}(m,d,\ell ,t+mx'\frac{M}{N})\ .
\end{align}
If $h$ and $h'$ are in different conjugacy classes of $C_{M_{24}}(g)$, we can choose the representative cocycle $\alpha$ in the class $[\alpha]$ in such a way that $x'=0$, so that
\begin{align}
\Phi_{g,h}(\sigma,\tau,z)= &p q^{\frac{1}{N}} y\prod_{(d,m,\ell)>0}\prod_{t=0}^{M-1}(1-e^{\frac{2\pi it}{M}}q^{\frac{m}{N}}y^\ell p^d)^{\hat c_{g,h}(d,m,\ell ,t)}\\
= &p q^{\frac{1}{N}} y\prod_{(d,m,\ell)>0}\prod_{t=0}^{M-1}(1-e^{\frac{2\pi it}{M}}q^{\frac{m}{N}}y^\ell p^d)^{\hat c_{g,h'}(m,d,\ell ,t)}\\
=&\Phi_{g,h'}(\frac{\tau}{N},N\sigma,z)\ .
\end{align} When $h$ and $h'$ are in the same conjugacy class of $C_{M_{24}}(g)$, then $x'$ cannot be eliminated in general. This corresponds to the case when $\varphi(h_\alpha)$ is conjugated with $Q^{x'}h_\alpha$ in the central extension $C^\alpha_{M_{24}}(g)$. In particular, since $\varphi(h_\alpha)^M=\varphi(h_\alpha^M)=e$, this implies that $Q^{x'}$ has order $M$, i.e. there is an integer $x$ such that
\be x'M=xN\ .
\ee Thus,
\begin{align}
\Phi_{g,h}(\sigma,\tau,z)= &p q^{\frac{1}{N}} y\prod_{(d,m,\ell)>0}\prod_{t=0}^{M-1}(1-e^{\frac{2\pi it}{M}}q^{\frac{m}{N}}y^\ell p^d)^{\hat c_{g,h}(d,m,\ell ,t)}\\
= &p q^{\frac{1}{N}} y\prod_{(d,m,\ell)>0}\prod_{t=0}^{M-1}(1-e^{\frac{2\pi it}{M}}q^{\frac{m}{N}}y^\ell p^d)^{\hat c_{g,h}(m,d,\ell ,t+mx)}\\
= &p q^{\frac{1}{N}} y\prod_{(d,m,\ell)>0}\prod_{t'=0}^{M-1}(1-e^{\frac{2\pi it'}{M}}e^{-\frac{2\pi ixm}{M}}q^{\frac{m}{N}}y^\ell p^d)^{\hat c_{g,h}(m,d,\ell ,t')}\\
= &p q^{\frac{1}{N}} y\prod_{(d,m,\ell)>0}\prod_{t'=0}^{M-1}(1-e^{\frac{2\pi it'}{M}}e^{-\frac{2\pi ix'm}{N}}q^{\frac{m}{N}}y^\ell p^d)^{\hat c_{g,h}(m,d,\ell ,t')}\\
=&e^{\frac{2\pi ix'}{N}}\Phi_{g,h}(\frac{\tau-x'}{N},N\sigma,z)\ .
\end{align} 
In fact, $x'\neq 0$ only when $g$ is in class 4B and $h$ in one of the classes 2B$_3$ or 4B$_1$ of $C_{M_{24}}(g)$. In both cases, the pair $g,h$ generates group 20 in the list of appendix \ref{app:tables} and $x'=2$.  For such pairs $(g,h)$, however, since $T^\alpha_L\phi_{g,h}=0$ for $L$ odd, the function $\Phi_{g,h}$ satisfies the identity
\be \Phi_{g,h}(\sigma,\tau,z)=-\Phi_{g,h}(\sigma+\frac{1}{2},\tau,z)\ .
\ee Using this identity, we conclude
\be \Phi_{g,h}(\sigma,\tau,z)=\Phi_{g,h}(\frac{\tau}{N},N\sigma,z)\ .
\ee
If the conjecture \ref{conj:relab} holds, then an S-duality property holds for all $\Phi_{g,h}$ such that $g$ is an element of $M_{23}\subset M_{24}$.

\medskip
\noindent {\bf Case 2:} {\it $g$ in one of the classes $2B$, $3B$, $4C$, $6B$, $12B$}

\vspace{.2cm}

\noindent The calculations in this case are rather technical and have therefore been relegated to appendix \ref{sec_case2}. Below we summarize the results.

 For a suitable choice of the representative 3-cocycle $\alpha$, we have 
\be \Phi_{g,h}(\sigma,\tau,z)=\Phi_{g,h^{-1}}(\frac{\tau}{N^2},N^2\sigma,z)\ ,
\ee if $h$ and $h^{-1}$ are not conjugated within $C_{M_{24}}(g)$, and
\be \Phi_{g,h}(\sigma,\tau,z)=\Phi_{g,h}(\frac{\tau}{N^2},N^2\sigma,z)\ ,
\ee otherwise. 
%In particular, the latter holds for a choice of the cocycle $\alpha$ such that
%\be \phi_{g,h}(\tau,z)=\phi_{g,h}^*(\tau,z)\ ,
%\ee i.e. such that all Fourier coefficients of $\phi_{g,h}$ are real.

\subsection{Automorphic properties}\label{sec_modular}

Using the results of the previous sections, we can now prove that the functions $\Phi_{g,h}(\sigma,\tau,z)$, defined by the analytic continuation of the infinite product \eqref{infprod} to the Siegel upper half-space $\mathbb{H}_2$ are automorphic forms under certain subgroups of $Sp(4,\RR)$ acting by
\be \left(\begin{matrix}
A & B\\ C & D
\end{matrix}\right)\cdot Z =(AZ+B)(CZ+D)^{-1}\ ,\qquad \left(\begin{matrix}
A & B\\ C & D
\end{matrix}\right)\in Sp(4,\RR)
\ee where \be Z=\left(\begin{matrix}
\tau & z\\ z & \sigma 
\end{matrix}\right)\ .\ee
Recall that the function $\Phi_{g,h}$ can be written as  
\be\label{onceagain} \Phi_{g,h}(\sigma,\tau,z)=p \psi_{g,h}(\tau,z)\exp[-\sum_{N=1}^\infty p^L(\T^\alpha_L \phi_{g,h})(\tau,z)]\ ,
\ee  where the right-hand side converges on a suitable domain in $\mathbb{H}_2$. 

Let us consider the action of various generators of $Sp(4,\RR)$ on $\Phi_{g,h}$.

\begin{itemize}
\item The Heisenberg subgroup $H(\ZZ)$ of $Sp(4,\ZZ)$ is defined as
\be
H(\ZZ)=\{[\zeta,\mu,\kappa]:=\begin{pmatrix}
 1 & 0 & 0 & \mu\\ 
 \zeta & 1 & \mu & \kappa\\ 
 0 & 0 & 1 & -\zeta\\ 
 0 & 0 & 0 & 1\\    
 \end{pmatrix},\zeta,\mu,\kappa\ \in \ZZ\}\ ,
\ee and acts by
\be [\zeta,\mu,\kappa]\cdot (\sigma,\tau,z)=(\sigma+\kappa+2\zeta z+\zeta^2\tau,\tau,z+\mu+\zeta\tau)\ .
\ee
By \eqref{onceagain}, using the elliptic properties of the Jacobi forms $\psi_{g,h}$  and $\T^\alpha_L\phi_{g,h}$, we conclude easily that every $\Phi_{g,h}$ is invariant under $H(\ZZ)$.
\item For many commuting pairs $g,h\in M_{24}$, the genera $\T^\alpha_L\phi_{g,h}$ vanish unless $L$ is an integer multiple of some $r\equiv r_{g,h}\in \ZZ$. Therefore, $\Phi_{g,h}$ is invariant up to a multiplier under $\sigma\mapsto \sigma+\frac{1}{r}$, i.e.
\be \Phi_{g,h}([0,0,\frac{\kappa}{r}]\cdot(\sigma,\tau,z))=e^{-\frac{2\pi i\kappa}{r}}\Phi_{g,h}(\sigma,\tau,z)\ ,
\ee where $[0,0,\frac{\kappa}{r}]\cdot(\sigma,\tau,z)=(\sigma+\frac{\kappa}{r},\tau,z)$. In particular, $r_{g,h}=1$ when $g,h$ generate a cyclic group and for groups 13, 23, 24 and 27, $r=3$ for groups 33 and 34, $r=4$ for groups 25 and 26 and $r=2$ in all the other cases. Notice that $r_{g,h}$ always divides $N\lambda$.\\
 More generally, the transformation $[0,0,\frac{\kappa}{N\lambda}]$ relates the functions $\Phi_{g,h}$ relative to two distinct choices of the cocycle $\alpha$, that differ from one each other by a $N\lambda$-root of unity.
\item The twisted twining genera $\T^\alpha_L\phi_{g,h}$ are Jacobi forms of weight $0$ and index $L$ under a group $\Gamma_{g,h}\subset SL(2,\ZZ)$  (see the tables in appendix \ref{app:tables}), up to a multiplier $\chi_{g,h}^L$. As noticed above, for each group $\langle g,h\rangle$, there is an integer $r\equiv r_{g,h}$ such that $\T^\alpha_L\phi_{g,h}$ vanishes unless $r|L$. Therefore, only the power $\chi_{g,h}^r$ needs to be a well-defined character of $\Gamma_{g,h}$.\\
On the other hand, the properties \eqref{etamodular} and \eqref{etaconj} imply that the eta products $\eta_{g,h}$ are modular forms of weight $w$ and multiplier $\upsilon_{g,h}$ under the same group $\Gamma_{g,h}$ associated with $\phi_{g,h}$. Equivalently, each $\psi_{g,h}$ is a weak Jacobi form of weight $w-2$, index $1$ and multiplier $\upsilon_{g,h}$ under $\Gamma_{g,h}$.\\
For each $(\gamma,k)\in SL(2,\ZZ)\times M_{24}$, let us choose $\mu_{g,h}(\gamma,k)\in \RR/\ZZ$ such that
\be e^{2\pi i r_{g,h}\mu_{g,h}(\gamma,k)}=\Bigl(\epsilon_{g,h}(\gamma,k)\Bigr)^{r_{g,h}}\ ,\qquad \gamma\in SL(2,\ZZ)\ .
\ee Then for any $\gamma=\left(\begin{smallmatrix}
a & b\\ c & d
\end{smallmatrix}\right)\in SL(2,\ZZ)$ and $k\in M_{24}$, we have
\be\upsilon_{g,h}(\gamma)\Phi_{(\gamma,k)\cdot(g,h)}\Bigl(\xi(\gamma)\cdot Z \Bigr)=\det (c\tau+d)^{w-2}\, \Phi_{g,h}(\sigma+\mu_{g,h}(\gamma,k),\tau,z)\ ,
\label{etatransf}
\ee  
where  $Z=\left(\begin{smallmatrix}
\tau & z\\ z & \sigma 
\end{smallmatrix}\right)$ and
\be \xi\left(\begin{matrix}
a & b\\ c & d
\end{matrix}\right):= \begin{pmatrix} a & 0 & b  & 0\\ 0 & 1 & 0 & 0 \\ c& 0 & d &0\\ 0 & 0& 0 & 1\end{pmatrix}\ ,\ee
so that 
\be \xi(\gamma)\cdot \begin{pmatrix}
\tau & z\\ z & \sigma 
\end{pmatrix}= \begin{pmatrix} \frac{a\tau+b}{c\tau+d} & \frac{z}{c\sigma+d}\\ \frac{z}{c\sigma+d} & \sigma-\frac{z^2}{c\sigma+d}\end{pmatrix}\ .\ee
In particular, by taking $(\gamma,k)\in M_{24}$ stabilizing $(g,h)$, we obtain
\be \upsilon'_{g,h}(\gamma,k)\Phi_{g,h}\Bigl([0,0,-\mu_{g,h}(\gamma,k)]\cdot\bigl(\xi(\gamma)\cdot Z\bigr) \Bigr)=\det (c\tau+d)^{w-2}\, \Phi_{g,h}(\sigma,\tau,z)\ ,
\ee where  $\gamma\in \Gamma_{g,h}$ and \be
\upsilon'_{g,h}(\gamma,k)=\upsilon_{g,h}(\gamma)e^{-2\pi i \mu_{g,h}(\gamma,k)}\ .
\ee
\item Finally, for several pairs $(g,h)$ of commuting elements of $M_{24}$, we have proved that
\be\label{Sdudu} \Phi_{g,h}(Z)=\Phi_{g,h'}(V_{N\lambda}\cdot Z)\ ,
\ee for a suitable $h'\in C_{M_{24}}(g)$ and a suitable choice of the cocycle $\alpha$, where
\be\label{viti} V_{t}=\frac{1}{\sqrt{t}}\begin{pmatrix}
 0 & t & 0 & 0\\ 
 1 & 0 & 0 & 0\\
 0 & 0 & 0 & 1\\
 0 & 0 & t & 0  
 \end{pmatrix}\ ,
\ee acts by
\be V_t\cdot \begin{pmatrix}
\tau & z\\ z & \sigma 
\end{pmatrix}=\begin{pmatrix}
t\sigma & z\\ z &  \frac{\tau}{t}
\end{pmatrix}\ .
\ee
 Notice that in each of the $55$ conjugacy classes of  abelian subgroups  $\langle g,h\rangle\subset M_{24}$, an identity of the form \eqref{Sdudu} has been proved for at least one pair of generators.
\end{itemize}

As discussed in \cite{GritsenkoNikulin2,GritsenkoClery}, for any integers $N,t>0$, the elements \be V_t;\qquad\qquad  \xi(\gamma)\ ,\quad \gamma\in \Gamma_0(N);\, \qquad \qquad [\zeta,\mu,\kappa/t]\ ,\quad \zeta,\mu,\kappa\in\ZZ\ ,
\ee generate the group
$\Gamma^+_t(N)=\langle \Gamma_t(N), V_t\rangle \subset  Sp(4,\RR)$  which is a normal double extension of the paramodular group
\be \Gamma_t(N)=\{ \left(\begin{matrix}
* & t* & * & * \\
* & * & * & t^{-1}*\\
N* & Nt* & * & * \\
Nt* & Nt* & t* & * 
\end{matrix}\right)\in Sp(4,\QQ),\ *\in \ZZ\}\ .
\label{para}
\ee 
 From the discussion above, it follows that every $\Phi_{g,h}$ is a modular function under some finite index subgroup $\Gamma^{(2)}_{g,h}$ of a paramodular  group $\Gamma_t\equiv \Gamma_t(1)$, for some suitable $t$. The image of $\Phi_{g,h}$ under the action of a generic element of $\Gamma_t$ is expected to be again a function $\Phi_{g',h'}$ for some (possibly different) commuting pair $g',h'\in M_{24}$, and defined with respect to a suitable choice of the cocycle $\alpha$. 
 
Under the action of $Sp(4,\ZZ)$, the functions $\Phi_{g,h}(\sigma,\tau,z)$, for all commuting $g,h\in M_{24}$, are mapped into one another, up to shifts and rescalings of the arguments, i.e., schematically, $\Phi_{g,h}(\gamma\cdot (\sigma,\tau,z))\sim \Phi_{g',h'} (\sigma/t+x,t\tau,z)$ for suitable $t\in\ZZ$, $x\in \RR$ (we dropped the automorphy factors). In particular, $\Phi_{g,h}$ is a Siegel modular form for some congruence subgroup of $Sp(4,\ZZ)$, as proved in the following theorem.

\begin{theorem}
For each pair of commuting elements $g,h\in M_{24}$ and for a suitable choice of the cocycle $\alpha$, the function $\Phi_{g,h}(Z)$ is a meromorphic Siegel modular function of weight $w-2$ and level $H$ for a suitable $H$, i.e. it is a meromorphic function on the upper half space $\HH_2$ such that
\be \Phi_{g,h}\bigl((AZ+B)(CZ+D)^{-1}\bigr)=\det (CZ+D)^{w-2} \Phi_{g,h}(Z)\ ,
\ee for all $\left(\begin{smallmatrix}
A & B\\ C & D
\end{smallmatrix}\right)\in \Gamma^{(2)}(H):=\ker (Sp(4,\ZZ)\to Sp(4,\ZZ/H\ZZ))$. Here, $w$ is the weight of the corresponding Mason's eta function $\eta_{g,h}$.
\end{theorem}

\medskip

\noindent \emph{Proof.} It is sufficient to prove the statement for those `special pairs' $(g,h)$ for which we proved that $\Phi_{g,h}(\sigma,\tau,z)=\Phi_{g,h'}(\tau/t,t\sigma,z)$ for a suitable choice of representative cocycle $\alpha$. Here, $h'$ is some element in $C_{M_{24}}(g)$ (possibly different from $h$) and $t=N\lambda$, where, as usual, $N$ denotes the order of $g$ and $\lambda$ the length of the shortest cycle of $g$ as a permutation of $24$ objects. Every other commuting pair of $M_{24}$ elements can be obtained by an $SL(2,\ZZ)$ transformation $(g,h)\cdot \gamma$ of one such `special pair' $(g,h)$, so that $\Phi_{(g,h)\cdot \gamma}$ is related to $\Phi_{g,h}$ by an $Sp(4,\ZZ)$ transformation $\xi(\gamma)$. Thus, if $\Phi_{g,h}$ is a modular function under $\Gamma^{(2)}(H)$, then $\Phi_{(g,h)\cdot \gamma}$ is a modular function under $\xi(\gamma)\Gamma^{(2)}(H)\xi(\gamma)^{-1}=\Gamma^{(2)}(H)$.

There is some integer $n$ such that the functions $\T^\alpha_L\phi_{g,h}$, $L\in \ZZ_{>0}$, and $\psi_{g,h}$ are weak Jacobi forms of weight $0$ and index $L$ (respectively, $w-2$ and $1$) with trivial multiplier under the congruence subgroup $\Gamma(n):=\ker (SL(2,\ZZ)\to SL(2,\ZZ/n\ZZ))$. We can choose $n$ so that also $\T^\alpha_L\phi_{g,h'}$ and $\psi_{g,h'}$ are weak Jacobi forms with trivial multiplier under the same group. Note that $n$ is necessarily a multiple of $t=N\lambda$, since this is the smallest integer for which $\phi_{g,h}(\tau+N\lambda,z)=\phi_{g,h}(\tau,z)$. Thus, both $\Phi_{g,h}(Z)$ and $\Phi_{g,h'}(Z)=\Phi_{g,h}(V_t\cdot Z)$ transform as modular forms of weight $w-2$ under $\xi(\Gamma(n))$ and $[\zeta,\mu,0]$, for all $\zeta,\mu\in\ZZ$. Equivalently, $\Phi_{g,h}$ transforms as a Siegel modular form of weight $w-2$ under the group generated by
\be \xi(\Gamma(n)),\qquad V_t\xi(\Gamma(n))V_t,\qquad [\zeta,\mu,0],\qquad V_t[\zeta,\mu,0]V_t\qquad \zeta,\mu\in\ZZ\ ,
\ee which is a subgroup of $Sp(4,\ZZ)$, since $t|n$. We will prove that this group contains $\Gamma^{(2)}(H)$, with $H=nt$. By Theorem 12.4 and Proposition 13.2 of \cite{Bass}, $\Gamma^{(2)}(nt)$ is generated by $\xi(\Gamma(nt))$ together with matrices of the form $(\begin{smallmatrix}
\II & ntB\\ 0 & \II
\end{smallmatrix})$, $(\begin{smallmatrix}
\II & 0\\ ntC & \II
\end{smallmatrix})$, with $B,C$ arbitrary $2\times 2$ integral symmetric matrices.  It is easy to check that every element of the form $(\begin{smallmatrix}
\II & ntB\\ 0 & \II
\end{smallmatrix})\in \Gamma^{(2)}(nt)$ is contained in the group generated 
by $[\zeta,\mu,0]$ and $\xi(\begin{smallmatrix}
1 & bn\\ 0 & 1
\end{smallmatrix})$, with $\zeta,\mu,b\in \ZZ$, and every element of the form $(\begin{smallmatrix}
\II & 0\\ ntC & \II
\end{smallmatrix})\in \Gamma^{(2)}(nt)$ is contained in the group generated by $\xi(\begin{smallmatrix}
1 & 0\\ cn & 1
\end{smallmatrix})$, $V_t\xi(\begin{smallmatrix}
1 & 0\\ cn & 1
\end{smallmatrix})V_t$ and $V_t[\zeta,0,0]\xi(\begin{smallmatrix}
1 & 0\\ n & 1
\end{smallmatrix})[-\zeta,0,0]V_t$, for all $\zeta,c\in \ZZ$. Clearly $\xi(\Gamma(nt))$ is a subgroup of $\xi(\Gamma(n))$, and this concludes the proof. \qed

\medskip
 
\noindent In section \ref{s:examples}, we will discuss the precise automorphic properties of some of these functions $\Phi_{g,h}$. We leave the detailed description of the groups $\Gamma^{(2)}_{g,h}$ for all pairs $g,h$ to a future work.

\subsection{Examples}\label{s:examples}

Having established that the $\Phi_{g,h}$ are indeed Siegel modular forms we now wish to analyze some specific examples in detail. We denote the pairs $(g,h)$ by the $M_{24}$-conjugacy class of $g$ and the $C_{M_{24}}(g)$-conjugacy class of $h$. See appendix \ref{app:tables} for more details.

\medskip

\noindent {\bf $\Phi_{2A,2A_{2,3,5}}$ (Groups 1, 2, 3)}

\medskip

\noindent Groups 1, 2 and 3 are $\ZZ_2\times\ZZ_2$ groups that contain three elements in class 2A of $M_{24}$. The corresponding $\Phi_{g,h}$ are identical
\be \Phi_{2A,2A_2}(\sigma,\tau,z)=\Phi_{2A,2A_3}(\sigma,\tau,z)=\Phi_{2A,2A_5}(\sigma,\tau,z)\ .
\ee The $3$-cocycle $\alpha$ can be chosen to have trivial restriction to these groups. The Hecke transforms of the twisted twining genera $\T_L\phi_{g,h}$ vanish unless $L$ is a multiple of $r=2$; in this case, they are Jacobi forms under $\Gamma_{g,h}=SL(2,\ZZ)$ with trivial multiplier $\chi_{g,h}^2=1$. The eta products for these elements are given by
\be \eta_{g,h}(\tau)=\eta(\tau)^{12}\ , 
\ee and are modular forms of weight $w=6$ under $SL(2,\ZZ)$ with multiplier (see eq. (\ref{etatransf}))
\be \upsilon_{g,h}(T)=-1\qquad \upsilon_{g,h}(S)=-1\ .
\ee Furthermore, from the results of section \ref{sec:Sdualityinvariance} we deduce that, since $g\in M_{23}$ (with $N=2$ and $\lambda=1$), $\Phi_{g,h}$ satisfies 
\begin{align}
\Phi_{2A,2A_2}(\sigma,\tau,z)=\Phi_{2A,2A_3}(\frac{\tau}{2},2\sigma,z)\ ,\qquad \Phi_{2A,2A_5}(\sigma,\tau,z)=\Phi_{2A,2A_5}(\frac{\tau}{2},2\sigma,z)\ .
\end{align}
Thus, $\Phi_{g,h}$ is a modular form of weight $w-2=4$ with a multiplier $\upsilon_{g,h}$ given above under the subgroup of $Sp(4,\RR)$ generated by the Heisenberg group $H(\ZZ)$, by $SL(2,\ZZ)$  and under $V_2$. These elements of $Sp(4,\RR)$ generate the paramodular group $\Gamma^+_2(1)$. Furthermore, $\Phi_{g,h}$ has a double zero at the rational quadratic divisor $z\to 0$ and at all modular images of this divisor. This allows to identify $\Phi_{g,h}$ as
\be \Phi_{g,h}(Z)=\Delta_2(Z)^2\ ,
\ee where $\Delta_2$ is the modular form of weight $2$ under $\Gamma^+_2(1)$ defined by Gritsenko and Nikulin in \cite{GritsenkoNikulin2}. Notice that in this case the function $\psi_{g,h}$ in (\ref{psighjacobi}) is given by 
\be
\psi_{g,h}(\tau, z)=-\vartheta_1(\tau, z)^2 \eta(\tau)^6,
\ee
which is the additive seed for $\Delta_2(Z)$ \cite{GritsenkoNikulin2}. Hence we conclude that in this case 
\be 
\Phi_{g,h}=\text{Add}[\psi_{g,h}],
\label{addliftex1}
\ee 
as claimed in section \ref{sec_addmult}. The Siegel modular form $(\Delta_2)^2$ Êalso appears in the context of umbral moonshine, where it corresponds to the Siegel modular form $\Phi^{(3)}(Z)$ (see section 2.6 of \cite{Cheng:2012tq}). In this context $\Phi^{(3)}$ arises as the multiplicative lift of the umbral Jacobi form $Z^{(3)}(\tau, z)$ of weight 0 and index 2. It is therefore interesting to ask how this relates to our construction of $\Phi_{g,h}$ as the multiplicative lift (\ref{infprod}). By comparing Fourier coefficients one can verify that we have the identity (see the table in appendix \ref{app:tables} for the result of the Hecke action)
\be
\T_2 \phi_{g,h}(\tau, z) = Z^{(3)}(\tau, z).
\ee
However, this does not necessarily imply that the two lifts are the same, since a priori our multiplicative lift in (\ref{infprod}) differ from that in \cite{Cheng:2012tq} since we are constructing it from the seed function $\phi_{g,h}$, which is a weak Jacobi form of weight 0 and index 1, using the  equivariant Hecke operator $\T_L$ (the twist by the 3-cocycle $\alpha$ is trivial in this case). Even so, by virtue of (\ref{addliftex1}), in the case at hand it turns out that the lifts do coincide and we thus have 
\be
\text{Mult}[\phi_{g,h}]=\Phi^{(3)}. 
\ee

\medskip

\noindent {\bf $\Phi_{2A,2A_4}$ (Group 7)}

\medskip

\noindent The function $\Phi_{2A,2A_4}$ of group $7$ satisfies
\be \Phi_{2A,2A_4}(V_2\cdot Z)=\Phi_{2A,4B_1}(Z)\ ,
\ee where $h'$ in class $4B_1$ of $C_{M_{24}}(g)$ is such that $h'^2=g$. Therefore, the right hand side is a $Sp(4,\ZZ)$ transformation of $\Phi_{e,4B}$, which is a Siegel modular form under $\Gamma_1(4)=\Gamma^{(2)}_0(4)$ \cite{Cheng:2012uy,Raum12}.

\medskip

\noindent {\bf $\Phi_{2A,2B_{1,2}}$ (Groups 8,9)}

\medskip

\noindent The functions $\Phi_{g,h}$ for groups 8 and 9 are identical
\be \Phi_{2A,2B_1}(Z)=\Phi_{2A,2B_2}(Z)\ .
\ee The twisted twining genera $\T^\alpha_L\phi_{g,h}$ vanish unless $L$ is a multiple of $r=2$; in this case, they are Jacobi forms under $\Gamma_{g,h}=\Gamma_0(2)$ with trivial multiplier. The eta products are given by
\be \eta_{g,h}(\tau)=\eta(\tau)^{4}\eta(2\tau)^{4}\ , 
\ee and are modular forms of weight $w=4$ under $\Gamma_0(2)$ with multiplier
\be \upsilon_{g,h}(\begin{smallmatrix}
1 & 1\\ 0 & 1
\end{smallmatrix})=-1\qquad \upsilon_{g,h}(\begin{smallmatrix}
1 & 0\\ 2 & 1
\end{smallmatrix})=-1\ .
\ee Therefore, $\Phi_{g,h}$ transforms as a modular form of weight $w-2=2$ under $\xi(\gamma)$, $\gamma\in\Gamma_0(2)$, with multiplier $\upsilon_{g,h}(\gamma)$.  Furthermore, for a suitable choice of the cocycle $\alpha$, we have
\begin{align}
\Phi_{2A,2B_{1,2}}(V_2\cdot Z)=\Phi_{2A,2B_{1,2}}(Z)\ .
\end{align}
We conclude $\Phi_{g,h}$ is a modular form of weight $2$ under the paramodular group $\Gamma_2(2)$. Assuming that it is holomorphic, then it can be identified as
\be \Phi_{g,h}(Z)=Q_1(Z)^2\ ,
\ee where $Q_1$ is the modular form of weight $1$ defined by Gritsenko and Clery in \cite{GritsenkoClery}. This conjecture is also supported by the fact that in this case $\psi_{g,h}$ is given by 
\be
\psi_{g,h}(\tau, z)=-\frac{\vartheta_1(\tau, z)^2}{\eta(\tau)^2}\eta(2\tau)^4,
\ee
which is the square of the additive seed for $Q_1$ (see eq. (16) in \cite{GritsenkoClery}).

\medskip

\noindent {\bf $\Phi_{2A,4B_4}$ (Group 12)}

\medskip

\noindent The function $\Phi_{2A,4B_4}$ of group $12$ satisfies
\be \Phi_{2A,4B_4}(V_2\cdot Z)=\Phi_{2A,8A_1}(Z)\ ,
\ee where $h'$ in class $8A_1$ of $C_{M_{24}}(g)$ is such that $h'^4=g$. Therefore, the right hand side is a $Sp(4,\ZZ)$ transformation of $\Phi_{e,8A}$, which is a Siegel modular form under $\Gamma_1(8)=\Gamma^{(2)}_0(8)$ \cite{Raum12}.

\medskip

\begin{samepage}
\noindent {\bf $\Phi_{2A,4B_{2,3,5}}$ (Groups 17,18,19)}\\[5pt]
\noindent The functions $\Phi_{g,h}$ for groups 17, 18, 19 are identical
\be \Phi_{2A,4B_2}(Z)=\Phi_{2A,4B_3}(Z)=\Phi_{2A,4B_5}(Z)\ .
\ee
The Jacobi forms $\T^\alpha_L\phi_{g,h}$ for these groups have exactly the same modular properties as the ones for groups 8 and 9; the eta products are also the same. Therefore, $\Phi_{g,h}$ is a modular function of weight $2$ under the paramodular group $\Gamma_2(2)$. 
\end{samepage}

\medskip

\noindent {\bf $\Phi_{4B,4B_{4,7}}$ (Groups 25,26)}

\medskip

\noindent The functions $\Phi_{g,h}$ for groups 25 and 26 are identical
\be \Phi_{4B,4B_4}(Z)=\Phi_{4B,4B_7}(Z)\ .
\ee The $3$-cocycle $\alpha$ can be chosen to be trivial when restricted to $\langle g,h\rangle$. The Hecke-transformed twisted twining genera $\T_L\phi_{g,h}$ vanish unless $L$ is a multiple of $r=4$; in this case, they are Jacobi forms under $\Gamma_{g,h}=SL(2,\ZZ)$ with trivial multiplier. The eta products are given by
\be \eta_{g,h}(\tau)=\eta(\tau)^{6}\ , 
\ee and are modular forms of weight $w=3$ under $SL(2,\ZZ)$ with multiplier 
\be \upsilon_{g,h}(\begin{smallmatrix}
1 & 1\\ 0 & 1
\end{smallmatrix})=-i\qquad \upsilon_{g,h}(\begin{smallmatrix}
1 & 0\\ 2 & 1
\end{smallmatrix})=-i\ .
\ee   Furthermore, 
\begin{align}
\Phi_{4B,4B_4}(V_4\cdot Z)=\Phi_{4B,4B_7}(Z)\ .
\end{align}
We conclude $\Phi_{g,h}$ is a modular function of weight $w-2=1$ under the paramodular group $\Gamma_4^+(1)$ and the multiplicity at  the rational quadratic divisor $z\to 0$ is non-negative. It follows that
\be \Phi_{g,h}(Z)=\Delta_{1/2}(Z)^2\ ,
\ee where $\Delta_{1/2}$ is the modular form of weight $1/2$ defined in \cite{GritsenkoNikulin2}. Also in this case the additive lift matches since 
\be
\psi_{g,h}(\tau, z)= -\vartheta_1(\tau, z)^2
\ee
is the square of the additive seed for $\Delta_{1/2}(Z)$, and hence $\text{Add}[\psi_{g,h}]=(\Delta_{1/2}(Z))^2$. 

The function $(\Delta_{1/2}(Z))^2$ also corresponds to the umbral Siegel modular form $\Phi^{(5)}(Z)$, which is the multiplicative lift of the umbral Jacobi form $Z^{(5)}(\tau, z)$ of weight 0 and index 4 \cite{Cheng:2012tq}. The relation between $Z^{(5)}(\tau, z)$ and the twisted twining genus $\phi_{g,h}$ is now:
\be
(\T_4\phi_{g,h})(\tau, z)=Z^{(4)}(\tau, z).
\ee 
Again, it is not a priori clear that the equivariant multiplicative lift of $\phi_{g,h}$, defined by (\ref{infprod}), will coincide with the ordinary multiplicative Borcherds lift of $Z^{(5)}(\tau, z)$ considered in \cite{Cheng:2012tq}. However, in the case at hand they do:
\be
\text{Mult}[\phi_{g,h}]=\Phi^{(5)}.
\ee

\medskip 

\noindent {\bf $\Phi_{3A,3A_3}$ (Group 33)}

\medskip

\noindent Consider the function $\Phi_{3A,3A_3}$ of groups 33.
 The twisted twining genera $\T^\alpha_L\phi_{g,h}$ vanish unless $L$ is a multiple of $r=3$; in this case, they are Jacobi forms under $\Gamma_{g,h}=SL(2,\ZZ)$ with trivial multiplier. The eta product is given by
\be \eta_{g,h}(\tau)=\eta(\tau)^{8}\ , 
\ee and is a modular form of weight $w=4$ under $SL(2,\ZZ)$ with multiplier 
\be \upsilon_{g,h}(\begin{smallmatrix}
1 & 1\\ 0 & 1
\end{smallmatrix})=e^{-\frac{2\pi i}{3}}\qquad \upsilon_{g,h}(\begin{smallmatrix}
0 & -1\\ 0 & 1
\end{smallmatrix})=1\ .
\ee   Furthermore, 
\begin{align}
\Phi_{3A,3A_3}(V_3\cdot Z)=\Phi_{3A,3A_3}(Z)\ .
\end{align}
We conclude $\Phi_{g,h}$ is a modular function of weight $w-2=2$ under the paramodular group $\Gamma_3^+(1)$ and the analysis of the multiplicities at its divisors shows that it is holomorphic with a double zero at $z=0$. It follows that
\be \Phi_{3A,3A_3}(Z)=\Delta_{1}(Z)^2\ ,
\ee where $\Delta_{1}$ is the modular form of weight $1$ defined  in \cite{GritsenkoNikulin2}. Also here we find that 
\be
\psi_{g,h}(\tau, z)= -\vartheta_1(\tau, z)^2 \eta(\tau)^2
\ee
is the square of the additive seed for $\Delta_1(Z)$. The function $\Delta_1(Z)^2$ coincides with the umbral Siegel modular form $\Phi^{(4)}(Z)$, which is the multiplicative lift of the umbral Jacobi form $Z^{(4)}(\tau, z)$ of weight 0 and index 3. In this case we have that, for a suitable choice of cocycle $\alpha$,
\be
(\T^\alpha_3 \phi_{g,h})(\tau, z)=Z^{(3)}(\tau, z), 
\ee
and the equivariant and multiplicative lifts coincide:
\be
\text{Mult}[\phi_{g,h}]=\Phi^{(4)}.
\ee

\medskip

\noindent {\bf $\Phi_{3A,3B_1}$ (Group 34)}

\medskip

Consider the function $\Phi_{3A,3B_1}$ of group 34.
 The twisted twining genera $\T^\alpha_L\phi_{g,h}$ vanish unless $L$ is a multiple of $r=3$; in this case, they are Jacobi forms under $\Gamma_{g,h}=\Gamma_0(3)$ with trivial multiplier. The eta product is given by
\be \eta_{g,h}(\tau)=\eta(\tau)^{2}\eta(3\tau)^{2}\ , 
\ee and is a modular form of weight $w=2$ under $SL(2,\ZZ)$ with multiplier 
\be \upsilon_{g,h}(\begin{smallmatrix}
1 & 1\\ 0 & 1
\end{smallmatrix})=e^{-\frac{2\pi i}{3}}\qquad \upsilon_{g,h}(\begin{smallmatrix}
1 & 0\\ 3 & 1
\end{smallmatrix})=e^{\frac{2\pi i}{3}}\ .
\ee   Furthermore, 
\begin{align}
\Phi_{3A,3B_1}(V_3\cdot Z)=\Phi_{3A,3B_1}(Z)\ .
\end{align}
We conclude $\Phi_{g,h}$ is a modular function of weight $w-2=0$ under the paramodular group $\Gamma_3^+(3)$. Since the weight vanishes, it must necessarily be meromorphic. 

\section{Conclusions}
\label{sec_conclusions}
\noindent In this paper we have proposed a second quantized version of (generalized) Mathieu moonshine, involving a class of Siegel modular forms $\Phi_{g,h}$ for discrete subgroups $\Gamma_{g,h}^{(2)}\subset Sp(4,\mathbb{R})$,  constructed from a multiplicative lift of the twisted twining genera $\phi_{g,h}$. For certain pairs of conjugacy classes of $M_{24}$ we were able to identify $\Phi_{g,h}$ with known Siegel modular forms. It would be interesting to extend these results and perform a more detailed investigation of all the modular groups $\Gamma_{g,h}^{(2)}$ and determine whether the remaining $\Phi_{g,h}$ coincide with known objects, or perhaps constitute new examples of Siegel modular forms. In \cite{Raum12}, Raum proved modularity for most of the cases $\Phi_{e,h}$ and found that not all of them are of a standard Borcherds product type, but in fact correspond to certain rescaled products of Borcherds modular forms. One would like to extend this analysis to determine whether the $\Phi_{g,h}$ for $g\neq e$ also contain such rescaled Borcherds products, or some generalization thereof.

As already mentioned in the introduction, an interesting by-product of our analysis is the fact that some of the Siegel modular forms $\Phi_{g,h}$ coincide with multiplicative lifts of the umbral Jacobi forms analyzed in \cite{Cheng:2012tq}. This might be a simple consequence of the constraints from modularity, but it might also indicate some deeper relation between umbral moonshine and generalized Mathieu moonshine which would be interesting to uncover. Could it be that some of the other products $\Phi_{g,h}$ also coincide with lifts of the more general $D_n$ or $E_n$ Niemeier-umbral Jacobi forms in \cite{Cheng:2013wca}?

It is natural to wonder whether our products $\Phi_{g,h}$ have interpretations in terms of denominator formulas of some generalized Kac-Moody algebras (GKMs). It is well-known that the inverse square root of the Igusa cusp form $\Phi_{10}=\Phi_{e,e}$  constitutes one side  of the denominator formula for a rank 3 GKM-algebra of hyperbolic type \cite{GritsenkoNikulin} (the other side corresponds to the additive lift in (\ref{addliftPhi10})). In the terminology of Borcherds \cite{Borcherds}, the other functions $\Phi_{e,h}$ then correspond to twisted denominator formulas for the same algebra. On the other hand, by analogy with generalized Monstrous moonshine \cite{CarnahanII,CarnahanIV}, we would expect that the functions $\Phi_{g,e}$ give denominator formulas for a class of rank 3 GKM-algebras. For some of the elements $g$ Êof small order these algebras have indeed been constructed in \cite{Cheng:2008fc,Cheng:2008kt,Govindarajan:2008vi,Govindarajan:2009qt} in the context of CHL-models (see also \cite{GritsenkoNikulin,GritsenkoNikulinSimplest,GritsenkoNikulin2} for earlier mathematical results). In this context the prefactor $ p q^{\frac{1}{N\lambda}} y$ in (\ref{infprod}) should  have an interpretation as the exponential of the Weyl vector $\rho$ of the algebra. If true one would expect that for fixed $g$, the associated twisted denominator formulas $\Phi_{g,h}$ all have the same prefactor involving the Weyl vector of the original GKM-algebra determined by the class $[g]$. In other words, the prefactor should be independent of the twining element $h$, and this is indeed what we find. In fact, the Weyl vectors extracted from $ p q^{\frac{1}{N\lambda}} y$ reduces to the ones in \cite{Cheng:2008fc,Cheng:2008kt} when the length of the shortest cycle $\lambda$ equals one.

As stressed in the introduction, we think that our results could have immediate applications to the understanding of dyon counting in CHL-orbifolds. In particular,  most of the functions $\Phi_{g,h}$ are expected to have interpretations as partition functions of twisted dyons, and we hope to investigate this relation in more detail in a future publication.

\section*{Acknowledgments}
\noindent We are especially grateful to Matthias Gaberdiel for collaboration in the initial stages of this work, and for many helpful discussions. We are also grateful to Geoff Mason for several useful discussions on $M_{24}$ moonshine for eta-products, and for sending us a printed copy of \cite{MasonEtas}; to Miranda Cheng for a discussion that drew our attention to the connection with umbral moonshine; and to Martin Raum for helpful discussions on infinite product representations of Siegel modular forms. In addition we thank Scott Carnahan,  Thomas Creutzig, Terry Gannon, Suresh Govindarajan, Jeff Harvey, Stefan Hohenegger, Gerald H\"ohn, Jim Lepowsky, Greg Moore, Sameer Murthy, Anne Taormina and Katrin Wendland  for inspiring discussions during the course of this work. We also thank the Simons Center and the organizers of the programme ``Mock Modular Forms, Moonshine, and String Theory'' for providing a stimulating research environment, and all the participants for many inspiring discussions. D.P. thanks the Albert Einstein Institute in Golm, and R.V. thanks the Department of Fundamental Physics at Chalmers for hospitality while this work was carried out. 

 \appendix

\section{Some group cohomology}\label{app:groupcoho}

In this appendix, we summarize some general results about group cohomology. See \cite{Karpilovsky1993} for more details.

For a finite group $G$, a 2-cochain $\beta \, :\, G\times G \, \rightarrow U(1)$ is closed (and hence
defines a cocycle) provided it satisfies 
\be 
\beta(g_1, g_2 g_3)\beta(g_2, g_3)= \beta(g_1g_2, g_3)\beta(g_1, g_2) 
\label{2cocycle}
\ee  
for $g_1, g_2, g_3\in G$. The second cohomology $H^2(G,U(1))$ then consists of the closed 2-cochains, 
modulo the ambiguity 
\be 
\beta(g_1, g_2) \rightarrow \beta(g_1, g_2) \frac{\gamma(g_1)\gamma(g_2)}{\gamma(g_1g_2)}\ , 
\label{betagauge}
\ee  
where $\gamma \, :\, G\, \to \, U(1)$ is an arbitrary 1-cochain, i.e.\ an arbitrary function 
$\gamma \, :\, G\, \to \, U(1)$.

A 3-cochain $\alpha$,
\be 
\alpha\, :\, G\times G\times G\, \rightarrow\, U(1)
\ee   
is closed provided it satisfies
\be 
\alpha(g_1, g_2, g_3)\, \alpha(g_1, g_2 g_3, g_4)\, \alpha(g_2, g_3, g_4)
=\alpha(g_1g_2, g_3, g_4)\, \alpha(g_1, g_2, g_3g_4)\ .
\label{3cocycle}
\ee  
In the cohomology group $H^3(G,U(1))$ closed 3-cochains are then identified modulo 
\be 
\alpha(g_1, g_2,g_3)\rightarrow \alpha(g_1, g_2, g_3) \,
\frac{\beta(g_1 g_2, g_3)\beta(g_1, g_2)}{\beta(g_1, g_2g_3)\beta(g_2, g_3)}\ .
\label{alphagauge}
\ee  
Note that the multiplying factor is trivial if $\beta$ is closed, i.e.\ if it satisfies the 2-cocycle condition 
\eqref{2cocycle}. In particular, for each cocycle $\alpha$ and element $x\in G$, the 3-cocycle $\alpha_x$, defined by
\be \alpha_x(g,h,k):=\alpha(x^{-1}gx,\; x^{-1}hx,\; x^{-1}kx)\ ,
\ee differs from $\alpha$ just by a 3-coboundary \cite{Bantay:1990yr}
\be\label{alphaconj} \alpha(g_1,g_2,g_3)=\alpha_x(g_1, g_2, g_3) \,
\frac{\eta_x(g_1 g_2, g_3)\eta_x(g_1, g_2)}{\eta_x(g_1, g_2g_3)\eta_x(g_2, g_3)},\ee where 
\be\label{etadef} \eta_z(x,y):=\frac{\alpha(x,y,z)\alpha(z,z^{-1}xz,z^{-1}y{z})}{\alpha(x,z,z^{-1}yz)}\ .
\ee

Given a 3-cocycle $\alpha$, we can define, for any $h\in G$, a map $c_h: G\times G \rightarrow U(1)$ via
\be 
c_g(h_1, h_2)= \frac{\alpha(g, h_1, h_2)\alpha(h_1, h_2,(h_1h_2)^{-1} g(h_1h_2))}{\alpha(h_1, h_1^{-1}gh_1, h_2)}\ .
\label{ch}
\ee 
It is shown in \cite{Dijkgraaf:1989pz} that $c_g$ defines a 2-cocycle of the stabilizer subgroup 
$C_G(g)\subseteq G$ (i.e.\ the subgroup of all elements $h_1,h_2$ which commute with $g$). When $h_1,h_2\in C_G(g)$, we have the simplified expression
\be c_g(h_1, h_2)= \frac{\alpha(g, h_1, h_2)\alpha(h_1, h_2,g)}{\alpha(h_1, g, h_2)}\ ,\qquad h_1,h_2\in C_G(g)\ .
\ee
 It is 
straightforward to check this by inserting \eqref{ch} into \eqref{2cocycle}, and making repeated 
use of the 3-cocycle condition \eqref{3cocycle}  (twice on each side of the equality sign) together with the 
fact that $g$ commutes with $h_1, h_2, h_3$.
\vspace{.4cm}

\noindent Under the `gauge transformation' \eqref{alphagauge}, $c_g$ transforms as 
\be 
c_g(h_1, h_2)\rightarrow \tilde c_g(h_1, h_2):=c_g(h_1, h_2) \frac{\gamma_g(h_1)\gamma_g(h_2)}{\gamma_g(h_1h_2)}\ ,\qquad h_1,h_2\in C_G(g)\ ,
\label{chgauge}
\ee  
where we defined the 1-cochain $\gamma_g$ by
\be 
\gamma_g(h)\equiv \frac{\beta(g,h)}{\beta(h,g)}\ .
\ee  
This is indeed of the form \eqref{betagauge}, and hence, 
for all $g\in G$, $c_g$  defines a map
\be 
c_g\, :\, H^3(G, U(1))\, \rightarrow H^2(C_G(g), U(1))\ .
\ee   In fact, if $c_g$ is the 2-cocycle associated to  a projective representation $\rho_g$ of $C_G(g)$, i.e.
\be \rho_g(h_1)\rho_g(h_2)=c_g(h_1,h_2)\rho_g(h_1h_2)\ ,
\ee then $\tilde{c}_g$ is the 2-cocycle associated with the projectively equivalent representation
\be\label{rhoredef} \tilde\rho_g(h):=\gamma_g(h)\rho_g(h)\ .
\ee In the context of holomorphic CFTs, the transformation \eqref{alphagauge} corresponds to a redefinition \eqref{rhoredef} of the projective representations $\rho_g$ of the centralizer $C_G(g)$ over the twisted sector $\H_g$, which induces the analogous transformation of the twisted twining partition functions
\be \label{phishift} Z_{g,h}\to \tilde{Z}_{g,h}=\gamma_g(h)Z_{g,h}\ .\ee Indeed, the new partition functions $\tilde Z_{g,h}$ satisfy the expected modular properties with respect to the new cocycle $\tilde\alpha$.

\medskip

In particular, it the case \eqref{alphaconj} of conjugation by $x\in G$, we have $\beta\equiv \eta_x^{-1}$, so that
\be \gamma_g(h)=\frac{\eta_x(h,g)}{\eta_x(g,h)}=\frac{c_g(x,x^{-1}hx)}{c_g(h,x)}\ ,\qquad h\in C_g(G)\ ,
\ee where  the latter equality follows from \cite{Roche:1990hs,Bantay:1990yr}. From this identity, one recovers
\be Z_{(\II,x^{-1})\cdot (g,h)}(\tau)\equiv Z_{x^{-1}gx,x^{-1}hx}(\tau)=\frac{c_g(x,x^{-1}hx)}{c_g(h,x)}Z_{g,h}(\tau)\ ,
\ee from which the formula \eqref{epsconj} for $\epsilon_{g,h}(\II,x)$ follows.

\medskip

One further useful identity is \cite{Bantay:1990yr}
\be\label{exchange} \frac{c_{x_1x_2}(z_1,z_2)}{c_{x_1}(z_1,z_2)c_{x_2}(z_1,z_2)}=\frac{c_{z_1}(x_1,x_2)c_{z_2}(x_1,x_2)}{c_{z_1z_2}(x_1,x_2)}\ ,\ee that holds for \emph{pairwise commuting}  $x_1,x_2,z_1,z_2\in G$.

\medskip

For most applications, only the restriction of a 3-cocycle $\alpha$ to subgroups of the form $\langle g,h\rangle\cong \ZZ_{N_1}\times\ZZ_{N_2}$ is needed. For such groups, we have $\H^3(\ZZ_{N_1}\times\ZZ_{N_2},U(1))=\ZZ_{N_1}\times\ZZ_{N_2}\times\ZZ_{\gcd (N_1,N_2)}$ and a set of normalized representatives for the generators are \cite{deWildPropitius:1995cf}
\begin{align}\label{icocci1}
&\alpha_{v_i}(g^{a_1}h^{a_2},g^{b_1}h^{b_2},g^{c_1}h^{c_2}):=e^{\frac{2\pi i v_i a_i}{N_i^2}([b_i]_{N_i}+[c_i]_{N_i}-[b_i+c_i]_{N_i})}\end{align}
where $v_i\in \ZZ/{N_i}\ZZ$, $i=1,2$, and
%\label{icocci2}&\alpha_{v_2}(g^{a_1}h^{a_2},g^{b_1}h^{b_2},g^{c_1}h^{c_2}):=e^{\frac{2\pi i v_2 %a_2}{M^2}([b_2]_{M}+[c_2]_M-[b_2+c_2]_M)} && v_2\in \ZZ/M\ZZ\ ,\\
\begin{align}
\label{icocci3}&\alpha_{v_{12}}(g^{a_1}h^{a_2},g^{b_1}h^{b_2},g^{c_1}h^{c_2}):=e^{\frac{2\pi i v_{12} a_1}{N_1N_2}([b_2]_{N_2}+[c_2]_{N_2}-[b_2+c_2]_{N_2})} && 
\end{align} 
where $v_{12}\in \ZZ/\gcd (N_1,N_2)\ZZ$ and $[\cdot]_x:\ZZ\to \{0,\ldots,x-1\}$ denotes the reduction modulo $x$. Notice that, with this choice for the generators, we have the simplified formula
\be c_x(y,z)=\alpha(x,y,z)\ , \qquad x,y,z\in \ZZ_{N_1}\times\ZZ_{N_2}\ .
\ee

\section{Modular properties of twisted twining partition functions}
\label{app:modtt}

\noindent In this appendix we include some details on the modular properties of twisted twining partition functions. In particular, we analyze the combined action of $SL(2,\mathbb{Z})$ and $G$ on the set of commuting pairs $(g,h)$ in $G$. Here we must also take into account the presence of a non-trivial 3-cocycle $\alpha$ which leads to a certain twisted action. We introduce a convenient ``twisted equivariant slash-operator" that simplifies many expressions since it combines (twisted) $SL(2,\mathbb{Z})$-equivariance with $G$-equivariance. Finally, we discuss a reformulated version of the cohomological obstructions found in \cite{Gaberdiel:2012gf}.

\subsection{$SL(2,\mathbb{Z})\times G$ - action on $\P_G$}
It is clear from (\ref{modtrans}) that modular transformations act on $\tau\in \mathbb{H}$ as well as on the set of commuting pairs $(g,h)\in G\times G$. In addition $G$ acts on itself by conjugation and thereby on the set of pairs $g,h$. In order to determine the subgroups $\Gamma_{g,h}\subset SL(2,\mathbb{Z})$ under which $Z_{g,h}$ (and $\phi_{g,h}$) are invariant for fixed $g,h\in G$, we must classify the orbits of the combined $SL(2,\mathbb{Z})\times G$-action on the commuting pair $(g,h)$. We are mainly interested in the case $G=M_{24}$ but we only make this specification at the end of the subsection.

For any finite group $G$, let $\P_G\subset G\times G$ be the set of commuting pairs of elements:
\be 
\P_G=\{(g,h)\in G\times G\mid gh=hg\} \ . 
\ee
The group $SL(2,\ZZ)\times G$ has a left action as a permutation over this set by
\begin{align}\label{groupa}
\Bigl(\left(\begin{matrix}
a & b\\ c & d
\end{matrix} \right),k\Bigr)\cdot (g,h) &:= (k\,g\,k^{-1},k\, h\,k^{-1})\left(\begin{matrix}
a & b\\ c & d
\end{matrix} \right)^{-1}=(kg^{d}h^{-c}k^{-1},kg^{-b}h^ak^{-1})\ ,
\end{align} 	where $(g,h)\in \P_G$, $\left(\begin{smallmatrix}
a & b\\ c & d
\end{smallmatrix}\right)\in SL(2,\ZZ)$ and $k\in G$. This action can be extended to an action of $GL(2,\ZZ)\times G$ in the obvious way.

\subsection{Twisted action on $\P_G$: generalized  permutation }
\label{sec_genperm}

As explained in section \ref{sec_twist}, in order to include the possibility of non-trivial multipliers in the modular properties of the  twisted twining genera one needs to consider the ``twisting'' of the action \eqref{groupa} by a 3-cocycle $\alpha$, representing a cohomology class in $H^3(G,U(1))$. This $\alpha$-twisted action is a generalized  permutation
\be\label{generpermut} \CC\P_G\times SL(2,\ZZ)\times G\to \CC\P_G\ ,
\ee
 on the complex vector space $\CC\P_G$ freely generated by the elements of $\P_G$. We define this action by the formula
\be (\gamma,k)_\alpha: (g,h)\mapsto \epsilon_{g,h}(\gamma,k)(kgk^{-1},khk^{-1})\gamma^{-1}\ .
\ee Here, $(\gamma,k)\in SL(2,\ZZ)\times G$ and $\epsilon_{g,h}(\gamma,k)\in U(1)$ is a phase which depends on the choice of  3-cocycle $\alpha$. More precisely, in terms of the 2-cocycle  $c_g(h_1, h_2)$ in (\ref{ciggi})  
 the phases $\epsilon_{g,h}(\gamma,k)$ are defined as
\begin{align}& \epsilon_{g,h}(\gamma_1\gamma_2,k_1k_2):=\epsilon_{(\gamma_2,k_2)\cdot(g,h)}(\gamma_1,k_1)\;\epsilon_{g,h}(\gamma_2,k_2)\ ,\\
& \epsilon_{g,h}(T,e):=\frac{1}{c_g(g,g^{-1}h)}\ ,\\
& \epsilon_{g,h}(S,e):=c_{g}(h^{-1},h)\ ,\\
& \epsilon_{g,h}(\II,k):=\frac{c_g(h,k^{-1})}{c_g(k^{-1},khk^{-1})}\ ,\qquad k\in G\ ,\label{epsconj}
\end{align}
where
\be\label{SL2Zgens} S:=\begin{pmatrix}
0 & -1\\ 1 & 0
\end{pmatrix}\, \qquad\qquad T:=\begin{pmatrix}
1 & 1\\ 0 & 1
\end{pmatrix}\ ,
\ee are the generators of $SL(2,\ZZ)$.

\subsection{Modular properties of twisted twining partition functions}
\label{sec_twistedslash}
We now wish to analyze the modular properties of the twisted twining partition function $Z_{g,h}$ in more detail. To this end we define the following equivariant slash operator
\be\label{slash} f(g,h;\tau)\rvert{(\gamma,k)}:= f\bigl((\gamma,k)\cdot(g,h);\,\gamma\cdot\tau\bigr)\ ,\qquad (\gamma,k)\in SL(2,\ZZ)\times G\ ,
\ee acting on modular functions $f:\P_G\times \HH_+\to\CC$. We can  think of the twisted twining partition functions $Z_{g,h}(\tau)$ as functions on $\P_G\times \HH_+$ that are equivariant under $SL(2,\ZZ)\times G$, up to a multiplier. In the simplest case when the multiplier system is trivial, this property can be expressed as follows in terms of the slash operator:
\be\label{slashinv} Z_{g,h}(\tau)\rvert{(\gamma,k)}=Z_{g,h}(\tau)\ .
\ee
In particular, for each fixed $(g,h)\in \P_G$, the function  $Z_{g,h}(\tau)$ is a modular function under some subgroup $\Gamma_{g,h}\in PSL(2,\ZZ)$, i.e.
\be Z_{g,h}(\gamma\cdot \tau)=Z_{g,h}(\tau)\ ,\qquad \gamma\in \Gamma_{g,h}\subseteq PSL(2,\ZZ)\ .
\ee
The group $\Gamma_{g,h}$ is the image $\pi(\tilde{\Gamma}_{g,h})$ of the stabilizer
\be\label{tildeGammagh} \tilde{\Gamma}_{g,h}:=\{(\gamma,k)\in SL(2,\ZZ)\times G\mid (g,h)\cdot (\gamma,k) =(g,h)\}\ ,
\ee under the homomorphism
\be \pi:SL(2,\ZZ)\times G\to PSL(2,\ZZ)\ .
\ee 
Similar properties hold for the twisted twining genera $\phi_{g,h}(\tau,z)$ of an $\N=(4,4)$ superconformal algebra: these are expected to be Jacobi forms of weight zero and index $1$ with respect to the same groups $\Gamma_{g,h}$. For this reason, we need to extend the definition of  the slash operator to an action on the space of functions $\psi:\P_G\times \HH_+\times \CC\to\CC$. Specifically, for  Jacobi forms of weight $0$ and index $m$ with respect to some $\Gamma\subset SL(2,\ZZ)$ we define
\be \psi(g,h;\tau,z)\rvert{\bigl(\gamma,k\bigr)}:= e^{-\frac{2\pi i m cz^2}{c\tau+d}}\psi\Bigl( k\,g^dh^{-c}\,k^{-1},\;k\,g^{-b}h^a\,k^{-1};\;\gamma\cdot  \tau ,\frac{z}{c\tau+d}\Bigr)\ ,\ee where $(\gamma,k)=\bigl((\begin{smallmatrix}
a & b\\ c & d
\end{smallmatrix}),k\bigr)\in SL(2,\ZZ)\times G$.
 Note that for theories with $\N=(4,4)$ superconformal symmetry, these genera are expected to be \emph{even} functions of $z$, so that the central element $C=S^2\in SL(2,\ZZ)$ acts trivially  and it makes sense to consider the action on such functions of the quotient $PSL(2,\ZZ)=SL(2,\ZZ)/\ZZ_2$. 

\noindent We are now ready to incorporate the $\alpha$-twist in the modular properties of the twisted twining genera. To this end we  define the $\alpha$-twisted generalization of the equivariant slash operator:
\be\label{slashalpha} f(g,h,\tau)\rvert_\alpha{(\gamma,k)}:=\epsilon_{g,h}(\gamma,k) f\bigl((\gamma,k)\cdot(g,h);\gamma\cdot\tau\bigr)\ ,\qquad (\gamma,k)\in SL(2,\ZZ)\times G\ ,
\ee for some $3$-cocycle $\alpha$ representing a class $[\alpha]\in H^3(G,U(1))$.
Similarly, we define the $\alpha$-twisted  slash operators on Jacobi forms of weight $0$ and index $m$ by
\be%\label{JacSlashAlpha} 
\psi(g,h;\tau,z)\rvert_\alpha\bigl(\gamma ,k\bigr):= \epsilon_{g,h}(\gamma ,k)e^{-\frac{2\pi i m cz^2}{c\tau+d}}\psi\Bigl( k\,g^dh^{-c}\,k^{-1},\;k\,g^{-b}h^a\,k^{-1};\;\gamma\cdot \tau,\frac{z}{c\tau+d}\Bigr)\ ,\ee
where $(\gamma, k)=\bigl((\begin{smallmatrix}
a & b\\ c & d
\end{smallmatrix}),k\bigr)\in SL(2,\ZZ)\times G$.

\subsection{Cohomological obstructions}

When $\alpha$ represents a  non-trivial class in $H^3(G, U(1))$, the partition functions $Z_{g,h}(\tau)$ are modular functions under $\Gamma_{g,h}\subseteq PSL(2,\ZZ)$ only up to some multiplier $\chi_{g,h}$, which depends on $\alpha$  (see section \ref{sec_twist}). In fact, by \eqref{tildeGammagh} and \eqref{epsconj}, the restriction of $\epsilon_{g,h}$ to $\tilde\Gamma_{g,h}$ is a group homomorphism $\tilde\Gamma_{g,h}\to U(1)$. If
\be\label{noobst} \epsilon_{g,h}(\gamma,k)=1\ ,\ee
for all $(\gamma,k)$ in 
\begin{align}
\tilde\Gamma_{g,h}\cap \ker\pi=&\{(e,k)\in SL(2,\ZZ)\times G\mid (k^{-1}gk,k^{-1}hk)=(g,h) \}\\ &\cup 
\{(S^2,k)\in SL(2,\ZZ)\times G\mid (k^{-1}gk,k^{-1}hk)=(g^{-1},h^{-1}) \}\ ,
\end{align} then the restriction $\epsilon_{g,h}:\tilde\Gamma_{g,h}\to U(1)$ induces a well-defined homomorphism $\chi_{g,h}:\Gamma_{g,h}\to U(1)$ on the image $\Gamma_{g,h}= \pi(\tilde\Gamma_{g,h})$. Explicitly, for each $\gamma \in \Gamma_{g,h}\subseteq PSL(2,\ZZ)$, one can choose a lift $(\gamma,k)\in \tilde\Gamma_{g,h}\subseteq SL(2,\ZZ)\times G$ and set
\be\label{defchi} \chi_{g,h}(\gamma):= \epsilon_{g,h}(\gamma,k)\ ,\qquad \text{for }\gamma=\pi(\gamma,k)\in \Gamma_{g,h}\ .
\ee By \eqref{noobst}, the definition is independent of the lift.  

On the contrary, if \eqref{noobst} is not satisfied for some $(\gamma,k)\in \tilde\Gamma_{g,h}\cap \ker\pi$, then eq.\eqref{slashalphainv} implies that $Z_{g,h}(\tau)$ must vanish identically. In this case, we say that the twisted twining partition function is \emph{obstructed}. Following the discussion in \cite{Gaberdiel:2012gf,Gaberdiel:2013nya}, we can distinguish between two kinds of obstructions:
\begin{enumerate} 
\item We say that, for a certain $(g,h)\in \P_G$, there is an obstruction of the \emph{first kind} if there is some element of the form $(e,k)\in \tilde\Gamma_{g,h}\cap \ker\pi$ for which \eqref{noobst} is not satisfied. 
\item We say that there is an obstruction of the \emph{second kind} if \eqref{noobst} is satisfied for all elements of the form $(e,k)\in \tilde\Gamma_{g,h}\cap \ker\pi$, but is violated by some element of the form $(S^2,k)\in \tilde\Gamma_{g,h}\cap \ker\pi$. By \eqref{epsconj}, this implies that \eqref{noobst} is false for \emph{all} elements of the form $(S^2,k)\in \tilde\Gamma_{g,h}\cap \ker\pi$.
\end{enumerate}
Although not phrased this way in \cite{Gaberdiel:2012gf} these obstructions are equivalent to the ones given there.

\section{Definition of twisted equivariant Hecke operators}\label{app:twistHecke}

\noindent In this appendix, we shall discuss the definition and properties of the $\alpha$-twisted equivariant Hecke operators.

\subsection{Twisted equivariant Hecke operator}

Let us denote by $\Mat(\ZZ)$ the ring of $2\times 2$ integral matrices with positive determinant, graded by the determinant,
\be
\Mat(\ZZ)= \bigcup_{L>0} \Mat_L(\ZZ)\ ,
\ee
where $\Mat_L(\ZZ)$ was defined in (\ref{MatN}).
For any $u\in \Mat_L(\ZZ)$ denote by $u^\vee\in \Mat_L(\ZZ)$ the dual  
\be \left(\begin{matrix}a & b\\ c & d\end{matrix}\right)^\vee := L \left(\begin{matrix}a & b\\ c & d\end{matrix}\right)^{-1}=\left(\begin{matrix}d & -b\\ -c & a\end{matrix}\right) \ ,\qquad \left(\begin{matrix}a & b\\ c & d\end{matrix}\right)\in \Mat_L(\ZZ)\ .
\ee 
Note that $\Mat_1(2,\ZZ)\equiv SL(2,\ZZ)$ is the group of invertible elements of $\Mat(\ZZ)$ and that each $\Mat_L(\ZZ)$ is a bimodule over $SL(2,\ZZ)$.

In order to define the $\alpha$-twisted equivariant Hecke operators $\T^\alpha_L$, we need to extend the action \eqref{generpermut} of $SL(2,\ZZ)\times G$ on $\CC\P_G$ to an action 
\be\label{ggpermut} \CC\P_G\times \Mat(\ZZ)\times G\to \CC\P_G\ ,
\ee
by
\be (u,k)_\alpha: (g,h)\mapsto \epsilon_{g,h}(u,k)(kgk^{-1},khk^{-1})u^{\vee}\ ,
\ee for a suitable $\epsilon_{g,h}(u,k)\in U(1)$, that reduces to the one discussed in section \ref{sec_genperm} when $u\in \Mat_1(\ZZ)= SL(2,\ZZ)$. Furthermore, the interpretation of $\phi_{g,h}$ as a section of $\L_{g,h}^\alpha$ suggests that the following composition law should be imposed
\be (u_1u_2,k_1k_2)_\alpha \cdot (g,h)=(u_1,k_1)_\alpha\cdot \bigl((u_2,k_2)_{\alpha^L}\cdot (g,h) \bigr)\ ,
\ee where $(u_1,k_1)\in \Mat_L(\ZZ)\times G$ and $(u_2,k_2)\in \Mat(\ZZ)\times G$. In terms of the phases $\epsilon_{g,h}$, this condition reads
\be\label{compositio} \epsilon_{g,h}((u_1,k_1)\cdots (u_n,k_n))=\prod_{i=1}^n \epsilon_{(u_{i+1},k_{i+1})\cdots (u_n,k_n)\cdot(g,h)}(u_i,k_i)^{|u_1\cdots u_{i-1}|}\ ,
\ee where $|u_i|:=\det u_i$.

Observe that $\Mat(\ZZ)$, as a multiplicative semigroup, is generated by $SL(2,\ZZ)$ together with the matrices of the form $\left(\begin{smallmatrix}
p & 0 \\0 & 1
\end{smallmatrix}\right)$ for $p$ prime. Therefore,  it is sufficient to specify the phases $\epsilon((\begin{smallmatrix}
p & 0 \\0 & 1
\end{smallmatrix}),e)$ for all primes $p$ and any other phase $\epsilon_{g,h}(u,k)$, $(u,k)\in \Mat(\ZZ)\times G$, is then determined by \eqref{compositio}. 
The $\alpha$-twisted slash operator (\ref{JacSlashAlpha}) can be trivially extended to an action with respect to $\gamma\in \Mat_L(\mathbb{Z})$, in which case it maps a Jacobi form $\phi$ of weight 0 and index $m$ to another Jacobi form $\phi |_\alpha(\gamma, k)$ of weight 0 and index $Lm$. 
Notice that, by \eqref{compositio}, the twisted slash operators satisfy the composition relation
 \be\label{slashcompo} \rvert_\alpha (u_1u_2,k_1k_2) = \rvert_{\alpha} (u_1,k_1)\rvert_{\alpha^L} (u_2,k_2)
 \ee
 
%\bigskip

We can now define the $\alpha$-twisted equivariant Hecke operators acting on the space of $SL(2,\ZZ)\times G$-equivariant Jacobi forms $\phi_{g,h}$ by
\begin{align}
\label{twistedHecke}
&\T^\alpha_L \phi_{g,h}(\tau,z):=\frac{1}{L}\sum_{u\in SL(2,\ZZ)\backslash \Mat_L(\ZZ)} \phi_{g,h}(\tau,z)\rvert_\alpha (u,e)\ ,\\
&\U^\alpha_L \phi_{g,h}(\tau,z):=\phi_{g,h}(\tau,z)\rvert_\alpha (\begin{smallmatrix}
L & 0\\ 0 & L
\end{smallmatrix},e)\ .
\end{align} 
One may check that these operators satisfy the Hecke algebra \eqref{hecke1}--\eqref{hecke3}. Furthermore, $\T^\alpha_L$ (respectively, $\U^\alpha_L$) maps the system of $\alpha$-twisted $SL(2,\ZZ)\times G$-equivariant (weak) Jacobi forms of weight $0$ and index $m$ to a system of $\alpha^L$-twisted (respectively, $\alpha^{L^2}$-twisted) $SL(2,\ZZ)\times G$-equivariant (weak) Jacobi forms of weight $0$ and index $Lm$ (respectively, index $L^2m$). The proofs of these properties are completely analogous to the case where $\alpha$ is trivial; in fact, these properties follow directly from the definition of the Hecke operators in terms of slash operators satisfying a composition law of the form \eqref{slashcompo}.

\subsection{Definition of slash operator for $\Mat(\ZZ)\times G$}

In this subsection, we will describe the extension of the $\alpha$-twisted slash operators to $\Mat(\ZZ)\times G$. As stressed in section \ref{s:tHecke}, it is sufficient to define such operators for matrices of the form $(\begin{smallmatrix}
 L & 0\\ 0 & 1
\end{smallmatrix})$, since, together with $SL(2,\ZZ)$ they generate the whole $\Mat(\ZZ)$. We will make an ansatz for the phase $\epsilon_{g,h}\bigl((\begin{smallmatrix}
 L & 0\\ 0 & 1
\end{smallmatrix}),e\bigr)$ and verify that it extends consistently to the whole $\Mat(\ZZ)$.

\medskip

Our ansatz for the phase $\epsilon_{g,h}\bigl((\begin{smallmatrix}
 L & 0\\ 0 & 1
\end{smallmatrix}),e\bigr)$ is based on the interpretation of the corresponding slash operator within the theory of symmetric orbifolds of conformal field theories. We recall that the twisted twining partition function $Z_{g,h}(\tau)$ in a holomorphic CFT is defined as a trace $\Tr_{\H_g}(\rho_g(h) q^{L_0-\frac{c}{24}})$, where $\H_g$ is the $g$-twisted sector and $\rho_g$ is the (possibly projective) representation of the centralizer $C_G(g)$ on $\H_g$. This partition function can be computed by a path integral on a torus $\CC/(\ZZ+\tau\ZZ)$ where the fields are required to have monodromies $g$ and $h$ along the cycles $-1$ and $\tau$, respectively. The function
\be Z_{g,h}(\tau)\rvert_\alpha \bigl((\begin{smallmatrix}
 L & 0\\ 0 & 1
\end{smallmatrix}),e\bigr)= \Upsilon^*_{(\begin{smallmatrix}
 L & 0\\ 0 & 1
\end{smallmatrix})}Z_{g,h}\ ,
\ee is associated with the $L$-fold covering $\CC/(\ZZ+L\tau\ZZ)$ and has a natural interpretation as a trace
\be Z_{g,h}(\tau)\rvert_\alpha \bigl((\begin{smallmatrix}
 L & 0\\ 0 & 1
\end{smallmatrix}),e\bigr)=\Tr_{\H_g}\bigl((\rho_g(h) q^{L_0-\frac{c}{24}})^L\bigr)\ .
\ee Using the product law \eqref{grouplaw} for the projective representation $\rho_g$,
 we obtain
\be Z_{g,h}(\tau)\rvert_\alpha \bigl((\begin{smallmatrix}
 L & 0\\ 0 & 1
\end{smallmatrix}),e\bigr) = \prod_{i=1}^{L-1}c_g(h,h^i)\Tr_{\H_g}(\rho_g(h^L) q^{LL_0-\frac{Lc}{24}})=\prod_{i=1}^{L-1}c_g(h,h^i) Z_{g,h^L}(L\tau)\ . 
\ee
This formula suggests the definition
\be
\epsilon_{g,h}\bigl((\begin{smallmatrix}
 L & 0\\ 0 & 1
\end{smallmatrix}),e\bigr)=\prod_{i=1}^{L-1}c_g(h,h^i)\ ,
\ee so that
\be f(g,h;\tau)\rvert_\alpha \bigl((\begin{smallmatrix}
 L & 0\\ 0 & 1
\end{smallmatrix}),e\bigr)=\prod_{i=1}^{L-1}c_g(h,h^i) f\Bigl((g,h)(\begin{smallmatrix}
 1 & 0\\ 0 & L
\end{smallmatrix});\,(\begin{smallmatrix}
 L & 0\\ 0 & 1
\end{smallmatrix})\cdot\tau\Bigr)\ ,
\ee

\bigskip

Any integer matrix of determinant $L$ can be written as a product of  elements in $SL(2,\ZZ)$ and matrices of the form $(\begin{smallmatrix}
 p & 0\\ 0 & 1
\end{smallmatrix})$ for prime $p$.
For each $(u,k)\in \Mat(\ZZ)\times G$, given a representation of $u$ as a word  $u=a_1\cdots a_n$ in the generators $a_1,\ldots,a_n$ of $\Mat(\ZZ)$, one can use the composition law
 \be\label{compo} f(g,h;\tau)\rvert_\alpha (u_1u_2,k_1k_2) = f(g,h;\tau)\rvert_{\alpha} (u_1,k_1)\;\rvert_{\alpha^{|u_1|}} (u_2,k_2)\ ,
 \ee where $|u|:=\det u$, to define the slash operator  \be f(g,h;\tau)\rvert_{\alpha} (a_1\cdots a_n,k):=f(g,h;\tau)\rvert_{\alpha} (\II,k)\; \rvert_{\alpha} (a_1,e)\;\rvert_{\alpha^{|a_1|}}(a_2,e)\;\cdots\; \rvert_{\alpha^{|a_1\cdots a_{n-1}|}}( a_n,e)\ .\ee One needs to check that this definition is consistent, i.e. that the operator $\rvert_{\alpha} (u,k)$ does not depend on representation of $(u,k)\in \Mat(\ZZ)\times G$ as a word in the generators and that the composition law is respected. The outline of the proof is given in the next subsection.

\bigskip

In particular, by \eqref{compo},  the identity
\be  \begin{pmatrix}
1 & 0\\ 0 & L
\end{pmatrix}=\begin{pmatrix}
0 & -1 \\ 1 & 0
\end{pmatrix}\begin{pmatrix}
L & 0\\ 0 & 1
\end{pmatrix}\begin{pmatrix}
0 & 1 \\ -1 & 0
\end{pmatrix}
\ee
yields
\be \epsilon_{g,h}\bigl((\begin{smallmatrix}
 1 & 0\\ 0 & L
\end{smallmatrix}),e\bigr)=\frac{c_g(h^{-1},h)^L\prod_{i=1}^{L-1} c_{h^{-1}}(g,g^i)}{c_{g^L}(h^{-1},h)}=\frac{1}{\prod_{i=1}^{L-1}c_h(g,g^i)}\ ,
\ee
where in the last step we use \eqref{exchange}, while from the identity
\be \begin{pmatrix}
a & b\\ 0 & d
\end{pmatrix}=\begin{pmatrix}
1 & 0\\ 0 & d
\end{pmatrix}\begin{pmatrix}
1 & 1\\ 0 & 1
\end{pmatrix}^b\begin{pmatrix}
a & 0\\ 0 & 1
\end{pmatrix}
\ee
we obtain the general formula
\begin{align}\label{epsHecke} \epsilon_{g,h}\bigl(\begin{smallmatrix}
 a & b\\ 0 & d
\end{smallmatrix}\bigr):=\epsilon_{g,h}\bigl((\begin{smallmatrix}
 a & b\\ 0 & d
\end{smallmatrix}),e\bigr)&= \frac{\prod_{i=1}^{a-1}c_g(h,h^i)^d}{\prod_{j=1}^{d-1}c_{g^{-b}h^a}(g,g^j)\prod_{k=1}^bc_g(g,g^{-k}h^a)^d}\ .
\end{align} Since for any coset of $SL(2,\ZZ)\backslash\Mat(\ZZ)$ we can choose a representative of the form $(\begin{smallmatrix}
 a & b\\ 0 & d
\end{smallmatrix})$, this formula is sufficient to determine the Hecke operators $\T^\alpha_L$ and $\U^\alpha_L$ for all $L$.
In particular, if the restriction of the 3-cocycle $\alpha$ to the group $\langle g,h\rangle\cong \ZZ_{N_1}\times \ZZ_{N_2}$ is given by
$\alpha = \alpha_{v_1}\alpha_{v_2}\alpha_{v_{12}}
$, in terms of the generators \eqref{icocci1}--\eqref{icocci3}, then ($N_1,N_2>1$)
\be
\epsilon_{g,h}\bigl((\begin{smallmatrix}
 a & b\\ 0 & d
\end{smallmatrix}),e\bigr)= e^{\frac{2\pi i}{N_1}\bigl(v_1b\sum_{j=1}^{d-1}\delta^{(N_1)}(j+1)-v_1d\sum_{j=1}^{b}\delta^{(N_1)}(1-j)+v_{12}d\sum_{j=1}^{a-1}\delta^{(N_2)}(j+1)\bigr)}\ .
\ee

\subsection{Proof of consistency}

In the rest of this section, we will show that our proposal for the twisted equivariant Hecke operators satisfies several consistency conditions. 
As a first consistency, let us consider the effect of a `gauge transformation' \eqref{alphagauge}, under which the normalized 3-cocycle $\alpha$ is multiplied by a 3-coboundary $\partial \beta$ (with $\beta(e,g)=\beta(g,e)=1$ to keep the normalization). Formally, this transformation  corresponds to a different choice of trivialization of the pull-back $\pi^*\L_{g,h}^\alpha$, where $\pi:\P\times \HH_+\times\CC\to\M$ is the covering map. As stressed in the previous appendix \ref{app:groupcoho}, for a holomorphic CFTs $\C$, the transformation \eqref{alphagauge} corresponds to a redefinition \eqref{rhoredef} of the projective representations $\rho_g$ of the centralizer $C_G(g)$ over the twisted sector $\H_g$ and induces the transformation \eqref{phishift} of the partition functions. For the tensor product $\C^{\otimes L}$ and the symmetric product $S^L \C$ theories, the twisted twining partition functions $Z_{g,h}$ are defined as traces over the representation induced by the (symmetrized)  $L$-tensor product of $\rho_g$, so that the appropriate transformation is
\be\label{ZNtransf} Z_{g,h}^{(L)}\mapsto \tilde Z_{g,h}^{(L)}:=\gamma_g(h)^LZ_{g,h}^{(L)}\ ,
\ee where $\gamma_g(h)=\beta(g,h)/\beta(h,g)$. Analogous properties hold for the twisted twining genera $\phi^{(L)}_{g,h}$ in superconformal field theories.
Using the identities $\gamma_e(g)=\gamma_g(e)=1$ and $\gamma_g(h)=\gamma_h(g)^{-1}$, it is easy to check that, under \eqref{rhoredef} and \eqref{chgauge},
\be\label{cobtransf}
\bigl(\gamma_g(h)f(g,h;\tau)\bigr)\rvert_{\tilde\alpha} \bigl((\begin{smallmatrix}
 a & b\\ 0 & d
\end{smallmatrix})\gamma,e\bigr)=\gamma_g(h)^L\Bigl(f(g,h;\tau)\rvert_{\alpha} \bigl((\begin{smallmatrix}
 a & b\\ 0 & d
\end{smallmatrix})\gamma,e\bigr)\Bigr)\ ,
\ee where $L=ad$, and similar properties hold for Jacobi forms. As a consequence, we obtain
\be \T^{\tilde \alpha}_L\tilde{Z}_{g,h}=\gamma_g(h)^L\T^\alpha_LZ_{g,h} \ ,
\ee 
and
\begin{align}
\T^{\tilde \alpha}_L\tilde{\phi}_{g,h}=\gamma_g(h)^L\T^\alpha_L\phi_{g,h}\ ,\qquad 
\U^{\tilde\alpha}_L\tilde{\phi}_{g,h}=\gamma_g(h)^{L^2}\U^\alpha_L\phi_{g,h}\ ,
\end{align} 
which indeed reproduce \eqref{ZNtransf} and the analogous formula for twisted twining genera. More generally, given \emph{any} representation of a matrix $u \in \Mat_L(\ZZ)$ as a word in the generators of $\Mat(\ZZ)$, it is easy to check that \eqref{cobtransf} holds for the slash operator $\rvert_\alpha (u,k)$, i.e
\be\label{ingener} \gamma_g(h)\tilde\epsilon_{g,h}(u,k)=\gamma_g(h)^L \epsilon_{g,h}(u,k)\ .
\ee
An easy corollary of this property is 
\be\label{kconsist}  f(g,h;\tau)\rvert_\alpha \bigl(u,k\bigr)\rvert_{\alpha^L} \bigl(\II,x\bigr)= f(g,h;\tau)\rvert_{\alpha} \bigl(\II,x\bigr)\rvert_\alpha \bigl(u,k\bigr)\ ,\qquad (u,k)\in \Mat_L(\ZZ)\times G,\ x\in G\ ,
\ee  which follows by considering \eqref{ingener} for the transformation \eqref{alphaconj}.  As a consequence of \eqref{kconsist}, one only needs to prove consistency of the slash operator for elements of the form $(u,e)$.

\bigskip

A set of generators of the multiplicative semigroup $\Mat(\ZZ)$ is given by the generators $S,T$ of $SL(2,\ZZ)$ (see eq.\eqref{SL2Zgens}), together with all matrices $(\begin{smallmatrix}
 p & 0\\ 0 & 1
\end{smallmatrix})$ for $p$ prime. To check the consistency of the definition of the slash operator, one needs to verify that, for each relation $a_1\cdots a_n=b_1\cdots b_m$ among the generators of $\Mat(\ZZ)$, the following identities hold for all $(g,h)\in \P_G$
\be f(g,h;\tau)\rvert_\alpha (a_1,e)\cdots \rvert_{\alpha^{|a_1\ldots a_{n-1}|}} (a_n,e)=
f(g,h;\tau)\rvert_\alpha (b_1,e)\cdots \rvert_{\alpha^{|b_1\ldots b_{m-1}|}} (b_m,e)\ .
\ee This is equivalent to checking the following identities for the phases
\be\label{epscosist} \prod_{i=1}^n \epsilon_{(g,h)a_{i+1}^\vee \cdots a_n^\vee}(a_i,e)^{|a_1\cdots a_{i-1}|}=
\prod_{i=1}^m \epsilon_{(g,h)b_{i+1}^\vee \cdots b_m^\vee}(b_i,e)^{|b_1\cdots b_{i-1}|}\ .
\ee

 For the relations $S^4=1$ and $(ST)^3=S^2$ within $SL(2,\ZZ)$, these conditions follow from  the identities in \cite{Roche:1990hs,Bantay:1990yr}; therefore, we will only consider the independent relations within $\Mat_L(\ZZ)$ for $L>1$.

\medskip

For each prime power $p$, the relations within $\Mat_{p}(\ZZ)$ can be obtained from the relations for $SL(2,\ZZ)$, together with relations of the form
\be \gamma_1 \begin{pmatrix}
 p & 0\\ 0 & 1
\end{pmatrix}=\begin{pmatrix}
 p & 0\\ 0 & 1
\end{pmatrix}\gamma_2\ ,
\ee for suitable $\gamma_1,\gamma_2\in SL(2,\ZZ)$. It is easy to check that such a relation exists if and only if $\gamma_2$ is an element of $\Gamma_0(p)$, which is generated by $T$, $S^2$ and $ST^pS=(\begin{smallmatrix}
1 & 0\\-p & 1
\end{smallmatrix})$. Therefore, the only independent relations within $\Mat_p(\ZZ)$ are
\be\label{Tprel} T^p \begin{pmatrix}
 p & 0\\ 0 & 1
\end{pmatrix}=\begin{pmatrix}
 p & 0\\ 0 & 1
\end{pmatrix}T\ ,
\ee 
\be\label{S2rel} S^2 \begin{pmatrix}
 p & 0\\ 0 & 1
\end{pmatrix}=\begin{pmatrix}
 p & 0\\ 0 & 1
\end{pmatrix}S^2\ ,
\ee
and
\be\label{STSrel} STS\begin{pmatrix}
 p & 0\\ 0 & 1
\end{pmatrix}=\begin{pmatrix}
 p & 0\\ 0 & 1
\end{pmatrix}ST^pS\ .
\ee The only remaining relations in $\Mat(\ZZ)$ are of the form
 \be\label{Nrel} \begin{pmatrix}
 p & 0\\ 0 & 1
 \end{pmatrix}\gamma_1 \begin{pmatrix}
p' & 0\\ 0 &1
\end{pmatrix}=\gamma_2\begin{pmatrix}
p' & 0\\ 0 &1
\end{pmatrix}\gamma_3 \begin{pmatrix}
 p & 0\\ 0 & 1
 \end{pmatrix}\gamma_4\ , 
\ee for all pairs of distinct primes $p,p'$ and suitable $\gamma_1,\gamma_2,\gamma_3,\gamma_4\in SL(2,\ZZ)$. In fact, it is easier to consider formula \eqref{epsHecke} as a definition of $\epsilon_{g,h}((\begin{smallmatrix}
a & b\\ 0 & d
\end{smallmatrix}),e)$ and verify the identities \eqref{epscosist} for the relations
\begin{align}
\label{ppleftrel} \begin{pmatrix}
 a & b\\ 0 & d
\end{pmatrix}\begin{pmatrix}
 p & 0\\ 0 & 1
\end{pmatrix}&=\begin{pmatrix}
 pa & b\\ 0 & d
\end{pmatrix}\ ,\\
\label{pprightrel} \begin{pmatrix}
 p & 0\\ 0 & 1
\end{pmatrix}\begin{pmatrix}
 a & b\\ 0 & d
\end{pmatrix}&=\begin{pmatrix}
 pa & pb\\ 0 & d
\end{pmatrix}\ ,
\end{align}  since the consistency conditions for the relations \eqref{Nrel} then follow. The consistency conditions are therefore
\begin{align}
\label{cTprel}\epsilon_{g,h^p}\bigl(\begin{smallmatrix}
 1 & p\\ 0 & 1
\end{smallmatrix}\bigr)\epsilon_{g,h}\bigl(\begin{smallmatrix}
 p & 0\\ 0 & 1
\end{smallmatrix}\bigr)&\stackrel{}{=}\epsilon_{g,g^{-1}h}\bigl(\begin{smallmatrix}
 p & 0\\ 0 & 1
\end{smallmatrix}\bigr)\epsilon_{g,h}\bigl(\begin{smallmatrix}
 1 & 1\\ 0 & 1
\end{smallmatrix}\bigr)^p\\ \label{cS2rel} \epsilon_{h^{-p},g}\bigl(\begin{smallmatrix}
0 & -1\\ 1 &0
\end{smallmatrix})
\epsilon_{g,h^{p}}\bigl(\begin{smallmatrix} 0 & -1\\ 1 &0 \end{smallmatrix})
\epsilon_{g,h}\bigl(\begin{smallmatrix} p & 0\\ 0 &1\end{smallmatrix}) 
&\stackrel{}{=}\epsilon_{g^{-1},h^{-1}}\bigl(\begin{smallmatrix}
p & 0\\ 0 &1
\end{smallmatrix})
\epsilon_{h^{-1},p}\bigl(\begin{smallmatrix} 0 & -1\\ 1 &0 \end{smallmatrix})^p
\epsilon_{g,h}\bigl(\begin{smallmatrix} 0 & -1\\ 1 &0\end{smallmatrix})^p\\
\label{cSTSrel}\epsilon_{g^p,h}\bigl(\begin{smallmatrix}
 1 & 1\\ 0 & 1
\end{smallmatrix}\bigr)\epsilon_{g,h}\bigl(\begin{smallmatrix}
 1 & 0\\ 0 & p
\end{smallmatrix}\bigr)
&\stackrel{}{=}
\epsilon_{g,g^{-p}h}\bigl(\begin{smallmatrix}
 1 & 0\\ 0 & p
\end{smallmatrix}\bigr)\epsilon_{g,h}\bigl(\begin{smallmatrix}
 1 & p\\ 0 & 1
\end{smallmatrix}\bigr)^p\\ 
\label{cppleftrel}\epsilon_{g,h^p}\bigl(\begin{smallmatrix}
 a & b\\ 0 & d
\end{smallmatrix}\bigr)\epsilon_{g,h}\bigl(\begin{smallmatrix}
 p & 0\\ 0 & 1
\end{smallmatrix}\bigr)^{ad}&\stackrel{}{=}\epsilon_{g,h}\bigl(\begin{smallmatrix}
 pa & b\\ 0 & d
\end{smallmatrix}\bigr)\\
\label{cpprightrel}\epsilon_{g^d,g^{-b}h^a}\bigl(\begin{smallmatrix}
 p & 0\\ 0 & 1
\end{smallmatrix}\bigr)\epsilon_{g,h}\bigl(\begin{smallmatrix}
 a & b\\ 0 & d
\end{smallmatrix}\bigr)^p&\stackrel{}{=}\epsilon_{g,h}\bigl(\begin{smallmatrix}
 pa & pb\\ 0 & d
\end{smallmatrix}\bigr)\ .
\end{align} The proof is a tedious but straightforward computation, consisting of a repeated use of the cocycle conditions. For example, the proof of \eqref{cppleftrel}:
\be\begin{split}
&\frac{\prod_{i=1}^{a-1}c_g(h,h^{ip})^d}{\prod_{j=1}^{d-1}c_{g^{-b}h^{ap}}(g,g^j)\prod_{k=1}^bc_g(g,g^{-k}h^{ap})^d}\prod_{i=1}^{p-1}c_g(h,h^i)^{ad}
\frac{\prod_{j=1}^{d-1}c_{g^{-b}h^{ap}}(g,g^j)\prod_{k=1}^bc_g(g,g^{-k}h^{ap})^d}{\prod_{i=1}^{pa-1}c_g(h,h^i)^d}\\
&=\Bigl(\frac{\prod_{i=1}^{a-1}c_g(h,h^{ip})\prod_{i=1}^{p-1}c_g(h,h^i)^{a}}{\prod_{i=1}^{pa-1}c_g(h,h^i)}\Bigr)^d=1
\end{split}\ee where the last equality follows from the $2$-cocycle condition on $c_g$. 
%(the easiest way to check this is to consider  a projective representation $\rho_g$ associated with $c_g$, and to rewrite this expression as a ratio of two phases, both corresponding to the mismatch between $\rho_g(h)^{pa}$ and $\rho_g(h^{pa})$).

\section{Some technical proofs}

\subsection{Proof of \eqref{identica2}.}\label{s:prova} In this section, we will prove eq.\eqref{identica2}. The sum $F_{g,h}(a,d,k,\ell)$ can be written as
\be F_{g,h}(a,d,k,\ell):=\sum_{n=0}^\infty \frac{1}{d}\sum_{b=0}^{d-1} e^{\frac{2\pi i bn}{N\lambda d}}c_{g^d_\alpha,g^{-b}_\alpha h^k_\alpha}(\frac{n}{N\lambda},\ell) q^{\frac{na}{N\lambda d}}\ ,\ee
in terms of the Fourier coefficients of the twisted twining genera $\phi_{g^d_\alpha,g^{-b}_\alpha h^k_\alpha}$ introduced in section \ref{s:centrext}.
Notice that the functions $\phi_{g_\alpha, h_\alpha}$ satisfy
\be\phi_{g^r_\alpha, g_\alpha^{-r}h_\alpha}(\tau+1,z)=\phi_{g^r_\alpha, h_\alpha}(\tau,z)\ ,
\ee so that 
\be\label{goodperio} c_{g_\alpha^d,g_\alpha^d h_\alpha}(r,\ell)=e^{\frac{2\pi i r}{N\lambda}}
c_{g_\alpha^d,h_\alpha}(r,\ell)\ ,
\ee for all $h_\alpha\in C^\alpha_{M_{24}}(g)$.
Set
\be e:=\gcd (d,N\lambda)\ ,\qquad f:=\frac{d}{e}\ ,
\ee and note that
\be\label{fNcomprime} \gcd(f,\frac{N\lambda}{e})=1\ ,\ee 
otherwise $e\gcd(f,\frac{N\lambda}{e})$ would be a common divisor of $d$ and $N\lambda$ greater than $e$. The order of $g_\alpha^d$ is $o(g_\alpha^{ef})=o(g_\alpha^e)=N\lambda/e$, so that $c_{g_\alpha^d,g_\alpha^{-b}h_\alpha^k}(\frac{n}{N\lambda},\ell)=0$ unless $e|n$. Thus, we can set $r:=n/e$ and obtain
\be F_{g,h}(a,d,k,\ell)=\sum_{r=0}^\infty \frac{1}{d}\sum_{b=0}^{d-1} e^{\frac{2\pi i bre}{N\lambda d}}c_{g_\alpha^d,g_\alpha^{-b}h_\alpha^k}(\frac{re}{N\lambda},\ell) q^{\frac{ar}{N\lambda f}}\ .\ee Notice that, by \eqref{goodperio}, the general term of the sum over $b$ is periodic under $b\to b+d$.
 We can set $b:=se+b'$ and replace the sum over $ b\in \ZZ/d\ZZ$ by a sum over $b'\in \ZZ/e\ZZ$ and $s\in \ZZ/f\ZZ$
\be F_{g,h}(a,d,k,\ell)=\sum_{r=0}^\infty \frac{1}{e}\sum_{b'=0}^{e-1}\frac{1}{f}\sum_{s=0}^{f-1} e^{\frac{2\pi i (b'+se)r}{N\lambda f}}c_{g_\alpha^{d},g_\alpha^{-b'}g_\alpha^{-se}h_\alpha^k}(\frac{re}{N\lambda},\ell) q^{\frac{ar}{N\lambda f}}\ .\ee Thus, the general term of the sum over $b'$ is periodic under $b'\to b'+e$ and we can sum over $N\lambda/e$ periods and divide by $N\lambda/e$
\be F_{g,h}(a,d,k,\ell)=\sum_{r=0}^\infty \frac{1}{N\lambda}\sum_{b'=0}^{N\lambda-1}\frac{1}{f}\sum_{s=0}^{f-1} e^{\frac{2\pi i (b'+se)r}{N\lambda f}}c_{g_\alpha^{d},g_\alpha^{-b'}g_\alpha^{-se}h_\alpha^k}(\frac{re}{N\lambda},\ell) q^{\frac{ar}{N\lambda f}}\ .\ee
 By \eqref{fNcomprime}, there are integers $x,y$ such that
\be\label{copprimi} xf+y \frac{N\lambda}{e}=1
\ee
 so that $se= sxd+syN\lambda$ and we obtain
\be\begin{split}
F_{g,h}(a,d,k,\ell)=&\sum_{r=0}^\infty \frac{1}{N\lambda}\sum_{b'=0}^{N\lambda-1}\frac{1}{f}\sum_{s=0}^{f-1} e^{\frac{2\pi i b'r}{N\lambda f}} e^{\frac{2\pi i s(xd+yN\lambda)r}{N\lambda f}}c_{g_\alpha^{d},g_\alpha^{-b'}g_\alpha^{-sdx}h_\alpha^k}(\frac{re}{N\lambda},\ell) q^{\frac{ar}{N\lambda f}}\\
=&\sum_{r=0}^\infty \sum_{b'=0}^{N\lambda-1}\frac{e^{\frac{2\pi i b'r}{N\lambda f}}}{N\lambda}\sum_{s=0}^{f-1}  \frac{e^{\frac{2\pi i syr}{f}}}{f}c_{g_\alpha^{d},g_\alpha^{-b'}h_\alpha^k}(\frac{re}{N\lambda},\ell) q^{\frac{ar}{N\lambda f}}\ ,
\end{split}\ee where in the last step we used \eqref{goodperio}. Since by \eqref{copprimi} $\gcd(y,f)=1$, the sum over $s$ just a projection
\be \sum_{s=0}^{f-1}  \frac{e^{\frac{2\pi i syr}{f}}}{f}=\begin{cases}1 & \text{if } f|r \ ,\\
0 & \text{otherwise}\ , \end{cases}
\ee so that, by setting $r=mf$, we obtain
\be F_{g,h}(a,d,k,\ell)
=\sum_{m=0}^\infty \sum_{b'=0}^{N\lambda-1}\frac{e^{\frac{2\pi i b'm}{N\lambda }}}{N\lambda}c_{g_\alpha^{d},g_\alpha^{-b'}h_\alpha^k}(\frac{md}{N\lambda},\ell) q^{\frac{am}{N\lambda }}\ ,\ee that is equivalent to \eqref{identica2}.

\subsection{Central extensions and special choices of the cocycle}\label{a:centrext}

In section \ref{s:centrext} a central extension $C^\alpha_{M_{24}}(g)$ of the centralizer $C_{M_{24}}(g)$ is defined, together with the representations $\tilde \rho_{g,r}$ on the $g^r$-twisted sector $\H_{g^r}$, for all $r\in \ZZ_{\ge 0}$. In this section, we will describe some special choices of the cocycle $\alpha$, for which these representations are particularly simple.

If two $3$-cocycles $\alpha$, $\alpha'$ differ by a coboundary $\partial\beta$, then there is an isomorphism $C^{\alpha'}_{M_{24}}(g)\stackrel{\cong}{\longrightarrow} C^\alpha_{M_{24}}(g)$ that relates the canonical lifts
 \be\label{isocobou} h_{\alpha'}\mapsto h_{\alpha}\,q(\nu_g(h))\ ,
 \ee where $\nu_g(h)\in\RR/\ZZ$ is defined by $e^{2\pi i\nu_g(h)}=\frac{\beta(g,h)}{\beta(h,g)}$. 

Under the shift $\alpha\to \alpha'$ by a coboundary $\partial\beta$, the cocycle $c_g$ transforms as
\be c_g(h,k)\to \frac{\beta(g,h)}{\beta(h,g)}\frac{\beta(g,k)}{\beta(k,g)}\frac{\beta(hk,g)}{\beta(g,hk)}c_g(h,k)\ ,\qquad h,k\in C_{M_{24}}(g)\ .\ee 
Correspondingly, the phases $f_{g,r}(h)$ of eq.\eqref{fgrh} transform as
\be f_{g,r}(h)\to f_{g,r}(h)\frac{\beta(h,g)^r}{\beta(g,h)^r}\frac{\beta(g^r,h)}{\beta(h,g^r)}\ ,\qquad h\in C_{M_{24}}(g)\ ,
\ee and, in particular,
\be f_{g,N}(h)\to f_{g,N}(h)\frac{\beta(h,g)^N}{\beta(g,h)^N}\ , \qquad h \in C_{M_{24}}(g)\ .
\ee Therefore, we can choose $\beta(g,h)/\beta(h,g)$ for each $h\in C_{M_{24}}(g)$, $h\not\in \langle g\rangle$, in such a way that
\be\label{easycoc} f_{g,N}(h)=1 \ ,\qquad \forall h\in C_{M_{24}}(g)\setminus \langle g\rangle\ .
\ee This condition determines $\beta(g,h)/\beta(h,g)$ for all $h\in C_{M_{24}}(g)\setminus \langle g\rangle$ up to $N$-th roots of unity. 
Notice that, in general,
\be \tilde\rho_{g,r+N}(h)=\frac{\rho_{g^r}(h)}{f_{g,r}(h)f_{g,N}(h)}=\frac{\tilde \rho_{g^r}(h)}{f_{g,N}(h)}\ ,\qquad h\in C_{M_{24}}(g)\ ,
\ee
so that, by imposing \eqref{easycoc}, we have
\be \tilde\rho_{g,r+N}(h)=\tilde\rho_{g^r}(h)\ ,\qquad h\in C_{M_{24}}(g)\setminus \langle g\rangle\ .
\ee
On the other hand, $f_{g,N}(g)$ depends only on the cohomology class $[\alpha]$. Since the restriction of $[\alpha]^N$ to $H^3(\langle g\rangle,U(1))\cong \ZZ_{N}$ is the trivial class, it follows that  $f_{g,N}(g)$ must be a $N$-th root of unity. In fact, for $M_{24}$, we have
\be f_{g,N}(g)=e^{-\frac{2\pi i}{\lambda}}\ ,
\ee where $\lambda|N$ is the length of the shortest cycle of $g$ in the $24$-dimensional permutation representation. It is also related to the spectrum of $L_0-\frac{c}{24}$ in the $g$-twisted sector, which takes values in $-\frac{1}{\lambda N}+\frac{1}{N}\ZZ$ and to the presence of a non-trivial multiplier system for the twining genus $\phi_{e,g}$. It is useful to distinguishes the cases when $\lambda=1$ and $\lambda\neq 1$.

\subsubsection{Case $\lambda=1$ (trivial multiplier)} This case occurs whenever $g$ is an element of some $M_{23}$ subgroup of $M_{24}$, i.e. when $g$ belongs to the classes
$$ 2A,\  3A,\ 4B,\ 5A,\ 6A,\ 7A,\ 7B,\ 8A,\ 11A,\ 14A,\ 14B,\ 15A,\ 15B,\ 23A,\ 23B\ .
$$ In this case the restriction of $[\alpha]$ to $H^3(\langle g\rangle,U(1))$ is the trivial class 
and one can choose $\beta(g^i,g^j)$ in such a way that
\be\label{triviac} c_{g^i}(g^j,g^k)=1\ .\ee
In particular,
\be f_{g,r}(g^i)=1\ ,
\ee for all $r,i$. By specializing \eqref{cigierre} to the case $r=N$, we obtain
\be c_g(h,k)^N=c_{g^N}(h,k)\frac{f_{g,N}(hk)}{f_{g,N}(h)f_{g,N}(k)}=1 \ ,\qquad h,k\in C(g)\ ,
\ee so that $c_g(h,k)$ is an $N$-th root of unity for all $h,k\in C_{M_{24}}(g)$, i.e. there is $\mu_g(h,k)\in \ZZ/N\ZZ$ such that
\be c_g(h,k)=e^{\frac{2\pi i \mu_g(h,k)}{N}}\ .
\ee
With this choice of cocycle, the central extension $C^\alpha_{M_{24}}(g)$ of section \ref{s:centrext} can be chosen to be finite
\be 1\to \langle Q\rangle\cong \ZZ_N \to  C^\alpha_{M_{24}}(g) \to  C_{M_{24}}(g)\to 1\ .
\ee where the central element $Q$ is related to the $U(1)$-generator $q(x)$ in section \ref{s:centrext} by
\be Q=q(1/N)\ .
\ee

\subsubsection{Case $\lambda\neq 1$ (non-trivial multiplier)}\label{a:casenontri} This case occurs when $g$ is in one of the $M_{24}$-classes
$$ 2B,\ 3B,\ 4A,\ 4C,\ 6B,\ 10A,\ 12A,\ 12B,\ 21A,\ 21B\ .
$$  Since $H^3(\langle g\rangle,U(1))\cong \ZZ_N$, we can choose the cocycle in such a way that 
\be\label{notsoeasy} c_{g^i}(g^j,g^k)^N=1\ ,
\ee for all $i,j,k\in \ZZ$. This condition fixes $\beta(g^i,g^j)$ up to $N$-th roots of unity. For $h,k\in C_{M_{24}}(g)$, with $h,k,hk\not\in \langle g\rangle$, we can apply again \eqref{cigierre} with $r=N$ and obtain
\be c_g(h,k)^N=1\ ,\qquad h,k,hk\in C_{M_{24}}(g)\setminus \langle g\rangle\ .
\ee If $\alpha$ satisfies \eqref{easycoc} and \eqref{notsoeasy},  then by \eqref{cigierre} we have
\be f_{g,N}(g^i)c_g(g^i,h)^N=1\ ,\qquad \frac{f_{g,N}(g^i)f_{g,N}(g^j)}{f_{g,N}(g^{i+j})}=1\ ,
\ee
so that
\be f_{g,N}(g^n)=f_{g,N}(g)^n=e^{-\frac{2\pi i n}{\lambda}}\ ,
\ee and
\be c_g(g^i,h)^{N\lambda}=1\ .
\ee Therefore, $c_g(h,k)$ is a $N\lambda$ roots of unity for all $h,k\in C_{M_{24}}(g)$ and we can define $\mu_g(h,k)\in \ZZ/N\lambda\ZZ$ such that
\be c_g(h,k)=e^{\frac{2\pi i \mu_g(h,k)}{N\lambda}}\ .
\ee The discussion is similar to the case $\lambda=1$. With this choice of cocycle, the central extension $C^\alpha_{M_{24}}(g)$ of section \ref{s:centrext} can be chosen to be finite
\be 1\to \langle q(\frac{1}{N\lambda})\rangle\cong \ZZ_{N\lambda} \to  C^\alpha_{M_{24}}(g) \to  C_{M_{24}}(g)\to 1\ .
\ee It is convenient to define also a central element $Q$ of order $N$ by
\be Q=q(1/N)\ ,
\ee so that the $g^r$-twisted sector is an eigenspace of $Q$ with eigenvalue $e^{\frac{2\pi ir}{N	}}$.

\subsection{S-duality for $g$ in classes $2B, 3B, 4C, 6B, 12B$}
\label{sec_case2}
Here we prove the S-duality transformation property of $\Phi_{g,h}$ for $g$ in the classes   2B, 3B, 4C, 6B, 12B. Let $g$ be an element of $M_{24}$ in one of these classes and let $N$ be its order. In all these cases, the restriction of the class $[\alpha]$ to $H^3(\langle g\rangle,U(1))$ is non-trivial and has order $\lambda =N$. Therefore,
\be \hat c_{g,h}(d,m,\ell ,t)=
\sum_{k=0}^{M-1}  \frac{e^{-\frac{2\pi itk}{M}}}{M}\sum_{b=1}^{N^2}\frac{e^{\frac{2\pi i bm}{N^2}}}{N^2} \Tr_{\H_{g^d}(\frac{md}{N^2},\ell)}\bigl(\tilde\rho_{g,d}(g)^{-b}\tilde\rho_{g,d}(h)^k(-1)^{F+\bar F}\bigr)\ .
\ee
In the $g^r$-twisted sector, for $r$ and $N$ coprime, the spectrum of $L_0-\frac{c}{24}$ takes values in $-\frac{1}{N^2}+\frac{1}{N}\ZZ$. More generally, if $\gcd (r,N)=e$, then in the $g^r$-twisted sector  the spectrum of $L_0-\frac{c}{24}$ takes values in $-\frac{1}{(N/e)^2}+\frac{1}{N/e}\ZZ$. Notice that $\hat c_{g,h}(d,m,\ell ,t)$ is a sum of traces in the $g^d$-twisted sector at level $\frac{md}{N^2}$, so that $\hat c_{g,h}(d,m,\ell ,t)= 0$ unless $\frac{md}{N^2}\in -\frac{1}{(N/e)^2}+\frac{1}{N/e}\ZZ$, where $e=\gcd(N,d)$. It is easy to verify that this condition is equivalent to
\be m\equiv -d\mod N\ ,
\ee where we used the property
\be \gcd(x,N)=1 \qquad \Leftrightarrow\qquad  x^2\equiv 1\mod N\ ,
\ee which holds for all $N$ that divide $24$.

As explained in appendix \ref{a:casenontri}, we can choose the cocycle $\alpha$ in such a way that 
\be f_{g,N}(h)\equiv \prod_{i=1}^{N-1} c_h(g,g^i)=1 \ ,\qquad h\in C_{M_{24}}(g)\setminus \langle g\rangle\ ,\ee while $f_{g,N}(g^k)$ depends only on the cohomology class $[\alpha]$ and equals \be f_{g,N}(g^k)= e^{-\frac{2\pi ik}{N}}\ .\ee
With this choice of cocycle, we have \be \tilde\rho_{g,x+N}(h)=\frac{\tilde\rho_{g,x}(h)}{f_{g,N}(h)}=\tilde\rho_{g,x}(h)\ ,\qquad h\in C_{M_{24}}(g)\setminus \langle g\rangle\ ,\ee while
\be \tilde\rho_{g,x+N}(g)=\frac{\tilde \rho_{g,x}(g)}{f_{g,N}(g)}=e^{\frac{2\pi i}{N}}\tilde\rho_{g,x}(g)\ .
\ee
 For $d\equiv -m\equiv 0 \mod N$, we have, for all $x\in \NN$,
\begin{align}
\hat c_{g,h}(d=Nu,m=Nv,\ell ,t)&=
\sum_{k=0}^{M-1}  \frac{e^{-\frac{2\pi itk}{M}}}{M}\sum_{b=1}^{N^2}\frac{e^{\frac{2\pi i bv}{N}}}{N^2} \Tr_{\H_{e}(uv,\ell)}\bigl(\tilde\rho_{g,Nu}(g)^{-b}\tilde\rho_{g,Nu}(h)^k(-1)^{F+\bar F}\bigr)\notag\\
&=
\sum_{k=0}^{M-1}  \frac{e^{-\frac{2\pi itk}{M}}}{M}\sum_{b=1}^{N^2}\frac{e^{\frac{2\pi i b(v-u)}{N}}}{N^2} \Tr_{\H_{e}(uv,\ell)}\bigl(g^{-b}h^k(-1)^{F+\bar F}\bigr)\ .
\end{align} Since the (untwisted) twining genera are invariant under charge conjugation, i.e.
\be \phi_{e,g^{-b}h^k}(\tau,z)=\phi_{e,g^{b}h^{-k}}(\tau,z)\ ,
\ee we obtain
\begin{multline}\label{cgh4C}
\hat c_{g,h}(Nu,Nv,\ell ,t)=
\sum_{k=0}^{M-1}  \frac{e^{-\frac{2\pi itk}{M}}}{M}\sum_{b=1}^{N^2}\frac{e^{\frac{2\pi i b(v-u)}{N}}}{N^2} \Tr_{\H_{e}(uv,\ell)}\bigl(g^{b}h^{-k}(-1)^{F+\bar F}\bigr)\\
=
\sum_{k=0}^{M-1}  \frac{e^{-\frac{2\pi itk}{M}}}{M}\sum_{b=1}^{N^2}\frac{e^{\frac{2\pi i b(u-v)}{N}}}{N^2} \Tr_{\H_{e}(uv,\ell)}\bigl(g^{-b}h^{-k}(-1)^{F+\bar F}\bigr)=\hat c_{g,h^{-1}}(Nv,Nu,\ell ,t)\ .
\end{multline}
Similarly, for $d\equiv x\equiv -m \mod N$, with $x=1,\ldots,N-1$, we obtain
\begin{multline}
\hat c_{g,h}(d=Nu+x,m=Nv+N-x,\ell ,t)
=\sum_{k=0}^{M-1}  \frac{e^{-\frac{2\pi itk}{M}}}{M}\sum_{b=1}^{N^2}\frac{e^{\frac{2\pi i b(Nv+N-x-Nu)}{N^2}}}{N^2}\\ \times\Tr_{\H_{g^x}(\frac{md}{N^2},\ell)}\bigl(\tilde\rho_{g,x}(g)^{-b}\tilde\rho_{g,x}(h)^k(-1)^{F+\bar F}\bigr)\ .
\end{multline} Using the relation
\be \phi_{g,h}(\tau,z)=\frac{1}{c_h(g,g^{-1})c_{g^{-1}}(h,h^{-1})}\phi_{g^{-1},h^{-1}}(\tau,z)\ ,
\ee and by 
\be \frac{\rho_{g^{-1}}(h^{-1})}{c_{g^{-1}}(h,h^{-1})}=\rho_{g^{-1}}(h)^{-1}\ ,
\ee we obtain
\begin{align} \Tr_{H_{g^x}(n,\ell)}&(\tilde \rho_{g,x}(h))=\frac{\Tr_{\H_{g^x}(n,\ell)}(\rho_{g^x}(h))}{\prod_{i=1}^{x-1}c_h(g,g^i)}=\frac{\Tr_{\H_{g^{-x}}(n,\ell)}(\rho_{g^{-x}}(h)^{-1})}{c_h(g^x,g^{-x})\prod_{i=1}^{x-1}c_h(g,g^i)}
\\&=\frac{\Tr_{\H_{g^{N-x}}(n,\ell)}(\tilde\rho_{g,N-x}(h)^{-1})}{c_h(g^x,g^{N-x})\prod_{i=1}^{x-1}c_h(g,g^i)\prod_{j=1}^{N-x-1}c_h(g,g^i)}=
\frac{\Tr_{\H_{g^{N-x}}(n,\ell)}(\tilde\rho_{g,N-x}(h)^{-1})}{f_{g,N}(h)}\ ,\notag
\end{align} where the last equality follows from
\be c_h(g^x,g^{N-x})\prod_{i=1}^{x-1}c_h(g,g^i)=\prod_{i=N-x}^Nc_h(g,g^i)\ ,
\ee that in turn is a consequence of the $2$-cocycle condition for $c_g$. It is easy to verify that this identity is compatible with the product in the central extension, i.e.
\be \Tr_{H_{g^x}(n,\ell)}(\tilde \rho_{g,x}(h)\tilde \rho_{g,x}(k))=\frac{\Tr_{\H_{g^{N-x}}(n,\ell)}(\tilde\rho_{g,N-x}(k)^{-1}\tilde\rho_{g,N-x}(h)^{-1})}{f_{g,N}(k)f_{g,N}(h)}\ .
\ee Using this relation, we obtain
\be\begin{split}
&\!\!\hat c_{g,h}(d=Nu+x,m=Nv+N-x,\ell ,t)
\\&=\sum_{k=0}^{M-1}  \frac{e^{-\frac{2\pi itk}{M}}}{M}\sum_{b=1}^{N^2}\frac{e^{\frac{2\pi i b(Nv-x-Nu)}{N^2}}}{N^2} \Tr_{\H_{g^{N-x}}(\frac{md}{N^2},\ell)}\bigl(\tilde\rho_{g,N-x}(g)^{b}\tilde\rho_{g,N-x}(h)^{-k}(-1)^{F+\bar F}\bigr)\\
&=\sum_{k=0}^{M-1}  \frac{e^{-\frac{2\pi itk}{M}}}{M}\sum_{b=1}^{N^2}\frac{e^{\frac{2\pi i b(Nu+x-Nv)}{N^2}}}{N^2}\Tr_{\H_{g^{N-x}}(\frac{md}{N^2},\ell)}\bigl(\tilde\rho_{g,N-x}(g)^{-b}\tilde\rho_{g,N-x}(h)^{-k}(-1)^{F+\bar F}\bigr)\\
&=\sum_{k=0}^{M-1}  \frac{e^{-\frac{2\pi itk}{M}}}{M}\sum_{b=1}^{N^2}\frac{e^{\frac{2\pi i b(Nu+x)}{N^2}}}{N^2} \Tr_{\H_{g^{N-x}}(\frac{md}{N^2},\ell)}\bigl(\tilde\rho_{g,Nv+N-x}(g)^{-b}\tilde\rho_{g,Nv+N-x}(h)^{-k}(-1)^{F+\bar F}\bigr)\\
&=\sum_{k=0}^{M-1}  \frac{e^{-\frac{2\pi itk}{M}}}{M}\sum_{b=1}^{N^2}\frac{e^{\frac{2\pi i bd}{N^2}}}{N^2} \Tr_{\H_{g^m}(\frac{md}{N^2},\ell)}\bigl(\tilde\rho_{g,m}(g)^{-b}\tilde\rho_{g,m}(h)^{-k}(-1)^{F+\bar F}\bigr)\ .
%\\
%&=\hat c_{g,h}(4v+4-x,4u+x,\ell ,-t)\ .
\end{split}\ee
Finally, we notice that 
\be \tilde\rho_{g,r}(h)^{-k}=c_g(h,h^{-1})^{-rk}\tilde\rho_{g,r}(h^{-1})^k\ ,
\ee where $c_g(h,h^{-1})$ is a $N$-th root of unity thanks to our choice of cocycle. We recall that imposing the condition $f_{g,N}(h)=1$ still leaves the possibility of modifying the cocycle $\alpha$ by a coboundary $\partial \beta$ with $\beta^N=1$. Under such a modification, $c_g(h,h^{-1})$ transforms as
\be\label{transfo} c_g(h,h^{-1})\to \frac{\beta(g,h)}{\beta(h,g)}\frac{\beta(g,h^{-1})}{\beta(h^{-1},g)}c_g(h,h^{-1})\ ,
\ee
We distinguish two cases. If $h^{-1}$ and $h$ are not conjugated within $C_{M_{24}}(g)$, then one can choose the cocycle $\alpha$ in such a way that
\be c_g(h,h^{-1})=1\ .
\ee With this choice, we obtain
\be \hat c_{g,h}(d,m,\ell ,t)=\hat c_{g,h^{-1}}(m,d,\ell ,t)\ ,
\ee so that
\be\begin{split}\label{inve} \Phi_{g,h}(\sigma,\tau,z)&=pq^{\frac{1}{N^2}}y\prod_{t=1}^M\prod_{(d,m,\ell)>0}(1-e^{\frac{2\pi it}{M}}q^{\frac{m}{N^2}}y^\ell p^d )^{\hat c_{g,h}(d,m,\ell ,t)}\\
&=pq^{\frac{1}{N^2}}y\prod_{t=1}^M\prod_{(d,m,\ell)>0}(1-e^{\frac{2\pi it}{M}}q^{\frac{m}{N^2}}y^\ell p^d )^{\hat c_{g,h^{-1}}(m,d,\ell ,t)}\\
&=\Phi_{g,h^{-1}}(\frac{\tau}{N^2},N^2\sigma,z)\ .
\end{split}\ee
In most cases, however, there is some $w\in C_{M_{24}}(g)$ such that $h^{-1}=w^{-1}hw$. 
%This property implies a `reality condition' on the Fourier coefficients of $\phi_{g,h}(\tau,z)$, namely
%\be\label{complconj}  \phi_{g,h}(\tau,z)=\frac{c_g(h,w) c_g(h,h^{-1})}{c_g(w,h^{-1})} \phi^*_{g,h}(\tau,z)=e^{-\frac{2\pi i n}{N}}\phi^*_{g,h}(\tau,z)\ ,
%\ee where
%\be\phi^*_{g,h}(\tau,z):=\overline{\phi_{g,h}(-\bar\tau,-\bar z)}=\sum_{r\in\QQ}\sum_{\ell\in\ZZ}\overline{c_{g,h}(r,\ell)}q^ry^\ell\ ,
%\ee is the Jacobi form with complex conjugate Fourier coefficients. 
%Thus, \eqref{complconj} implies that all Fourier coefficients of $\phi_{g,h}$ have the same argument. 
At the level of the central extension $C^\alpha_{M_{24}}(g)$, the relation $hw=wh^{-1}$ leads to a relation among the lifts
\be h_\alpha^{-1}=\frac{c_g(w,h^{-1})}{c_g(h,w) c_g(h,h^{-1})}w_\alpha^{-1}h_\alpha w_\alpha=w_\alpha^{-1}Q^n h_\alpha w_\alpha\ ,
\ee where 
$n\in \ZZ/N\ZZ$ is such that
\be e^{-\frac{2\pi i n}{N}}=\frac{c_g(h,w) c_g(h,h^{-1})}{c_g(w,h^{-1})}\ ,
\ee and we used the fact that, with our choice of cocycle, the right-hand side is a $N$-th root of unity. Since $Q^n h_\alpha$ is conjugated with $h_\alpha^{-1}$ within $C^\alpha_{M_{24}}(g)$, it must have the same order $M$, so that $nM\equiv 0\mod N$ or, equivalently,
\be  nM=sN\ ,
\ee for some $s\in \ZZ/M\ZZ$. Therefore,
\be\begin{split}
\hat c_{g,h}(&d,m,\ell ,t)\\
&=\sum_{k=0}^{M-1}  \frac{e^{-\frac{2\pi itk}{M}}}{M}\sum_{b=1}^{N^2}\frac{e^{\frac{2\pi i bd}{N^2}}}{N^2} \Tr_{\H_{g^m}(\frac{md}{N^2},\ell)}\bigl(\tilde\rho_{g,m}(g)^{-b}\tilde\rho_{g,m}(h)^{k}\tilde\rho_{g,m}(Q^n)^{k}(-1)^{F+\bar F}\bigr)\\
&=\sum_{k=0}^{M-1}  \frac{e^{-\frac{2\pi itk}{M}}}{M}\sum_{b=1}^{N^2}\frac{e^{\frac{2\pi i bd}{N^2}}}{N^2} e^{\frac{2\pi inmk}{N}}\Tr_{\H_{g^m}(\frac{md}{N^2},\ell)}\bigl(\tilde\rho_{g,m}(g)^{-b}\tilde\rho_{g,m}(h)^{k}(-1)^{F+\bar F}\bigr)\\
&=\sum_{k=0}^{M-1}  \frac{e^{-\frac{2\pi i(t-sm)k}{M}}}{M}\sum_{b=1}^{N^2}\frac{e^{\frac{2\pi i bd}{N^2}}}{N^2} \Tr_{\H_{g^m}(\frac{md}{N^2},\ell)}\bigl(\tilde\rho_{g,m}(g)^{-b}\tilde\rho_{g,m}(h)^{k}(-1)^{F+\bar F}\bigr)\\
&=\hat c_{g,h}(m,d,\ell ,t-sm)\ .
\end{split}\ee
Using this identity, we obtain
\be\begin{split}\label{strana} \Phi_{g,h}(\sigma,\tau,z)&=pq^{\frac{1}{N^2}}y\prod_{t=1}^M\prod_{(d,m,\ell)>0}(1-e^{\frac{2\pi it}{M}}q^{\frac{m}{N^2}}y^\ell p^d )^{\hat c_{g,h}(d,m,\ell ,t)}\\
&=pq^{\frac{1}{N^2}}y\prod_{t=1}^M\prod_{(d,m,\ell)>0}(1-e^{\frac{2\pi it}{M}}q^{\frac{m}{N^2}}y^\ell p^d )^{\hat c_{g,h}(m,d,\ell ,t-sm)}\\
&=pq^{\frac{1}{N^2}}y\prod_{t'=1}^M\prod_{(d,m,\ell)>0}(1-e^{\frac{2\pi ism}{M}}e^{\frac{2\pi it'}{M}}q^{\frac{m}{N^2}}y^\ell p^d )^{\hat c_{g,h}(m,d,\ell ,t')}\\
&=pq^{\frac{1}{N^2}}y\prod_{t'=1}^M\prod_{(d,m,\ell)>0}(1-e^{\frac{2\pi inm}{N}}e^{\frac{2\pi it'}{M}}q^{\frac{m}{N^2}}y^\ell p^d )^{\hat c_{g,h}(m,d,\ell ,t')}\\
&=e^{-\frac{2\pi in}{N}}\Phi_{g,h}(\frac{\tau}{N^2}+\frac{n}{N},N^2\sigma,z)\\
&=e^{\frac{2\pi in}{N}}\Phi_{g,h}(\frac{\tau}{N^2},N^2\sigma-Nn,z)\ ,
\end{split}\ee where we used the property that the exponent $\hat c_{g,h}(m,d,\ell ,t')$ is non-zero only for $d\equiv -m\mod N^2$, so that
\be e^{-\frac{2\pi ix}{N^2}}\Phi_{g,h}(\sigma+\frac{x}{N^2},\tau,z)=e^{\frac{2\pi ix}{N^2}}\Phi_{g,h}(\sigma,\tau-x,z)\ ,\qquad x\in\ZZ\ .
\ee The identity \eqref{strana} can be simplified by a suitable choice of the $3$-cocycle.
Let us shift the cocycle $\alpha$ by a $3$-coboundary $\partial \beta$ such that $\frac{\beta(g,h)}{\beta(h,g)}=e^{\frac{2\pi i r}{N^2}}$ for some $r\in\ZZ$, so that, by \eqref{newPhi}, the new form $\Phi'$ satisfies
\begin{align} \Phi'_{g,h}&(\sigma,\tau,z)=e^{-\frac{2\pi ir}{N^2}}\Phi_{g,h}(\sigma+\frac{r}{N^2},\tau,z)
=e^{\frac{2\pi i(Nn-r)}{N^2}}\Phi_{g,h}(\frac{\tau}{N^2},N^2\sigma+r-Nn,z)\\
&=e^{\frac{2\pi i(r-Nn)}{N^2}}\Phi_{g,h}(\frac{\tau}{N^2}+\frac{Nn-r}{N^2},N^2\sigma,z)=e^{\frac{2\pi i(2r-Nn)}{N^2}}\Phi'_{g,h}(\frac{\tau}{N^2}+\frac{Nn-2r}{N^2},N^2\sigma,z)\ .\notag
\end{align} By choosing $r=\frac{n(N+N^2)}{2}$, we finally obtain
\be \Phi'_{g,h}(\sigma,\tau,z)=\Phi'_{g,h}(\frac{\tau}{N^2}-n,N^2\sigma,z)=\Phi'_{g,h}(\frac{\tau}{N^2},N^2\sigma,z)\ .
\ee

\section{Modular forms and Jacobi forms}\label{app:moduJaco}

The Dedekind $\eta$ function and  Jacobi theta functions are defined as
\be\begin{split}
\eta(\tau) & =  q^{\frac{1}{24}} \prod_{n=1}^{\infty} (1 - q^n) \\
\vartheta_1(\tau,z) & = -iq^{\frac{1}{8}} y^{\frac{1}{2}}\, \prod_{n=1}^\infty(1-q^n)(1-yq^n)(1-y^{-1}q^{n-1})
 \\
\vartheta_2(\tau,z) & = 2\, q^{\frac{1}{8}} \cos(\pi z)\, \prod_{n=1}^{\infty} (1-q^n)\, (1+yq^n)
(1+y^{-1} q^n)    \\
\vartheta_3(\tau,z) & =    \prod_{n=1}^{\infty} (1-q^n) \, (1+yq^{n-1/2})(1+y^{-1}q^{n-1/2}) \\
\vartheta_4(\tau,z) & =    \prod_{n=1}^{\infty} (1-q^n) \, (1-yq^{n-1/2}) (1-y^{-1}q^{n-1/2}) \ \ .
\end{split}\ee
The standard weak Jacobi forms $\phi_{0,1}$ and $\phi_{-2,1}$ of index $1$ and weight $0$ and $2$ can be defined as \cite{EichlerZagier}
\be \label{A.6}
\phi_{0,1}(\tau,z)=4\sum_{i=2}^4\frac{\vartheta_i(\tau,z)^2}{\vartheta_i(\tau,0)^2}\ ,\qquad \qquad \phi_{-2,1}(\tau,z)=-\frac{\vartheta_1(\tau,z)^2}{\eta(\tau)^6}\ .
\ee
Every weak Jacobi forms of weight $0$ and index $1$ under a congruence subgroup $\Gamma\subseteq SL(2,\ZZ)$ is given by
\be A\phi_{0,1}(\tau,z)+F(\tau)\phi_{-2,1}(\tau,z)\ ,
\ee
where
 $A$ is a constant and $F(\tau)$ is a modular form of weight $2$ for $\Gamma$.  
 
The Eisenstein series
\be \psi^{(N)}=q\frac{\partial}{\partial q}\log\frac{\eta(N\tau)}{\eta(\tau)}=E_2(\tau)- N E_2(N\tau)\ ,
\ee 
where
\be E_2(\tau)=-\frac{1}{24}+\sum_{n=1}^\infty \Bigl(\sum_{d|n} d\Bigr) q^n\ ,
\ee
are modular forms of weight $2$ under $\Gamma_0(N)\subset SL(2,\ZZ)$. We will also need the newforms
\begin{align}
f_{23,a}(\tau)=&q-q^3-q^4-2 q^6+2 q^7-q^8+2 q^9+2 q^{10}+\cdots\\
 f_{23,b}(\tau)=&-q^2+2 q^3+q^4-2 q^5-q^6-2 q^7+2 q^8+2 q^{10}+\cdots
\end{align} that are modular (cusp) forms of weight $2$ for $\Gamma_0(23)$.

\section{Tables}\label{app:tables}

In the following tables, we collect information about the $55$ abelian
subgroups $\langle g,h\rangle$ of $M_{24}$ and the corresponding twisted twining genera.

\medskip

Table \ref{t:cycl} describes the $21$ cyclic subgroups $\langle g\rangle$, $g\in M_{24}$ and the corresponding twining genera. For each group we report the conjugacy class of the generators, the orbits in the $24$-dim representation, the twisted twining genera for some  pairs of elements in the group. The twining genus $\phi_{e,g}$ is a weak Jacobi form of weight $0$ and index $1$ for the group 
\be 
\Gamma_0(N):=\{ \left(\begin{smallmatrix}
a & b\\ c & d
\end{smallmatrix}\right) \in SL(2,\ZZ)\mid c\equiv 0 \mod N\}\ ,
\ee where $N$ is the order of the $g$.
Thus, $\phi_{e,g}$ is given by $\frac{\Tr_{\bf 24}(g)}{12}\phi_{0,1}(\tau,z)+F_{e,g}(\tau)\phi_{-2,1}(\tau,z)$, where $\phi_{0,1}(\tau,z)$, $\phi_{-2,1}(\tau,z)$  are the standard generators of the ring of weak Jacobi forms of index $1$ and $\Tr_{\bf 24}(g)$ and $F_{e,g}(\tau)$ are reported in the last two columns. The table is divided into two parts, the first containing the groups whose generators are in some $M_{23}$ subgroup of $M_{24}$ (equivalently, the groups have some fixed points in the $24$-dimensional permutation representation).

\bigskip

Table \ref{t:noncycl} contains information about the $34$ conjugacy classes of non-cyclic abelian groups of $M_{24}$ (see \cite{Gaberdiel:2012gf}). For each such $\langle g,h\rangle\subset M_{24}$ we have described the structure as 
an abelian group, i.e.\ as $\ZZ_m\times\ZZ_n$, the $M_{24}$ classes of all its elements (excluding the identity), 
the order of the centralizer $C(g,h)$ and its index $|N(g,h)|/|C(g,h)|$ in the normalizer of 
$\langle g,h\rangle$ in $M_{24}$, and  the lengths of the orbits of 
$\langle g,h\rangle\subset M_{24}$ when acting as a group of permutations of $24$ objects. Furthermore, we give a classification of all modular groups $\Gamma_{g,h}\subset PSL(2,\mathbb{Z})$ and twisted twining genera $\phi_{g,h}$ for commuting pairs $g,h\in M_{24}$. 
The latter functions are denoted as $\phi_{[g],[h]}$, where the first subscript is the $M_{24}$-class of $g$ and the second is $C_{M_{24}}(g)$-class of $h$; the names of the classes follow the conventions of \cite{Gaberdiel:2012gf}. Finally, for each of the $34$ groups we provide the explicit expressions of the twisted twining genera $\phi_{g,h}$ and $\T^\alpha_L\phi_{g,h}$, $L=2,3,4$, for a pair $(g,h)$ of generators and for a choice of cocycle $\alpha$ satisfying \eqref{easycoc} and \eqref{triviac} (when $g$ is in $M_{23}\subset M_{24}$) or \eqref{notsoeasy} (otherwise).

Group $27$ is the only case where the pairs $(g,h)$ and $(g^{-1},h^{-1})$ are not conjugated within $M_{24}$. Thus, in this case, charge conjugation gives the identities $\phi_{{\rm 2B},{\rm 8A_1}}=\phi_{{\rm 2B},{\rm 8A_2}}$, $\phi_{{\rm 8A},{\rm 2B_1}}=\phi_{{\rm 8A},{\rm 2B_2}}$, and so on, and the respective functions are denoted in the following tables by $\phi_{{\rm 2B},{\rm 8A_{1,2}}}$, $\phi_{{\rm 8A},{\rm 2B_{1,2}}}$, etcetera.

Most of the modular groups $\Gamma_{g,h}$ are of the form $\Gamma(1)=SL(2,\ZZ)$, $\Gamma_0(N)$ 
 or conjugates of $\Gamma_0(N)$ in $SL(2,\ZZ)$. The exceptions are  the group in case $32$, where 
\be 
\Gamma_{\rm  2B,10A}=\bigcup_{i\in \ZZ/3\ZZ, j\in \ZZ/4\ZZ}\left( \begin{smallmatrix}
1 & 1\\ -5 & -4
\end{smallmatrix} \right)^i\left( \begin{smallmatrix}
-3 & -1\\ 10 & 3
\end{smallmatrix} \right)^j\Gamma_{2,10}\ ,\ee
is a subgroup of index $12$ in $SL(2,\ZZ)$ and 
\be
\Gamma_{2,10}:=\{ \left(\begin{smallmatrix}
a & b\\ c & d
\end{smallmatrix}\right) \in SL(2,\ZZ)\mid a\equiv 1,\ b\equiv 0\mod 2,\  c\equiv 0,\ d\equiv 1 \mod 10\}\ ,
\ee is the group of elements $\gamma\in SL(2,\ZZ)$ such that $(g,h)\cdot\gamma=(g,h)$;
 the group in case $12$, with
\be 
\Gamma_{\rm  \rm 2A,4A}=\{ \left(\begin{smallmatrix}
a & b\\ c & d
\end{smallmatrix}\right) \in SL(2,\ZZ)\mid b\equiv 0\mod 2, c\equiv 0 \mod 4\}\ ,
\ee 
which is a conjugate of $\Gamma_0(8)$ in $SL(2,\RR)$; and the group in case $22$, with
\be 
\Gamma_{\rm 4A,4C}=\langle \left(\begin{smallmatrix}
-1 & 1\\ -2 & 1
\end{smallmatrix}\right)\;, \  \left(\begin{smallmatrix}
1 & 2\\ 0 & 1
\end{smallmatrix} \right) \;  , \
\left(\begin{smallmatrix} 
1 & 0\\ 4 & 1
\end{smallmatrix} \right) \; , \
\left( \begin{smallmatrix}
3 & -2\\ -4 & 3
\end{smallmatrix} \right) \rangle\ .
\ee

\newpage

%\advance\voffset by -2.3 cm % Shift top margin
%\advance\hoffset by -1 cm % Shift top margin
\newgeometry{margin=1cm}
\thispagestyle{empty}
\begin{table}
 \newcolumntype{L}{>{$}l<{$}}
  \newcolumntype{C}{>{$}c<{$}}
\rowcolors{2}{gray!11}{}
\scalebox{0.95}{\renewcommand*{\arraystretch}{1.3}
\begin{tabular}{cCCCC}
$g$ &  \text{Orbits on }{\bf 24} & \text{Functions} & \Tr_{\bf 24}(g) & F_{e,g}(\tau)  \\\hline
1A & 1^{24} & & 24 & 0 \\ 
2A & 1^8\cdot 2^8 & \phi_{2A,2A_1} & 8 & 32\,\psi^{(2)}(\tau)\\
3A & 1^6\cdot 3^6 & \phi_{3A,3A_{1,2}} & 6 & 18\,\psi^{(3)}(\tau)\\
4B & 1^4\cdot 2^2\cdot 4^4 & \phi_{4B,4B_{2,6}},\,\phi_{4B,2A_1},\,\phi_{2A,4B_1} & 4 & -8\,\psi^{(2)}(\tau)+16\,\psi^{(4)}(\tau)\\
5A & 1^4\cdot 5^4 & \phi_{5A,5A_{1,2,3,4}} & 4 & 10\,\psi^{(5)}(\tau)\\
6A & 1^2\cdot 2^2\cdot 3^2 \cdot 6^2 & \begin{matrix}\phi_{6A,6A_{5,6}},\,\phi_{6A,3A_{1,2}},\,\phi_{6A,2A_3},\\ \phi_{3A,6A_{1,2}},\phi_{3A,2A_{1}},\,\phi_{2A,6A_{1}},\,\phi_{2A,3A_{1}} \end{matrix}& 2 &  -4\,\psi^{(2)}(\tau)-6\,\psi^{(3)}(\tau)+12\,\psi^{(6)}(\tau)\\
7AB & 1^3\cdot 7^3 & \phi_{7A,7A_{1,2,3}},\,\phi_{7A,7B_{1,2,3}} & 3 &7\,\psi^{(7)}(\tau)\\
8A & 1^2\cdot 2\cdot 4 \cdot 8^2 & \begin{matrix}\phi_{8A,8A_{1,3,4,5}},\,\phi_{8A,4B_{2,4}},\\ \phi_{8A,2A_1},\,\phi_{4B,8A_{1,4}},\,\phi_{2A,8A_{1}} \end{matrix}& 2 & -4\,\psi^{(4)}(\tau)+8\,\psi^{(8)}(\tau)\\
11A & 1^2\cdot 11^2 & & 2 & \frac{22}{5}\bigl(\psi^{(11)}(\tau) -\eta(\tau)^2\eta(11\tau)^2\bigr)\\
14AB & 1\cdot 2\cdot 7\cdot 14 & \begin{matrix}\phi_{2A,7AB_1},\,\phi_{2A,14AB_1},\, \phi_{7AB,2A_1},\\
\phi_{7AB,14A_{1,2,3}},\,\phi_{7AB,14B_{1,2,3}}\end{matrix}& 1 & \begin{array}{c}\frac{1}{3}\bigl(-2\,\psi^{(2)}(\tau)- 7\,\psi^{(7)}(\tau)+14\,\psi^{(14)}(\tau)\\
-14\,\eta(\tau)\eta(2\tau)\eta(7\tau)\eta(14\tau)\bigr)\end{array}\\
15AB & 1\cdot 3\cdot 5\cdot 15 & \begin{matrix}\phi_{3A,5A_{1,2}},\,\phi_{3A,15A_{1,2}},\,
\phi_{3A,15B_{1,2}} \\
 \phi_{5A,3A_{1,2}},\, \phi_{5A,15A_{1,2,3,4}},\,\phi_{5A,15B_{1,2,3,4}}\end{matrix} & 1 &
\begin{array}{c} \frac{1}{4}\bigl(-3\,\psi^{(3)}(\tau)- 5\,\psi^{(5)}(\tau)+15\,\psi^{(15)}(\tau)
\\
 -15\,\eta(\tau)\eta(3\tau)\eta(5\tau)\eta(15\tau)\bigr)\end{array}\\
23AB & 1\cdot 23 & & 1& 
        \frac{23}{11}\bigl(\psi^{(23)}(\tau)- f_{23,a}(\tau)+3f_{23,b}(\tau)\bigr)      \\ \hline
2B & 2^{12} & \phi_{2B,2B_3} & 0 &2\,{\eta(\tau)^8\over \eta(2\tau)^4}\\
3B & 3^8 & \phi_{3B,3B_{1,2}}& 0 & 2\,{\eta(\tau)^6\over \eta(3\tau)^2}\\
4A & 2^4\cdot 4^4 & \phi_{4A,4A_{6,8}},\, \phi_{4A,2A_{4}},\, \phi_{2A,4A_{1}} & 0 &2\,{\eta(2\tau)^8\over \eta(4\tau)^4}\\
4C & 4^6 & \phi_{4C,4C_{1,2}},\, \phi_{4C,2B_{1}},\, \phi_{2B,4C_{1}}& 0 & 2\,{\eta(\tau)^4 \, \eta(2\tau)^2\over
        \eta(4\tau)^2}\\
6B & 6^4 & \begin{matrix}\phi_{6B,6B_{4,6}},\,\phi_{6B,3B_{1,2}},\,\phi_{6B,2B_3},\\ \phi_{3B,6B_{1,2}},\phi_{3B,2B_{1}},\,\phi_{2B,6B_{1}},\,\phi_{2B,3B_{1}} \end{matrix}& 0 & 2\,{\eta(\tau)^2 \, \eta(2\tau)^2 \, \eta(3\tau)^2\over
        \eta(6\tau)^2}\\
10A & 2^2\cdot 10^2 & \begin{matrix} \phi_{10A,10A_{4,6,8,12}},\,\phi_{10A,2B_{1}},\,\phi_{10A,5A_{1,2,3,4}},\\ \phi_{2B,10A_{2}},\, \phi_{2B,5A_1},\, \phi_{5A,10A_{1,2,3,4}},\,\phi_{5A,2B_1}\end{matrix}& 0 & 2\,{\eta(\tau)^3 \, \eta(2\tau) \, \eta(5\tau)\over
        \eta(10\tau)}\\
12A & 2\cdot 4\cdot 6\cdot 12 & \begin{matrix}
\phi_{2A,12A_{1}},\,  \phi_{3A,12A_{1,2}},\, \phi_{3A,4A_{1}},\, \phi_{4A,12A_{1,2}},\\ \phi_{4A,6A_{1}},\, \phi_{4A,3A_{1}},\, \phi_{6A,12A_{1,2}},\, \phi_{6A,4A_{1}}
\end{matrix}& 0 & 2\,{\eta(\tau)^3 \, \eta(4\tau)^2 \,\eta(6\tau)^3\over
        \eta(2\tau) \, \eta(3\tau)
        \,\eta(12\tau)^2}\\
12B & 12^2 & \begin{matrix}
\phi_{2B,12B_{1}},\,  \phi_{3B,12B_{1,2}},\, \phi_{3B,4C_{1}},\, \phi_{4C,12B_{1,2}},\\ \phi_{4C,6B_{1}},\, \phi_{4C,3B_{1}},\, \phi_{6B,12B_{1,2}},\, \phi_{6B,4C_{1}}
\end{matrix} & 0 & 2\,{\eta(\tau)^4 \, \eta(4\tau) \, \eta(6\tau)\over
        \eta(2\tau) \, \eta(12\tau)}\\
21AB & 3\cdot 21 & \begin{matrix}
\phi_{3B,21A_{1,2}},\, \phi_{3B,21B_{1,2}},\, \phi_{3B,7A_{1}},\, \phi_{3B,7B_{1}},\\ \phi_{7AB,21A_{1,2,3}},\, \phi_{7AB,21B_{1,2,3}},\, \phi_{7AB,3B_{1}} 
\end{matrix} & 0 & 
        \frac{7}{3} \,
        \frac{
          \eta(\tau)^3 \,
          \eta(7 \, \tau)^3
        }{
          \eta(3 \, \tau) \, \eta(21 \, \tau)
        }
        - \frac{1}{3} \,
        \frac{ \eta(\tau)^6}{
          \eta(3 \, \tau)^2}   
\end{tabular}}
\caption{\label{t:cycl}\small  The cyclic subgroups of $M_{24}$. 
}\end{table}

\restoregeometry

\newpage
\newgeometry{top=3.5cm,left=2cm}
\begin{landscape}
%\advance\voffset by 1.35cm % Shift top margin
\begin{table}
 \newcolumntype{L}{>{$}l<{$}}
  \newcolumntype{C}{>{$}c<{$}}
\rowcolors{2}{gray!11}{}
{\renewcommand*{\arraystretch}{1.3}
\begin{tabular}{CCLCCCCCCCCCCCCCCCCCCCC}%{cclcccccccccccccccccccc}
\# &  \text{Structure} & \text{Elements} & |C(g,h)| & \frac{|N(g,h)|}{|C(g,h)|} & \text{Orbits on }{\bf 24}  & \text{Max subgr.} & \text{Functions}& \Gamma_{g,h}  \\\hline
1.&  \ZZ_2\times\ZZ_2 & ({\rm 2A})^3& 1536 &6&  1^ 8 \cdot    4^ 4   &  & \phi_{\rm 2A,2A_2} & \Gamma(1)  \\
2.&  \ZZ_2\times\ZZ_2 & ({\rm 2A})^3&1536 &6&  2^ {12}   &  & \phi_{\rm 2A,2A_3} & \Gamma(1)  \\
3.&  \ZZ_2\times\ZZ_2 & ({\rm 2A})^3&128 &6&  1^ 4 \cdot    2^ 6 \cdot    4^ 2   & & \phi_{\rm 2A,2A_5} & \Gamma(1)  \\
4.&  \ZZ_2\times\ZZ_2 & ({\rm 2B})^3& 3840 &6&  4^ 6 & & \phi_{\rm 2B,2B_2} & \Gamma(1) \\
5.&  \ZZ_2\times\ZZ_2 & ({\rm 2B})^3& 96 &6& 4^ 6 & & \phi_{\rm 2B,2B_4} & \Gamma(1)   \\
6.&  \ZZ_2\times\ZZ_2 &({\rm 2B})^3&  64 &6&  4^ 6 & & \phi_{\rm 2B,2B_6} & \Gamma(1)\\ 
7.&  \ZZ_2\times\ZZ_2 & ({\rm 2A})^2({\rm 2B})& 256 &2& 2^ 8 \cdot    4^ 2 & & \phi_{\rm 2B,2A_1},\ \phi_{\rm 2A,2B_3},\ \phi_{\rm 2A,2A_4} & \Gamma_0(2) \\
8.&  \ZZ_2\times\ZZ_2 & ({\rm 2A})({\rm 2B})^2& 512 &2&2^ 4 \cdot    4^ 4  & &  \phi_{\rm 2A,2B_1},\ \phi_{\rm 2B,2A_2},\ \phi_{\rm 2B,2B_1} & \Gamma_0(2)  \\
 9.&  \ZZ_2\times\ZZ_2 &({\rm 2A})({\rm 2B})^2& 128 &2& 2^ 4 \cdot    4^ 4  & & \phi_{\rm 2A,2B_2},\ \phi_{\rm 2B,2A_3},\ \phi_{\rm 2B,2B_5} & \Gamma_0(2)   \\
 10.&  \ZZ_2\times\ZZ_4 & ({\rm 2A})^3({\rm 4A})^4& 64 & 8& 2^ 4 \cdot    8^ 2 &   1  & \phi_{\rm 2A,4A_2},\ \phi_{\rm 4A,2A_2},\ \phi_{\rm 4A,4A_3} & \Gamma_0(2)  \\[-1pt]
 11.&  \ZZ_2\times\ZZ_4 & ({\rm 2A})^3({\rm 4A})^4& 64 &8&  4^ 6 & 2  & \phi_{\rm 2A,4A_3},\ \phi_{\rm 4A,2A_3},\ \phi_{\rm 4A,4A_7} & \Gamma_0(2)   \\
 12.&  \ZZ_2\times\ZZ_4 & \begin{matrix}({\rm 2A})^2({\rm 2B})\\({\rm 4A})^2({\rm 4B})^2\end{matrix}&  32 &2& 2^ 2 \cdot    4^ 3 \cdot    8^ 1 &  7 & {\renewcommand*{\arraystretch}{0.6}\begin{matrix}
\phi_{\rm 2A,4A_4},\phi_{\rm 2A,4B_4},\phi_{\rm 2B,4A_1},\\\phi_{\rm 2B,4B_1},\phi_{\rm 4A,2A_1},\phi_{\rm 4A,2B_1},\\
\phi_{\rm 4B,2A_4},\phi_{\rm 4B,2B_2},\phi_{\rm 4B,4A_1},\\\phi_{\rm 4B,4A_2},\phi_{\rm 4A,4B_3},\phi_{\rm 4A,4B_4}
\end{matrix}} & \Gamma_{\rm 2A,4A} \\
 13.&  \ZZ_2\times\ZZ_4 & ({\rm 2A})({\rm 2B})^2({\rm 4A})^4& 64&8& 4^ 2 \cdot    8^ 2   &  8  & \phi_{\rm 2B,4A_2},\ \phi_{\rm 4A,2B_3},\ \phi_{\rm 4A,4A_5} & \Gamma_0(2) \\
14.&  \ZZ_2\times\ZZ_4 & ({\rm 2A})({\rm 2B})^2({\rm 4A})^4&32 &8& 4^ 2 \cdot    8^ 2 &  8  & \phi_{\rm 2B,4A_3},\ \phi_{\rm 4A,2B_2},\ \phi_{\rm 4A,4A_2} & \Gamma_0(2)  \\
15.&  \ZZ_2\times\ZZ_4 & ({\rm 2A})({\rm 2B})^2({\rm 4C})^4& 32 &4& 4^ 2 \cdot    8^ 2 &  8  & 
{\renewcommand*{\arraystretch}{0.6}\begin{matrix}
\phi_{\rm 2A,4C_1},\phi_{\rm 4C,2A_2},\phi_{\rm 4C,2B_3},\\\phi_{\rm 2B,4C_2},
\phi_{\rm 4C,4C_4},\phi_{\rm 4C,4C_6}
\end{matrix}}
& \Gamma_0(4)  \\
16.&  \ZZ_2\times\ZZ_4 &({\rm 2A})({\rm 2B})^2({\rm 4C})^4& 16&4& 4^ 2 \cdot    8^ 2  &   9  & {\renewcommand*{\arraystretch}{0.6}\begin{matrix}
\phi_{\rm 2A,4C_2},\phi_{\rm 4C,2A_1},\phi_{\rm 4C,2B_2},\\\phi_{\rm 2B,4C_3},
\phi_{\rm 4C,4C_3},\phi_{\rm 4C,4C_5}
\end{matrix}}& \Gamma_0(4) \\
17.&  \ZZ_2\times\ZZ_4 & ({\rm 2A})^3({\rm 4B})^4& 64 &8& 1^ 4 \cdot    2^ 2 \cdot    8^ 2  &  1 & \phi_{\rm 2A,4B_2},\ \phi_{\rm 4B,2A_2},\ \phi_{\rm 4B,4B_8} & \Gamma_0(2)  \\
18.&  \ZZ_2\times\ZZ_4 &({\rm 2A})^3({\rm 4B})^4& 64 &  8& 2^ 4 \cdot    4^ 4 &   2 &  \phi_{\rm 2A,4B_3},\ \phi_{\rm 4B,2A_5},\ \phi_{\rm 4B,4B_9} & \Gamma_0(2) \\[-1pt]
19.&  \ZZ_2\times\ZZ_4 &({\rm 2A})^3({\rm 4B})^4& 16 & 8& 1^ 2 \cdot    2^ 3 \cdot    4^ 2 \cdot    8^ 1&  3  & \phi_{\rm 2A,4B_5},\ \phi_{\rm 4B,2A_3},\ \phi_{\rm 4B,4B_3} & \Gamma_0(2)   
\end{tabular}}\caption{\label{t:noncycl} The non-cyclic abelian groups $\langle g,h\rangle$.}\end{table}

\end{landscape}
\restoregeometry

\newpage

\begin{landscape}
\advance\voffset by 1cm % Shift top margin
\begin{table}
 \newcolumntype{L}{>{$}l<{$}}
  \newcolumntype{C}{>{$}c<{$}}
\rowcolors{2}{gray!11}{}
{\renewcommand*{\arraystretch}{1.3}
\begin{tabular}{CCLCCCCCCCCCCCCCCCCCCCC}%{cclcccccccccccccccccccc}
\# &  \text{Structure} & \text{Elements} & |C(g,h)| & \frac{|N(g,h)|}{|C(g,h)|} & \text{Orbits on }{\bf 24}  & \text{Max subgr.} & \text{Functions}& \Gamma_{g,h}  \\\hline
 20.&  \ZZ_2\times\ZZ_4 & ({\rm 2A})({\rm 2B})^2({\rm 4B})^4&64 &8& 2^ 4 \cdot    8^ 2 &  8   &\phi_{\rm 2B,4B_2},\ \phi_{\rm 4B,2B_3},\ \phi_{\rm 4B,4B_1} & \Gamma_0(2) \\
21.&  \ZZ_2\times\ZZ_4 & ({\rm 2A})({\rm 2B})^2({\rm 4B})^4& 16 & 8&  2^ 4 \cdot    8^ 2 &  9  & \phi_{\rm 2B,4B_3},\ \phi_{\rm 4B,2B_1},\ \phi_{\rm 4B,4B_5} & \Gamma_0(2)  \\
22.&  \ZZ_4\times\ZZ_4 & {\renewcommand*{\arraystretch}{.8}\begin{matrix}
({\rm 2A})({\rm 2B})^2 \\({\rm 4A})^4({\rm 4C})^8
\end{matrix}}& 16 &16&  8^ 1 \cdot   16^ 1 &  13, 15, 15 &  {\renewcommand*{\arraystretch}{0.6}\begin{matrix}
\phi_{\rm 4A,4C_1},\phi_{\rm 4A,4C_2},\phi_{\rm 4C,4A_1},\\\phi_{\rm 4C,4A_2},\phi_{\rm 4C,4C_7},
\phi_{\rm 4C,4C_8}
\end{matrix}}& \Gamma_{\rm 4A,4C}   \\
23.&  \ZZ_4\times\ZZ_4 & ({\rm 2A})^3({\rm 4A})^8({\rm 4B})^4&16& 32& 2^ 2 \cdot    4^ 1 \cdot   16^ 1 &    10, 10, 17 &  \phi_{\rm 4B,4A_3},\ \phi_{\rm 4A,4B_1},\ \phi_{\rm 4A,4A_1} & \Gamma_0(2)    \\
24.&  \ZZ_4\times\ZZ_4 & ({\rm 2A})^3({\rm 4A})^8({\rm 4B})^4& 16 &32&  4^ 2 \cdot    8^ 2 &   11, 11, 18   & \phi_{\rm 4B,4A_4},\ \phi_{\rm 4A,4B_2},\ \phi_{\rm 4A,4A_4} & \Gamma_0(2)   \\
25.&  \ZZ_4\times\ZZ_4 & ({\rm 2A})^3({\rm 4B})^{12}&16& 96&  1^ 4 \cdot    4^ 1 \cdot   16^ 1  &   17, 17, 17 &  \phi_{\rm 4B,4B_4} & \Gamma(1) \\[-1pt]
26.&  \ZZ_4\times\ZZ_4 & ({\rm 2A})^3({\rm 4B})^{12}&16& 96&  4^ 6 &  18, 18, 18 &   \phi_{\rm 4B,4B_7} & \Gamma(1) \\
27.&  \ZZ_2\times\ZZ_8 &{\renewcommand*{\arraystretch}{.8}\begin{matrix}
({\rm 2A})({\rm 2B})^2\\ ({\rm 4B})^4({\rm 8A})^8
\end{matrix}}  &16 &8& 2^ 2 \cdot    4^ 1 \cdot   16^ 1 & 20  &  {\renewcommand*{\arraystretch}{0.6}\begin{matrix}
\phi_{\rm 2B,8A_{1,2}},\phi_{\rm 8A,2B_{1,2}},\phi_{\rm 4B,8A_{2,3}},\\ \phi_{\rm 8A,4B_{1,3}},
\phi_{\rm 8A,8A_{2,8}},\phi_{\rm 8A,8A_{6,7}}
\end{matrix}} & \Gamma_0(4)\\
28.&  \ZZ_2\times\ZZ_6 & ({\rm 2A})^3({\rm 3A})^2({\rm 6A})^6& 12& 12& 1^ 2 \cdot    3^ 2 \cdot    4^ 1 \cdot   12^ 1 &  1 &  {\renewcommand*{\arraystretch}{0.6}\begin{matrix}\phi_{\rm 2A,6A_2},\phi_{\rm 6A,2A_1},\\\phi_{\rm 6A,6A_1},\phi_{\rm 6A,6A_2}\end{matrix}} & \Gamma_0(3) \\
29.&  \ZZ_2\times\ZZ_6 & ({\rm 2A})^3({\rm 3A})^2({\rm 6A})^6& 12 &12&  2^ 3 \cdot    6^ 3  &   2  &  {\renewcommand*{\arraystretch}{0.6}\begin{matrix}\phi_{\rm 2A,6A_3},\phi_{\rm 6A,2A_2},\\\phi_{\rm 6A,6A_3},\phi_{\rm 6A,6A_4}\end{matrix}} & \Gamma_0(3)\\
30.&  \ZZ_2\times\ZZ_6 & ({\rm 2B})^3({\rm 3B})^2({\rm 6B})^6&12 &12&  12^ 2 &  4   & {\renewcommand*{\arraystretch}{0.6}\begin{matrix}\phi_{\rm 2B,6B_2},\phi_{\rm 6B,2B_1},\\\phi_{\rm 6B,6B_2},\phi_{\rm 6B,6B_5}\end{matrix}} & \Gamma_0(3) \\
31.&  \ZZ_2\times\ZZ_6 & ({\rm 2B})^3({\rm 3B})^2({\rm 6B})^6& 12 &12& 12^ 2 &  5  &  {\renewcommand*{\arraystretch}{0.6}\begin{matrix}\phi_{\rm 2B,6B_3},\phi_{\rm 6B,2B_2},\\\phi_{\rm 6B,6B_1},\phi_{\rm 6B,6B_3}\end{matrix}}  & \Gamma_0(3) \\
32.&  \ZZ_2\times\ZZ_{10} & ({\rm 2B})^3({\rm 5A})^4({\rm 10A})^{12} & 20 &12&4^ 1 \cdot   20^ 1 & 4 &  {\renewcommand*{\arraystretch}{0.6}\begin{matrix} \phi_{\rm 2B,10A_1},\phi_{\rm 2B,10A_3},\phi_{\rm 10A,10A_1},\\\phi_{\rm 10A,10A_2},\phi_{\rm 10A,10A_3},\phi_{\rm 10A,10A_5},
\\ \phi_{\rm 10A,10A_7},\phi_{\rm 10A,10A_9},\phi_{\rm 10A,10A_{10}},\\\phi_{\rm 10A,10A_{11}},
\phi_{\rm 10A,2B_2},\phi_{\rm 10A,2B_3}
\end{matrix}}  & \Gamma_{\rm 2B,10A}  \\
33.&  \ZZ_3\times\ZZ_3 & ({\rm 3A})^8 & 9 &48&  1^ 3 \cdot    3^ 4 \cdot    9^ 1&  & \phi_{\rm 3A,3A_3} & \Gamma(1)  \\
34.&  \ZZ_3\times\ZZ_3 & ({\rm 3A})^2({\rm 3B})^6& 9 &12&  3^ 2 \cdot    9^ 2 &  & {\renewcommand*{\arraystretch}{0.6}\begin{matrix}\phi_{\rm 3A,3B_1},\phi_{\rm 3B,3A_1},\\\phi_{\rm 3B,3B_3},\phi_{\rm 3B,3B_4}\end{matrix}} &\Gamma_0(3)
\end{tabular}}\centering{ Table \ref*{t:noncycl} (continued)}\end{table}
\normalsize
\end{landscape}

\newpage

\thispagestyle{empty}

\begin{landscape}
\advance\voffset by 2.7cm % Shift top margin

\begin{table}
\newcolumntype{L}{>{$}l<{$}}
  \newcolumntype{C}{>{$}c<{$}}
   \newcolumntype{m}{>{\bgroup\footnotesize$}c<{$\egroup}}
 \newcolumntype{R}{>{$}r<{$}}
\rowcolors{2}{gray!11}{}
\scalebox{0.85}{\renewcommand*{\arraystretch}{1.3}
\begin{tabular}{CCCCCC}
\# &   {\Large{ (g,h)}} & \phi_{g,h} & 2\T^\alpha_2\phi_{g,h} & 3\T^\alpha_3\phi_{g,h} & 4\T^\alpha_4\phi_{g,h}\\\hline
 \begin{matrix} 1.\\ 2.\\3.\end{matrix} & 
 \begin{matrix} ({2A},{2A_2})\\ ({2A},{2A_3}) \\  ({2A},{2A_5})\end{matrix}
 & 0& %\frac{1}{2} \left(
 \phi _{{2A},{e}}(2 \tau ,2 z)+\phi
   _{{e},{2A}}\left(\frac{\tau }{2},z\right)+\phi _{{e},{2A}}\left(\frac{\tau +1}{2},z\right)%\right) 
   & 0 &
  \begin{matrix} \phi _{{2A},{e}}(4 \tau ,4 z)+\phi _{{e},{2A}}\left(\frac{2\tau+1}{2} ,2 z\right)\\+\phi _{{e},{e}}(\tau ,2 z)+\sum_{i=0}^3\phi _{{e},{2A}}\left(\frac{\tau+i }{4},z\right)\end{matrix}\\[7pt]
 \begin{matrix}
 4. \\ 5.\\ 6.
 \end{matrix} &
 \begin{matrix}
  ({2B},{2B_2})\\ ({2B},{2B_4})\\ ({2B},{2B_6})
 \end{matrix}
  & 0& %\frac{1}{2} \left(
  -\phi _{{2B},{e}}(2 \tau ,2 z)+\phi
   _{{e},{2B}}\left(\frac{\tau }{2},z
   \right)-\phi _{{e},{2B}}\left(\frac{\tau +1}{2},z\right)%\right)  
   & 0 
   & \begin{matrix} \phi _{{2B},{e}}(4 \tau ,4 z)-\phi _{{e},{2B}}\left(\frac{2\tau+1}{2} ,2 z\right)\\+\phi _{{e},{e}}(\tau ,2 z)+\sum_{i=0}^3\phi _{{e},{2B}}\left(\frac{\tau+i }{4},z\right)\end{matrix}\\[7pt]
 \begin{matrix} 7.\end{matrix} &\begin{matrix}  ({2A},{2A_4})\end{matrix} &  0&%\frac{1}{2} \left(
 -\phi _{{2A},{e}}(2 \tau ,2 z)+\phi
   _{{e},{2A}}\left(\frac{\tau }{2},z\right)+\phi _{{e},{2B}}\left(\frac{\tau +1}{2},z\right)%\right)  
   & 0 
   &\begin{matrix} \phi _{{2A},{e}}(4 \tau ,4 z)+\phi _{{e},{2A}}\left(\frac{2\tau+1}{2} ,2 z\right)+\phi _{{e},{e}}(\tau ,2 z)\\+\sum_{i=0}^1\left(\phi _{{e},{2A}}\left(\frac{\tau+2i }{4},z\right)+\phi _{{e},{2B}}\left(\frac{\tau+2i+1 }{4},z\right)\right)\end{matrix}\\[7pt]
  \begin{matrix}
  8. \\ 9.
\end{matrix}    & \begin{matrix}
  ({2A},{2B_1}) \\ ({2A},{2B_2})
\end{matrix}  & 0& %\frac{1}{2} \left(
 \phi _{{2A},{e}}(2 \tau ,2 z)+\phi
   _{{e},{2B}}\left(\frac{\tau }{2},z\right)+\phi _{{e},{2B}}\left(\frac{\tau +1}{2},z\right)%\right)
     & 0
     &\begin{matrix} \phi _{{2A},{e}}(4 \tau ,4 z)+\phi _{{e},{2A}}\left(\frac{2\tau+1}{2} ,2 z\right)\\+\phi _{{e},{e}}(\tau ,2 z)+\sum_{i=0}^3\phi _{{e},{2B}}\left(\frac{\tau+i }{4},z\right)\end{matrix}\\[7pt]
 \begin{matrix}
 10. \\ 11.
\end{matrix}   &
\begin{matrix}
 ({2A},{4A_2}) \\  ({2A},{4A_3})
\end{matrix}  & 0& %\frac{1}{2} \left(%\phi _{{2A},{2A}}(2 \tau ,2 z)+
 \phi _{{e},{4A}}\left(\frac{\tau }{2},z\right)+\phi _{{e},{4A}}\left(\frac{\tau +1}{2},z\right)%\right)
   & 0
   &\begin{matrix} \phi _{{2A},{e}}(4 \tau ,4 z)+\phi _{{e},{2A}}\left(\frac{2\tau+1}{2} ,2 z\right)\\+\phi _{{e},{2A}}(\tau ,2 z)+\sum_{i=0}^3\phi _{{e},{4A}}\left(\frac{\tau+i }{4},z\right)\end{matrix}\\[7pt]
 \begin{matrix}
 12. 
\end{matrix}   & \begin{matrix}  ({2A},{4B_4}) \end{matrix} & 0& %\frac{1}{2} \left(%\phi _{{2A},{2A}}(2 \tau ,2 z)+
 \phi   _{{e},{4A}}\left(\frac{\tau +1}{2},z\right)+\phi _{{e},{4B}}\left(\frac{\tau }{2},z\right)%\right) 
  & 0
  &\begin{matrix} -\phi _{{2A},{e}}(4 \tau ,4 z)+\phi _{{e},{2B}}\left(\frac{2\tau+1}{2} ,2 z\right)+\phi _{{e},{2A}}(\tau ,2 z)\\+\sum_{i=0}^1\left(\phi _{{e},{4B}}\left(\frac{\tau+2i }{4},z\right)+\phi _{{e},{4A}}\left(\frac{\tau+2i+1 }{4},z\right)\right)\end{matrix}\\[7pt]
 \begin{matrix} 13. \end{matrix} & \begin{matrix}
  ({2B},{4A_2})
\end{matrix}   &  4\frac{\eta(2\tau)^2}{\eta(\tau)^4}\vartheta_1(\tau,z)^2 &%\frac{1}{2} \left(%\phi _{{2B},{2A}}(2 \tau ,2 z)+
 \phi   _{{e},{4A}}\left(\frac{\tau }{2},z\right)-\phi _{{e},{4A}}\left(\frac{\tau +1}{2},z\right)%\right) 
  & \begin{matrix}%\frac{1}{3} \left(
  \phi _{{2B},{4A_2}}(3 \tau ,3 z)+\phi _{{2B},{4A_2}}\left(\frac{\tau }{3},z\right)\\ +i\phi _{{2B},{4A_2}}\left(\frac{\tau+1 }{3},z\right)-\phi_{{2B},{4A_2}}\left(\frac{\tau+2 }{3},z\right)%\right)
  \end{matrix}
  &\begin{matrix} \phi _{{2B},{e}}(4 \tau ,4 z)-\phi _{{e},{2B}}\left(\frac{2\tau+1}{2} ,2 z\right)\\+\phi _{{e},{2A}}(\tau ,2 z)+\sum_{i=0}^3\phi _{{e},{4A}}\left(\frac{\tau+i }{4},z\right)\end{matrix}
 \\[7pt]
 \begin{matrix} 14.\end{matrix} & \begin{matrix}  ({2B},{4A_3}) \end{matrix} &  0&%\frac{1}{2} \left(%\phi _{{2B},{2A}}(2 \tau ,2 z)+
 \phi   _{{e},{4A}}\left(\frac{\tau }{2},z\right)-\phi _{{e},{4A}}\left(\frac{\tau +1}{2},z\right)%\right) 
 & 0
  &\begin{matrix} \phi _{{2B},{e}}(4 \tau ,4 z)-\phi _{{e},{2B}}\left(\frac{2\tau+1}{2} ,2 z\right)\\+\phi _{{e},{2A}}(\tau ,2 z)+\sum_{i=0}^3\phi _{{e},{4A}}\left(\frac{\tau+i }{4},z\right)\end{matrix}\\[7pt]
 \begin{matrix}
 15. \\16.
\end{matrix}  &
 \begin{matrix}
  ({2A},{4C_1}) \\  ({2A},{4C_2})
\end{matrix}   &  0&%\frac{1}{2} \left(%\phi _{{2A},{2B}}(2 \tau ,2 z)+
 \phi   _{{e},{4C}}\left(\frac{\tau }{2},z\right)+\phi _{{e},{4C}}\left(\frac{\tau +1}{2},z\right)%\right) 
  & 0
  &\begin{matrix} \phi _{{2A},{e}}(4 \tau ,4 z)+\phi _{{e},{2B}}\left(\frac{2\tau+1}{2} ,2 z\right)\\+\phi _{{e},{2B}}(\tau ,2 z)+\sum_{i=0}^3\phi _{{e},{4C}}\left(\frac{\tau+i }{4},z\right)\end{matrix}\\[7pt]
% 16.  &  ({2A},{4C_2}) &  0&%\frac{1}{2} \left(%\phi _{{2A},{2B}}(2 \tau ,2 z)+
% \phi   _{{e},{4C}}\left(\frac{\tau }{2},z\right)+\phi _{{e},{4C}}\left(\frac{\tau +1}{2},z\right)%\right) 
%  & 0&\\
 \begin{matrix}
 17. \\ 18. \\ 19.
\end{matrix}   & 
 \begin{matrix}
  ({2A},{4B_2}) \\  ({2A},{4B_3})\\  ({2A},{4B_5})
\end{matrix}    &  0&%\frac{1}{2} \left(%\phi _{{2A},{2A}}(2 \tau ,2 z)+
 \phi  _{{e},{4B}}\left(\frac{\tau }{2},z\right)+\phi _{{e},{4B}}\left(\frac{\tau +1}{2},z\right)%\right) 
  & 0
  &\begin{matrix} \phi _{{2A},{e}}(4 \tau ,4 z)+\phi _{{e},{2A}}\left(\frac{2\tau+1}{2} ,2 z\right)\\+\phi _{{e},{2A}}(\tau ,2 z)+\sum_{i=0}^3\phi _{{e},{4B}}\left(\frac{\tau+i }{4},z\right)\end{matrix}
 \end{tabular}}\\[5pt]
\centering{ Table \ref*{t:noncycl} (continued)}
\end{table}

\end{landscape} 
%\restoregeometry

\newpage

% Shift top margin
\thispagestyle{empty} 

\begin{landscape}
\advance\voffset by 2.7cm

\begin{table}
 \newcolumntype{L}{>{$}l<{$}}
  \newcolumntype{C}{>{$}c<{$}}
   \newcolumntype{m}{>{\bgroup\footnotesize$}c<{$\egroup}}
 \newcolumntype{R}{>{$}r<{$}}
\rowcolors{2}{gray!11}{}
\scalebox{0.85}{\renewcommand*{\arraystretch}{1.3}
\begin{tabular}{CCCCCC}
\# &   {\Large{ (g,h)}} & \phi_{g,h} & 2\T^\alpha_2\phi_{g,h} & 3\T^\alpha_3\phi_{g,h} & 4\T^\alpha_4\phi_{g,h}\\\hline
 \begin{matrix}
 20.\\ 21.
\end{matrix}   &\begin{matrix}
 ({2B},{4B_2})\\  ({2B},{4B_3})
\end{matrix}  & 0& %\frac{1}{2} \left(%\phi _{{2B},{2A}}(2 \tau ,2 z)+
 \phi_{{e},{4B}}\left(\frac{\tau }{2},z\right)-\phi _{{e},{4B}}\left(\frac{\tau +1}{2},z\right)%\right) 
  & 0 &\begin{matrix} \phi _{{2B},{e}}(4 \tau ,4 z)-\phi _{{e},{2B}}\left(\frac{2\tau+1}{2} ,2 z\right)\\+\phi _{{e},{2A}}(\tau ,2 z)+\sum_{i=0}^3\phi _{{e},{4B}}\left(\frac{\tau+i }{4},z\right)\end{matrix}\\[7pt]
% 21. &  ({2B},{4B_3}) & 0& %\frac{1}{2} \left(%\phi _{{2B},{2A}}(2 \tau ,2 z)+
% \phi   _{{e},{4B}}\left(\frac{\tau }{2},z\right)-\phi _{{e},{4B}}\left(\frac{\tau +1}{2},z\right)%\right)
%   & 0 \\
 \begin{matrix}
 22. 
\end{matrix}   &\begin{matrix}  ({4A},{4C_1})\end{matrix} & 0& %\frac{1}{2} \left(%\phi _{{2A},{4C}}\left(\frac{\tau }{2},z\right)+\phi
  % _{{2A},{4C}}\left(\frac{\tau +1}{2},z\right)+
  \phi _{{4A},{2B_3}}(2 \tau ,2 z)%\right) 
   & 0
   &\begin{matrix} -\phi _{{4A},{e}}(4 \tau ,4 z)+\phi _{{2A},{4A_1}}\left(\frac{2\tau+1}{2} ,2 z\right)\\+\sum_{i=0}^3(-1)^i\phi _{{e},{4C}}\left(\frac{\tau+i }{4},z\right)\end{matrix}\\[7pt]
 \begin{matrix}
 23. \\24.
\end{matrix}  &\begin{matrix}
  ({4B},{4A_3})\\ ({4B},{4A_4})
\end{matrix}  & 2\sqrt{2}\frac{\eta(2\tau)^2}{\eta(\tau)^4}\vartheta_1(\tau,z)^2 
 & 0%\frac{1}{2} \left(\phi _{{2A},{4A}}\left(\frac{\tau }{2},z\right)+\phi
  % _{{2A},{4A}}\left(\frac{\tau +1}{2},z\right)+\phi _{{4B},{2A}}(2 \tau ,2 z)\right) 
  &\begin{matrix}%\frac{1}{3} \left(
  \phi _{{4B},{4A_3}}(3 \tau ,3 z)+\phi _{{4B},{4A_3}}\left(\frac{\tau }{3},z\right)\\ +i\phi _{{4B},{4A_3}}\left(\frac{\tau+1 }{3},z\right)-\phi_{{4B},{4A_3}}\left(\frac{\tau+2 }{3},z\right)%\right)
  \end{matrix} 
  &\begin{matrix} \phi _{{4B},{e}}(4 \tau ,4 z)+\phi _{{2A},{4B_1}}\left(\frac{2\tau+1}{2} ,2 z\right)\\+\sum_{i=0}^3\phi _{{e},{4A}}\left(\frac{\tau+i }{4},z\right)\end{matrix}\\[7pt]
% 24. &  ({4B},{4A_4}) &2\sqrt{2}\frac{\eta(2\tau)^2}{\eta(\tau)^4}\vartheta_1(\tau,z)^2
% & 0%\frac{1}{2} \left(\phi _{{2A},{4A}}\left(\frac{\tau }{2},z\right)+\phi
%   %_{{2A},{4A}}\left(\frac{\tau +1}{2},z\right)+\phi _{{4B},{2A}}(2 \tau ,2 z)\right) 
%   &\begin{matrix}%\frac{1}{3} \left(
%  \phi _{{4B},{4A_4}}(3 \tau ,3 z)+\phi _{{4B},{4A_4}}\left(\frac{\tau }{3},z\right)\\ +i\phi _{{4B},{4A_4}}\left(\frac{\tau+1 }{3},z\right)-\phi_{{4B},{4A_4}}\left(\frac{\tau+2 }{3},z\right)%\right)
%  \end{matrix}
%  &\\
 \begin{matrix}
 25.\\ 26.
\end{matrix}     &\begin{matrix}
 ({4B},{4B_4})\\ ({4B},{4B_7})
\end{matrix}  & 0& 0%\frac{1}{2} \left(\phi _{{2A},{4B}}\left(\frac{\tau }{2},z\right)+\phi
  % _{{2A},{4B}}\left(\frac{\tau +1}{2},z\right)+\phi _{{4B},{2A}}(2 \tau ,2 z)\right)
   & 0 
   &\begin{matrix} \phi _{{4B},{e}}(4 \tau ,4 z)+\phi _{{2A},{4B_1}}\left(\frac{2\tau+1}{2} ,2 z\right)\\+\sum_{i=0}^3\phi _{{e},{4B}}\left(\frac{\tau+i }{4},z\right)\end{matrix}\\[7pt]
% 26.  & ({4B},{4B_7}) & 0
% & 0%\frac{1}{2} \left(\phi _{{2A},{4B}}\left(\frac{\tau }{2},z\right)+\phi
%  % _{{2A},{4B}}\left(\frac{\tau +1}{2},z\right)+\phi _{{4B},{2A}}(2 \tau ,2 z)\right) 
%   & 0\\
 \begin{matrix} 27.\end{matrix}  & \begin{matrix} ({2B},{8A_1})\end{matrix}  & 2\frac{\eta(2\tau)^2}{\eta(\tau)^4}\vartheta_1(\tau,z)^2
 & %\frac{1}{2} \left(%\phi _{{2B},{4B}}(2 \tau ,2 z)+
 \phi  _{{e},{8A}}\left(\frac{\tau }{2},z\right)-\phi _{{e},{8A}}\left(\frac{\tau +1}{2},z\right)%\right)
  & \begin{matrix}%\frac{1}{3} \left(
  \phi _{{2B},{8A_{1,2}}}(3 \tau ,3 z)+\phi _{{2B},{8A_{1,2}}}\left(\frac{\tau }{3},z\right)\\ +i\phi _{{2B},{8A_{1,2}}}\left(\frac{\tau+1 }{3},z\right)-\phi_{{2B},{8A_{1,2}}}\left(\frac{\tau+2 }{3},z\right)%\right)
  \end{matrix}
  &\begin{matrix} -\phi _{{e},{4B}}\left(\frac{2\tau+1}{2} ,2 z\right)+\phi _{{e},{4B}}\left(\tau ,2 z\right)\\+\sum_{i=0}^3\phi _{{e},{8A}}\left(\frac{\tau+i }{4},z\right)\end{matrix}\\[7pt]
\begin{matrix}
28. \\ 29.
\end{matrix}  & \begin{matrix}
 ({2A},{6A_2})\\ ({2A},{6A_3})
\end{matrix} & 0& %\frac{1}{2} \left(
 \phi _{{2A},{3A_1}}(2 \tau ,2 z)+\phi
   _{{e},{6A}}\left(\frac{\tau }{2},z\right)+\phi _{{e},{6A}}\left(\frac{\tau +1}{2},z\right)%\right) 
    & 0
    &\begin{matrix} \phi _{{2A},{3A}}(4 \tau ,4 z)+\phi _{{e},{6A}}\left(\frac{2\tau+1}{2} ,2 z\right)\\+\phi _{{e},{3A}}(\tau ,2 z)+\sum_{i=0}^3\phi _{{e},{6A}}\left(\frac{\tau+i }{4},z\right)\end{matrix}\\[7pt]
% 29.  & ({2A},{6A_3}) & 0& %\frac{1}{2} \left(
% \phi _{{2A},{3A_1}}(2 \tau ,2 z)+\phi
%   _{{e},{6A}}\left(\frac{\tau }{2},z\right)+\phi _{{e},{6A}}\left(\frac{\tau +1}{2},z\right)%\right) 
%    & 0&\\
 \begin{matrix}
 30. \\ 31.
\end{matrix}    & \begin{matrix}
 ({2B},{6B_2})\\  ({2B},{6B_3})
\end{matrix}  & 0& %\frac{1}{2} \left(
 \phi _{{2B},{3B_1}}(2 \tau ,2 z)+\phi
   _{{e},{6B}}\left(\frac{\tau }{2},z\right)-\phi _{{e},{6B}}\left(\frac{\tau +1}{2},z\right)%\right) 
    & 0
    &\begin{matrix} \phi _{{2B},{3B}}(4 \tau ,4 z)-\phi _{{e},{6B}}\left(\frac{2\tau+1}{2} ,2 z\right)\\+\phi _{{e},{3B}}(\tau ,2 z)+\sum_{i=0}^3\phi _{{e},{6B}}\left(\frac{\tau+i }{4},z\right)\end{matrix}\\[7pt]
% 31.  & ({2B},{6B_3}) & 0& %\frac{1}{2} \left(
% \phi _{{2B},{3B_1}}(2 \tau ,2 z)+\phi
%   _{{e},{6B}}\left(\frac{\tau }{2},z\right)-\phi _{{e},{6B}}\left(\frac{\tau +1}{2},z\right)%\right)
%     & 0&\\
\begin{matrix} 32.\end{matrix} &\begin{matrix} ({2B},{10A_1})\end{matrix} & 0& %\frac{1}{2} \left(
\phi _{{2B},{5A_1}}(2 \tau ,2 z)+\phi
   _{{e},{10A}}\left(\frac{\tau }{2},z\right)-\phi _{{e},{10A}}\left(\frac{\tau +1}{2},z\right)%\right) 
    & 0
    &\begin{matrix} \phi _{{2B},{5A}}(4 \tau ,4 z)-\phi _{{e},{10A}}\left(\frac{2\tau+1}{2} ,2 z\right)\\+\phi _{{e},{5A}}(\tau ,2 z)+\sum_{i=0}^3\phi _{{e},{10A}}\left(\frac{\tau+i }{4},z\right)\end{matrix}\\[7pt]
 \begin{matrix} 33.\end{matrix}  &\begin{matrix} ({3A},{3A_3})\end{matrix} & 0& 0%\frac{1}{2} \left(\phi _{{3A},{3A}}\left(\frac{\tau }{2},z\right)+\phi
   %_{{3A},{3A}}(2 \tau ,2 z)+\phi _{{3A},{3A}}\left(\frac{\tau +1}{2},z\right)\right)   
  & \phi _{{3A},{e}}(3 \tau ,3 z)+\sum_{k=0}^2\phi _{{e},{3A}}\left(\frac{\tau+k }{3},z\right) & 0 \\[7pt]
\begin{matrix}
34.
\end{matrix}  &\begin{matrix} ({3A},{3B_1})\end{matrix} & 0& 0 & \phi _{{3A},{e}}(3 \tau ,3 z)+\sum_{k=0}^2\phi _{{e},{3B}}\left(\frac{\tau+k }{3},z\right) & 0
 %\frac{1}{2} \left(\phi _{{3B},{3A}}\left(\frac{\tau }{2},z\right)+\phi
%   _{{3B},{3A}}(2 \tau ,2 z)+e^{-\frac{2 i \pi }{3}} \phi _{{3B},{3B}}\left(\frac{\tau
%   +1}{2},z\right)\right) 
   \end{tabular}}\\[8pt]
\centering{ Table \ref*{t:noncycl} (continued)}\end{table}

\end{landscape}

\newpage

\newgeometry{left=2cm}
\begin{landscape}
\advance\voffset by 1cm % Shift top margin

\begin{table}
\newcolumntype{L}{>{$}l<{$}}
  \newcolumntype{C}{>{$}c<{$}}
   \newcolumntype{m}{>{\bgroup\footnotesize$}c<{$\egroup}}
 \newcolumntype{R}{>{$}r<{$}}
\rowcolors{2}{gray!11}{}
\begin{minipage}[t]{.6\linewidth}
\scalebox{0.85}{\renewcommand*{\arraystretch}{1.3}
\begin{tabular}{CCCCCC}
\text{Group} &   {\Large{ (g,h)}} & \eta_{g,h} & w\\\hline
 \begin{matrix} 1,\ 2,\ 3\end{matrix} & 
 \begin{matrix} ({2A},{2A_{2,3,5}})\end{matrix}
 & 2^{12} & 6\\[7pt]
 \begin{matrix}
 4,\ 5,\ 6
 \end{matrix} &
 \begin{matrix}
  ({2B},{2B_{2,4,6}})
 \end{matrix}
  & 4^6 & 3\\[7pt]
 \begin{matrix} 7\end{matrix} &\begin{matrix}  ({2A},{2A_4})\\({2A},{2B_3})\\({2B},{2A_1})\end{matrix} &  
 \begin{matrix}1^4\cdot 2^2\cdot 4^4\\ 2^{14}/1^4\\ 4^{14}/8^4\end{matrix}&
 5\\[7pt]
  \begin{matrix}
  8,\ 9
\end{matrix}    & \begin{matrix}
  ({2A},{2B_{1,2}}),\, ({2B},{2A_{2,3}})\\
  ({2B},{2B_{1,5}})
\end{matrix}  & \begin{matrix}
 2^44^4\\  4^{16}/2^48^4
\end{matrix}& 4\\[7pt]
 \begin{matrix}
 10,\  11
\end{matrix}   &
\begin{matrix}
 ({2A},{4A_{2,3}}),\,  ({4A},{2A_{2,3}}) \\  ({4A},{4A_{3,7}})
\end{matrix}  & \begin{matrix}
  4^6\\ 8^{18}/4^6\cdot 16^6
\end{matrix} & 3\\[7pt]
 \begin{matrix}
 12 
\end{matrix}   & \begin{matrix}  ({2A},{4B_4})\\ ({2A},{4A_4})\\ ({2B},{4B_1})\\ ({2B},{4A_1})\\(4B,4A_2) \end{matrix} &\begin{matrix}
1^2\cdot 2\cdot 4\cdot 8^2\\ 2^7\cdot 8^2/1^2\cdot 4\\2^2\cdot 8^7/4\cdot 16^2\\4^5\cdot 8^5/2^2\cdot 16^2\\ \text{irrational}
\end{matrix} & 3\\[7pt]
 \begin{matrix} 13,\ 14 \end{matrix} & \begin{matrix}
  ({2B},{4A_{2,3}}),\, ({4A},{2B_{3,2}})\\ ({4A},{4A_{5,2}})
\end{matrix}   &  \begin{matrix}
 4^2\cdot 8^2\\ 8^8/ 4^2\cdot 16^2
\end{matrix} & 2
 \\[7pt]
 \begin{matrix}
 15,\ 16
\end{matrix}  &
 \begin{matrix}
  ({2A},{4C_{1,2}}),\,({4C},{2A_{2,1}}) \\ ({2B},{4C_{2,3}}),\, ({4C},{2B_{3,2}})\\ ({4C},{4C_{4,6,3,5}})
\end{matrix}   & \begin{matrix}
 4^2\cdot 8^2 \\ 8^8/ 4^2\cdot 16^2 \\  \text{irrational}
\end{matrix} & 2 \\[7pt]
 \begin{matrix}
 17,\ 18,\ 19
\end{matrix}   & 
 \begin{matrix}
  ({2A},{4B_{2,3,5}}),\,  ({4B},{2A_{2,5,3}})\\  ({4B},{4B_{8,9,3}})
\end{matrix}    &  \begin{matrix}
 2^4\cdot 4^4\\ 4^{16}/2^4\cdot 8^4
\end{matrix} & 4
 \end{tabular}}
\end{minipage}
 \begin{minipage}[t]{.45\linewidth}
 \newcolumntype{L}{>{$}l<{$}}
  \newcolumntype{C}{>{$}c<{$}}
   \newcolumntype{m}{>{\bgroup\footnotesize$}c<{$\egroup}}
 \newcolumntype{R}{>{$}r<{$}}
\rowcolors{2}{gray!11}{}
\scalebox{0.85}{\renewcommand*{\arraystretch}{1.3}
\begin{tabular}{CCCCCC}
\text{Group} &   {\Large{ (g,h)}} & \eta_{g,h} & w\\\hline
 \begin{matrix}
 20,\ 21
\end{matrix}   &\begin{matrix}
 ({2B},{4B_{2,3}}),\,({4B},{2B_{3,1}}),\,({4B},{4B_{1,5}}),
\end{matrix}  & 4^6 & 3\\[7pt]
 \begin{matrix}
 22 
\end{matrix}   &\begin{matrix}  ({4A},{4C_{1,2}}),\, ({4C},{4A_{1,2}})\\({4C},{4C_{7,8}})\end{matrix} & \begin{matrix}  8\cdot 16\\ 16^4/8\cdot 32\end{matrix} & 1 \\[7pt]
 \begin{matrix}
 23,\ 24
\end{matrix}  &\begin{matrix}
  ({4B},{4A_{3,4}}),\, ({4A},{4B_{1,2}})\\ ({4A},{4A_{1,4}})
\end{matrix}  & \begin{matrix}
 4^2\cdot 8^2\\ 8^8/ 4^2\cdot 16^2
\end{matrix} & 2 \\[7pt]
 \begin{matrix}
 25,\  26
\end{matrix}     &\begin{matrix}
 ({4B},{4B_{4,7}})
\end{matrix}  & 4^6 & 3\\[7pt]
 \begin{matrix} 27\end{matrix}  & \begin{matrix} ({2B},{8A_{1,2}}),\, ({8A},{2B_{1,2}})\\ ({4B},{8A_{2,3}}),\, ({8A},{4B_{1,3}})\\ ({8A},{8A_{2,6,7,8}})\end{matrix}  &
 \begin{matrix}
  4^2\cdot 8^2\\ 4^2\cdot 8^2\\  8^8/ 4^2\cdot 16^2
\end{matrix}  & 2\\[7pt]
\begin{matrix}
28,\  29
\end{matrix}  & \begin{matrix}
 ({2A},{6A_{2,3}}),\, ({6A},{2A_{1,2}})\\({6A},{6A_{1,2,3,4}})
\end{matrix} & \begin{matrix}
 2^3\cdot 6^3\\ \text{irrational}
\end{matrix} & 3\\[7pt]
 \begin{matrix}
 30,\ 31
\end{matrix}    & \begin{matrix}
 ({2B},{6B_{2,3}}),\,  ({6B},{2B_{1,2}})\\({6B},{6B_{2,5,1,3}})
\end{matrix}  & \begin{matrix}
  12^2\\ \text{irrational}
\end{matrix} & 1\\[7pt]
\begin{matrix} 32\end{matrix} &\begin{matrix} ({2B},{10A_{1,3}}),\,
({10A},2B_{2,3})\\ ({10A},10A)\end{matrix} &\begin{matrix}
 4\cdot 20\\  \text{irrational}
\end{matrix} & 1\\[7pt]
 \begin{matrix} 33\end{matrix}  &\begin{matrix} ({3A},{3A_3})\end{matrix} & 3^8 & 4  \\[7pt]
\begin{matrix}
34
\end{matrix}  &\begin{matrix} ({3A},{3B_1}),\,({3B},{3A_1})\\ ({3B},{3B_{3,4}})\end{matrix} & \begin{matrix}
 3^2\cdot 9^2\\  \text{irrational}
\end{matrix} & 2\\[7pt]\\[37pt]
   \end{tabular}}
   \end{minipage}
{} \\[7pt]
\caption{\small The eta products of \cite{MasonGen}. For each pair $g,h$, a cycle shape of the form $\prod_\ell \ell^{i(\ell)}$ is given: the corresponding eta-product is $\eta_{g,h}(\tau)=\prod_\ell \eta(\ell\tau/N\lambda)^{i(\ell)}$, where $N$ is the order of $g$ and $\lambda$ is the length of the shortest cycle. In the last column, $w$ is the modular weight.}\label{TableMason2}\end{table}

 \end{landscape}

\restoregeometry 
 
 \newpage

\begin{table}
\newcolumntype{L}{>{$}l<{$}}
  \newcolumntype{C}{>{$}c<{$}}
   \newcolumntype{m}{>{\bgroup\footnotesize$}c<{$\egroup}}
 \newcolumntype{R}{>{$}r<{$}}
 \begin{minipage}{.5\textwidth}
\begin{tabular}{CC|CC}
\langle g,h\rangle & h & \langle g,h'\rangle & h'\\\hline 
 \ZZ_2A & 1A & \ZZ_2A & 1A \\ \ZZ_2A & Q & \ZZ_2A & g \\  \ZZ_2A & g & \ZZ_2A & Q \\ \ZZ_2A & Qg & \ZZ_2A & Qg \\     \ZZ_4A & 4A_1 & \ZZ_4A & 4A_1 \\  \ZZ_4A & 4A_1 & \ZZ_4A & 4A_1 \\  \ZZ_4B & 4B_1 & 7 & 2A_4 \\  \ZZ_4B & 4B_1 & 7 & 2B_3 \\   \ZZ_{6}A & 3A_1 & \ZZ_{6}A & 3A_1 \\ \ZZ_{6}A & 3A_1 & \ZZ_{6}A & 6A_1 \\ \ZZ_{6}A & 6A_1 & \ZZ_{6}A & 3A_1 \\    \ZZ_{6}A & 6A_1 & \ZZ_{6}A & 6A_1\\ \ZZ_8A & 8A_1 & 12 & 4B_4 \\  \ZZ_8A & 8A_1 & 12 & 4A_4 \\   \ZZ_{12}A & 12A_1 & \ZZ_{12}A & 12A_1 \\  \ZZ_{12}A & 12A_1 & \ZZ_{12}A & 12A_1 \\  \ZZ_{14}AB & 7A_1 & \ZZ_{14}AB & 7B_1 \\  \ZZ_{14}AB & 7A_1 & \ZZ_{14}AB & 14A_1 \\  \ZZ_{14}AB & 7B_1 & \ZZ_{14}AB & 7A_1  \\ \ZZ_{14}AB & 7B_1 & \ZZ_{14}AB & 14B_1 \\\ZZ_{14}AB & 14A_1 & \ZZ_{14}AB & 7A_1 \\    \ZZ_{14}AB & 14B_1 & \ZZ_{14}AB & 7B_1 
  \end{tabular}
 \end{minipage}
 \begin{minipage}{.5\textwidth}
  \begin{tabular}{CC|CC}
\langle g,h\rangle & h & \langle g,h'\rangle & h'\\\hline
  \ZZ_{14}AB & 14A_1 & \ZZ_{14}AB & 14B_1 \\  \ZZ_{14}AB & 14B_1 & \ZZ_{14}AB & 14A_1 \\  1 & 2A_2 & 2 & 2A _3\\   2 & 2A_3 & 1 & 2A_2 \\  3 & 2A_5 & 3 & 2A_5 \\ 7 & 2A_4 & \ZZ_4B & 4B_1 \\  7 & 2B_3 & \ZZ_4B_1 & 4B \\ 
   8 & 2B_1 & 8 & 2B_1 \\    9 & 2B_2 & 9 & 2B_2 \\ 
   10 & 4A_2 & 11 & 4A_3 \\  11 & 4A_3 & 10 & 4A_2 \\  12 & 4A_4 & \ZZ_8A & 8A_1 \\  12 & 4B_4 & \ZZ_8A & 8A_1 \\   15 & 4C_1 & 15 & 4C_1 \\ 16 & 4C_2 & 16 & 4C_2 \\  17 & 4B_2 & 18 & 4B_3 \\ 18 & 4B_3 & 17 & 4B_2 \\ 19 & 4B_5 & 19 & 4B_5 \\ 28 & 6A_2 & 29 & 6A_3 \\ 
   28 & 6A_2 & 29 & 6A_3 \\    29 & 6A_3 & 28 & 6A_2 \\  29 & 6A_3 & 28 & 6A_2 
\end{tabular}\end{minipage}
\bigskip
\caption{Relabeling for $g$ in class 2A. Each line corresponds to a conjugacy class in the central extension $C_{M_{24}}^\alpha(g)$. For each such class, we report the corresponding class in $C_{M_{24}}(g)$ with representative $h$, the group $\langle g,h\rangle$ generated by $g$ and $h$, the `relabeled' class in $C_{M_{24}}(g)$ with representative $h'$ and the group $\langle g,h'\rangle$ generated by $g$ and $h'$.}\label{t:relabeling}
\end{table}

\newpage

\newgeometry{left=2cm}
\begin{landscape}
\begin{table}
\newcolumntype{L}{>{$}l<{$}}
  \newcolumntype{C}{>{$}c<{$}}
   \newcolumntype{m}{>{\bgroup\footnotesize$}c<{$\egroup}}
 \newcolumntype{R}{>{$}r<{$}}
 \begin{minipage}{.3\linewidth}
\begin{tabular}{CC|CC}
\langle g,h\rangle & h & \langle g,h'\rangle & h'\\\hline 
 \ZZ_4B & 1A & \ZZ_4B & 1A \\  \ZZ_4B & Q^2 & \ZZ_4B & g^2 \\  \ZZ_4B & Q & \ZZ_4B & g \\  \ZZ_4B & Q^3 & \ZZ_4B & g^3 \\  \ZZ_4B & g^2 & \ZZ_4B & Q^2 \\  \ZZ_4B & g^2Q^2 & \ZZ_4B & g^2Q^2 \\  \ZZ_4B & g^2Q & \ZZ_4B & gQ^2 \\
   \ZZ_4B & g^2Q^3 & \ZZ_4B & g^3Q^2 \\  \ZZ_4B & gQ^2 & \ZZ_4B & g^2Q \\  \ZZ_4B & g & \ZZ_4B & Q \\  \ZZ_4B & gQ & \ZZ_4B & gQ \\  \ZZ_4B & gQ^3 & \ZZ_4B & g^3Q \\  \ZZ_4B & g^3Q^2 & \ZZ_4B & g^2Q^3 \\  \ZZ_4B & g^3 & \ZZ_4B & Q^3 \\
   \ZZ_4B & g^3Q & \ZZ_4B & gQ^3 \\  \ZZ_4B & g^3Q^3 & \ZZ_4B & g^3Q^3 \\    \ZZ_8A & 8A_1 & 12 & 2B_2 \\  \ZZ_8A & 8A_1 & 12 & 2A_4 \\  \ZZ_8A & 8A_1 & 12 & 4A_1 \\  \ZZ_8A & 8A_1 & 12 & 4A_2 \\  \ZZ_8A & 8A_4 & 12 & 2B_2 \\
   \ZZ_8A & 8A_4 & 12 & 2A_4 \\  \ZZ_8A & 8A_4 & 12 & 4A_1 \\  \ZZ_8A & 8A_4 & 12 & 4A_2 \\  12 & 2B_2 & \ZZ_8A & 8A_1 \\  12 & 2B_2 & \ZZ_8A & 8A_4 \\  12 & 2A_4 & \ZZ_8A & 8A_1 
\end{tabular}\end{minipage}   
\begin{minipage}{.3\linewidth}
\begin{tabular}{CC|CC}
\langle g,h\rangle & h & \langle g,h'\rangle & h'\\\hline 
 12 & 2A_4 & \ZZ_8A & 8A_4 \\  12 & 4A_1 & \ZZ_8A & 8A_1 \\  12 & 4A_1 & \ZZ_8A & 8A_4 \\  12 & 4A_2 & \ZZ_8A & 8A_1 \\
   12 & 4A_2 & \ZZ_8A & 8A_4 \\17 & 2A_2 & 18 & 2A_5 \\  17 & 2A_2 & 18 & 4B_9   \\  17 & 4B_8 & 18 & 2A_5 \\17 & 4B_8 & 18 & 4B_9 \\
     18 & 2A_5 & 17 & 2A_2 \\  18 & 2A_5 & 17 & 4B_8 \\  18 & 4B_9 & 17 & 2A_2 \\  18 & 4B_9 & 17 & 4B_8 \\  19 & 2A_3 & 19 & 2A_3 \\  19 & 2A_3 & 19 & 4B_3 \\
   19 & 4B_3 & 19 & 2A_3 \\  19 & 4B_3 & 19 & 4B_3 \\20 & 2B_3 & 20 & 2B_3 \\  20 & 2B_3 & 20 & 2B_3 \\  20 & 2B_3 & 20 & 4B_1 \\  20 & 2B_3 & 20 & 4B_1 \\  20 & 4B_1 & 20 & 2B_3 \\  20 & 4B_1 & 20 & 2B_3 \\  20 & 4B_1 & 20 & 4B_1 \\
   20 & 4B_1 & 20 & 4B_1 \\ 21 & 2B_1 & 21 & 2B_1 \\21 & 2B_1 & 21 & 4B_5 
   \end{tabular}
   \end{minipage}
   \begin{minipage}{.3\linewidth}
\begin{tabular}{CC|CC}
\langle g,h\rangle & h & \langle g,h'\rangle & h'\\\hline    
   21 & 4B_5 & 21 & 2B_1 \\21 & 4B_5 & 21 & 4B_1 \\23 & 4A_3 & 24 & 4A_4 \\
   23 & 4A_3 & 24 & 4A_4 \\  23 & 4A_3 & 24 & 4A_4 \\  23 & 4A_3 & 24 & 4A_4 \\ 24 & 4A_4 & 23 & 4A_3 \\  24 & 4A_4 & 23 & 4A_3 \\  24 & 4A_4 & 23 & 4A_3 \\  24 & 4A_4 & 23 & 4A_3 \\  25 & 4B_4 & 26 & 4B_7 \\  25 & 4B_4 & 26 & 4B_7 \\  25 & 4B_4 & 26 & 4B_7 \\  25 & 4B_4 & 26 & 4B_7 \\     26 & 4B_7 & 25 & 4B_4 \\
   26 & 4B_7 & 25 & 4B_4 \\  26 & 4B_7 & 25 & 4B_4 \\  26 & 4B_7 & 25 & 4B_4
   \\27 & 8A_2 & 27 & 8A_2 \\  27 & 8A_2 & 27 & 8A_2 \\  27 & 8A_2 & 27 & 8A_2 \\  27 & 8A_2 & 27 & 8A_2  \\
     27 & 8A_3 & 27 & 8A_3 \\  27 & 8A_3 & 27 & 8A_3 \\  27 & 8A_3 & 27 & 8A_3 \\  27 & 8A_3 & 27 & 8A_3 \\        \phantom{ 21} & \phantom{ 4B} & \phantom{21} & \phantom{4B} 
\end{tabular}
\end{minipage}
\bigskip
\caption{Relabeling for $g$ in class 4B.}\label{t:relabeling2}
\end{table}

\end{landscape}

\restoregeometry


\clearpage

\newpage
\begin{thebibliography}{99}


\bibitem{Eguchi:2010ej}
T.~Eguchi, H.~Ooguri and Y.~Tachikawa,
``Notes on the K3 Surface and the Mathieu group $M_{24}$,''
 Exper.\ Math.\  {\bf 20}, 91 (2011)
{\tt [arXiv:1004.0956 [hep-th]]}.
%%CITATION = ARXIV:1004.0956;%%

\bibitem{Cheng:2010pq}
M.C.N.~Cheng,
``K3 Surfaces, N=4 dyons, and the Mathieu group $M_{24}$,"
Commun.\ Number\ Theory\ Phys. {\bf 4}, 623  (2010) 
{\tt [arXiv:1005.5415 [hep-th]]}.
%%CITATION = ARXIV:1005.5415;%%



\bibitem{Gaberdiel:2010ch}
M.R.~Gaberdiel, S.~Hohenegger and R.~Volpato,
``Mathieu twining characters for K3,"
JHEP {\bf 1009}, 058 (2010) 
{\tt [arXiv:1006.0221 [hep-th]]}.
%%CITATION = JHEPA,1009,058;%%

\bibitem{Gaberdiel:2010ca}
M.R.~Gaberdiel, S.~Hohenegger and R.~Volpato,
``Mathieu Moonshine in the elliptic genus of K3,"
JHEP {\bf 1010}, 062 (2010) 
{\tt [arXiv:1008.3778 [hep-th]]}.
%%CITATION = JHEPA,1010,062;%%

\bibitem{Eguchi:2010fg}
T.~Eguchi and K.~Hikami,
``Note on twisted elliptic genus of K3 surface,"
Phys.\ Lett.\  B {\bf 694}, 446 (2011) 
{\tt [arXiv:1008.4924 [hep-th]]}.
%%CITATION = PHLTA,B694,446;%%

%\cite{Taormina:2010pf}
\bibitem{Taormina:2010pf}
  A.~Taormina and K.~Wendland,
  ``The Symmetries of the tetrahedral Kummer surface in the Mathieu group $M_{24}$,''
  arXiv:1008.0954 [hep-th].
  %%CITATION = ARXIV:1008.0954;%%




%\cite{Cheng:2011ay}
\bibitem{Cheng:2011ay}
M.C.N.~Cheng and J.F.R.~Duncan,
``On Rademacher sums, the largest Mathieu group, and the holographic modularity of Moonshine,'' Commun. Number Theory Phys. {\bf 6} (3), 697--758 (2012)
{\tt  arXiv:1110.3859 [math.RT]}.
%%CITATION = ARXIV:1110.3859;%%

%\cite{Gaberdiel:2011fg}
\bibitem{Gaberdiel:2011fg}
M.R.~Gaberdiel, S.~Hohenegger and R.~Volpato,
``Symmetries of K3 sigma models,'' Commun.Num.Theor.Phys. {\bf 6}
1--50 (2012)
{\tt  arXiv:1106.4315 [hep-th]}.
%%CITATION = ARXIV:1106.4315;%%

%\cite{Taormina:2011rr}
\bibitem{Taormina:2011rr}
  A.~Taormina and K.~Wendland,
  ``The overarching finite symmetry group of Kummer surfaces in the Mathieu group $M_{24}$,''
  JHEP {\bf 1308} (2013) 125
 {\tt [arXiv:1107.3834 [hep-th]]}.
  %%CITATION = ARXIV:1107.3834;%%


\bibitem{Volpato:2012qe}
R.~Volpato, ``Mathieu Moonshine and symmetries of K3 sigma models,''
Fortsch.Phys. {\bf 60}, 1112-1117 (2012) {\tt [arXiv:1201.6172]}.


\bibitem{Gaberdiel:2012um} 
M.R.~Gaberdiel and R.~Volpato,
``Mathieu moonshine and orbifold K3s,''
in \emph{Conformal Field Theory, Automorphic Forms and Related Topics}, Contrib. Math. Comput. Sci. {\bf 8}, Springer-Verlag, Berlin - Heidelberg - New York (2014)
{\tt [arXiv:1206.5143 [hep-th]]}.
%%CITATION = ARXIV:1206.5143;%%

\bibitem{GannonMathieu}
T.~Gannon, 
``Much ado about Mathieu,"
arXiv:1211.5531 [math.RT].

%\cite{Taormina:2013jza}
\bibitem{Taormina:2013jza}
  A.~Taormina and K.~Wendland,
  ``Symmetry-surfing the moduli space of Kummer K3s,''
  arXiv:1303.2931 [hep-th].
  %%CITATION = ARXIV:1303.2931;%%

%\cite{Taormina:2013mda}
\bibitem{Taormina:2013mda}
  A.~Taormina and K.~Wendland,
  ``A twist in the M24 moonshine story,''
  arXiv:1303.3221 [hep-th].
  %%CITATION = ARXIV:1303.3221;%%

%\cite{Gaberdiel:2013psa}
\bibitem{Gaberdiel:2013psa}
  M.~R.~Gaberdiel, A.~Taormina, R.~Volpato and K.~Wendland,
  ``A K3 sigma model with $\mathbb{Z}_2^8:M_{20}$ symmetry,''
  JHEP {\bf 1402} (2014) 022
  {\tt [arXiv:1309.4127 [hep-th]]}.
  %%CITATION = ARXIV:1309.4127;%%

%\cite{Creutzig:2013mqa}
\bibitem{Creutzig:2013mqa}
  T.~Creutzig and G.~Hoehn,
  ``Mathieu Moonshine and the Geometry of K3 Surfaces,'' Commun.\ Num.\ TheorPhys.\  {\bf 08} (2014) 295
 {\tt [arXiv:1309.2671 [math.QA]]}.
  %%CITATION = ARXIV:1309.2671;%%


\bibitem{Gaberdiel:2012gf}
  M.~R.~Gaberdiel, D.~Persson, H.~Ronellenfitsch and R.~Volpato,
  ``Generalized  Mathieu Moonshine,'' Commun. Num. Theor. Phys. {\bf 7}, 145-223 (2013)
{\tt   [arXiv:1211.7074 [hep-th]]}.
  %%CITATION = ARXIV:1211.7074;%%

\bibitem{Gaberdiel:2013nya}
  M.~R.~Gaberdiel, D.~Persson and R.~Volpato,
  ``Generalized  Moonshine and Holomorphic Orbifolds,''
  arXiv:1302.5425 [hep-th].
  %%CITATION = ARXIV:1302.5425;%%
  
\bibitem{Norton}
S.~Norton, ``Generalized moonshine'',
Proc.\ Sympos.\ Pure Math.\ {\bf 47},
% The Arcata Conference on Representations of Finite Groups (Arcata, Calif., 1986), 
209–210, Amer.\ Math.\ Soc., Providence, RI (1987).

\bibitem{Cheng:2013kpa} 
  M.~C.~N.~Cheng, X.~Dong, J.~Duncan, J.~Harvey, S.~Kachru and T.~Wrase,
  ``Mathieu Moonshine and N=2 String Compactifications,''
  JHEP {\bf 1309}, 030 (2013)
 {\tt [arXiv:1306.4981 [hep-th]]}.
  
\bibitem{Harrison:2013bya}
  S.~Harrison, S.~Kachru and N.~M.~Paquette,
  ``Twining Genera of (0,4) Supersymmetric Sigma Models on K3,''
  JHEP {\bf 1404} (2014) 048
 {\tt [arXiv:1309.0510 [hep-th]]}.
  %%CITATION = ARXIV:1309.0510;%%  
  
\bibitem{Harvey:2013mda} 
  J.~A.~Harvey and S.~Murthy,
  ``Moonshine in Fivebrane Spacetimes,''
  JHEP {\bf 1401} (2014) 146
 {\tt [arXiv:1307.7717 [hep-th].]}  

\bibitem{Dijkgraaf:1996xw}
R.~Dijkgraaf, G.W.~Moore, E.P.~Verlinde and H.L.~Verlinde,
``Elliptic genera of symmetric products and second quantized strings,''
Commun.\ Math.\ Phys.\  {\bf 185}, 197 (1997)
{\tt [arXiv:hep-th/9608096]}.
  %%CITATION = HEP-TH/9608096;%%


%\cite{Sen:2009md}
\bibitem{Sen:2009md}
  A.~Sen,
  ``A Twist in the Dyon Partition Function,''
  JHEP {\bf 1005} (2010) 028
  {\tt [arXiv:0911.1563 [hep-th]]}.
  %%CITATION = ARXIV:0911.1563;%%

%\cite{Sen:2010ts}
\bibitem{Sen:2010ts}
  A.~Sen,
  ``Discrete Information from CHL Black Holes,''
  JHEP {\bf 1011} (2010) 138
 {\tt  [arXiv:1002.3857 [hep-th]]}.
  %%CITATION = ARXIV:1002.3857;%%
  
  %\cite{Govindarajan:2010fu}
\bibitem{Govindarajan:2010fu}
  S.~Govindarajan,
  ``BKM Lie superalgebras from counting twisted CHL dyons,''
  JHEP {\bf 1105} (2011) 089
  {\tt [arXiv:1006.3472 [hep-th]]}.
  %%CITATION = ARXIV:1006.3472;%%

%\cite{Govindarajan:2011em}
\bibitem{Govindarajan:2011em}
  S.~Govindarajan,
  ``Unravelling Mathieu Moonshine,''
  Nucl.\ Phys.\ B {\bf 864} (2012) 823
  {\tt [arXiv:1106.5715 [hep-th]]}.
  %%CITATION = ARXIV:1106.5715;%%

\bibitem{Cheng:2012uy}
M.C.N.~Cheng and J.F.R.~Duncan,
``The largest Mathieu group and (mock) automorphic forms,'' Proc. Sympos. Pure Math. {\bf 85}, 53--82 (2012)
{\tt arXiv:1201.4140 [math.RT]}.
%%CITATION = ARXIV:1201.4140;%%

\bibitem{Raum12}
M.~Raum,
``Powers of M24-twisted Siegel product expansions are modular,"
{\tt arXiv:1208.3453v2 [math.NT]}.

%\cite{Cheng:2012tq}
\bibitem{Cheng:2012tq}
  M.~C.~N.~Cheng, J.~F.~R.~Duncan and J.~A.~Harvey,
  ``Umbral Moonshine,'' Commun.\ Num.\ TheorPhys.\  {\bf 08} (2014) 101
 {\tt [ arXiv:1204.2779 [math.RT]]}.
  %%CITATION = ARXIV:1204.2779;%%

\bibitem{MasonM24}
G.~Mason, ``$M_{24}$ and certain automorphic forms'',
Contemp. Math. {\bf 45}, p. 223-244, (1985).

\bibitem{MasonEtas}
G.~Mason, ``Elliptic systems and the eta-function,"
Notas\ Soc.\ Mat.\ Chile {\bf 8}, 37 (1989).


\bibitem{MasonGen}
G.~Mason, ``On a system of elliptic modular forms attached to the largest 
Mathieu group'', Nagoya Math.\ J.\ {\bf 118}, 177 (1990).

   
   


\bibitem{MasonGelliptic}
G.~Mason, ``$G$-elliptic systems and the genus zero problem for $M_{24}$'',
Bull. Amer. Math. Soc. {\bf 25}, no. 1, p. 45-53, (1991).
  

\bibitem{ConwayNorton}
J.H.~Conway and S.P.~Norton, 
``Monstrous Moonshine'',
Bull.\ London Math.\ Soc.\ {\bf 11}, 308 (1979).

\bibitem{FLM}
I.~Frenkel, J.~Lepowsky and A.~Meurman, 
``Vertex operator algebras and the Monster'',
Pure and Applied Mathematics {\bf 134},  Academic Press, Inc., Boston, MA, (1988).

\bibitem{Borcherds}
R.E.~Borcherds,   
``Monstrous moonshine and monstrous Lie superalgebras''
Invent.\ Math.\ {\bf 109}, 405 (1992).







\bibitem{Tuite:1994ni} 
M.P.~Tuite,
``Generalized moonshine and Abelian orbifold constructions,'' Contemp. Math. {\bf 193}, 353--368 (1996)
%conference={
%      title={Moonshine, the Monster, and related topics (South Hadley, MA,
%      1994)},
%   },
%      publisher={Amer. Math. Soc.},
%      place={Providence, RI},
{\tt arXiv:hep-th/9412036}.
%%CITATION = HEP-TH/9412036;%%

\bibitem{Dong:1997ea}
C.-y.~Dong, H.-s.~Li and G.~Mason,
``Modular invariance of trace functions in orbifold theory,''
Commun.\ Math.\ Phys.\  {\bf 214}, 1 (2000)  {\tt  [arXiv:q-alg/9703016]}.
  %%CITATION = Q-ALG/9703016;%%

\bibitem{Hohn}
G.~H\"ohn, ``Generalized moonshine for the baby Monster,''
Habilitation thesis, Mathematisches Institut, Universit\"at Freiburg (2003),
{\tt http://www.math.ksu.edu/~gerald/papers/baby8.ps}.


\bibitem{CarnahanI}
S.~Carnahan, ``Generalized Moonshine I: Genus zero functions,'' Algebra Number Theory {\bf 4} (6), 649--679 (2010)
{\tt arXiv:0812.3440 [math.RT]}.

\bibitem{CarnahanII}
S.~Carnahan, ``Generalized Moonshine II: Borcherds products,'' Duke Math. J. {\bf 161} (5), 893--950 (2012)
{\tt arXiv:0908.4223 [math.RT]}.

\bibitem{CarnahanIII}
S.~Carnahan, ``Generalized Moonshine III: Equivariant intertwining operators,''
to appear.

\bibitem{CarnahanIV}
S.~Carnahan, ``Generalized Moonshine IV: Monstrous Lie algebras,''
{\tt arXiv:1208.6254 [math.RT]}.


\bibitem{Dijkgraaf:1989pz}
R.~Dijkgraaf and E.~Witten,
``Topological gauge theories and group cohomology,''
Commun.\ Math.\ Phys.\  {\bf 129}, 393 (1990).

\bibitem{Roche:1990hs}
P.~Roche, V.~Pasquier and R.~Dijkgraaf,
``Quasi-Hopf algebras, group cohomology and orbifold models,''
Nucl.\ Phys.\ Proc.\ Suppl.\  {\bf 18B}, 60 (1990).
%%CITATION = NUPHZ,18B,60;%%  


\bibitem{Bantay:1990yr}
P.~Bantay,
``Orbifolds and Hopf algebras,''
Phys.\ Lett.\  B {\bf 245}, 477 (1990).


\bibitem{Coste:2000tq}
A.~Coste, T.~Gannon and P.~Ruelle,
``Finite group modular data,''
Nucl.\ Phys.\  B {\bf 581}, 679 (2000)
{\tt [arXiv:hep-th/0001158]}.

\bibitem{Gannontalk}
T.~Gannon, talk at `Mathieu Moonshine', ETH Z\"urich, July 2011.


\bibitem{Gritsenko:1999fk}
V.~Gritsenko,
``Elliptic genus of Calabi-Yau manifolds and Jacobi and Siegel modular forms,'' St. Petersburg Math. J. {\bf 11} (5), 781--804 (2000)
{\tt arXiv:math/9906190}.
%%CITATION = MATH/9906190;%%

\bibitem{GritsenkoNikulin2}
V.~A.~Gritsenko and V.~V.~Nikulin, ``Automorphic Forms and Lorentzian Kac-Moody Algebras. II'', Internat. J. Math. {\bf 9} (2), 201--275 (1998) {\tt [arXiv:alg-geom/9611028]}.

\bibitem{Dijkgraaf:1996it}
 R.~Dijkgraaf, E.P.~Verlinde and H.L.~Verlinde,
``Counting dyons in N=4 string theory,''
Nucl.\ Phys.\ B {\bf 484}, 543 (1997) 
{\tt [arXiv:hep-th/9607026]}.
%%CITATION = HEP-TH/9607026;%%


\bibitem{Dixon:1986jc}
  L.~J.~Dixon, J.~A.~Harvey, C.~Vafa and E.~Witten,
  ``Strings on Orbifolds,''
  Nucl.\ Phys.\ B {\bf 261} (1985) 678; ``Strings on Orbifolds. 2.,''
  Nucl.\ Phys.\ B {\bf 274} (1986) 285.

\bibitem{Dolan:1989vr}
  L.~Dolan, P.~Goddard and P.~Montague,
  ``Conformal Field Theory of Twisted Vertex Operators,''
  Nucl.\ Phys.\ B {\bf 338} (1990) 529.

\bibitem{Ganter}
N.~Ganter, 
``Hecke operators in equivariant elliptic cohomology and generalized moonshine'',
%Groups and symmetries: from the Neolithic Scots to John McKay, 173-209, 
CRM Proc.\ Lecture Notes {\bf 47}, 173, 
Amer.\ Math.\ Soc., Providence, RI (2009) {\tt [arXiv:0706.2898 [math.AT]]}.


\bibitem{GKV}
V. Ginzburg, M. Kapranov, E. Vasserot, ``Elliptic algebras and equivariant elliptic cohomology I.''
[arXiv:q-alg/9505012]

\bibitem{EichlerZagier}
M.~Eichler and D.~Zagier,
{\it The Theory of Jacobi Forms},
Birkh\"auser (1985).



\bibitem{Borcherds2}
R.~Borcherds, ``Automorphic forms on ${\rm O}_{s+2,2}({\bf R})$ and infinite
   products," Invent. Math. {\bf 120} (1), 161
(1995).

%\cite{Dabholkar:2006xa}
\bibitem{Dabholkar:2006xa}
  A.~Dabholkar and S.~Nampuri,
  ``Spectrum of dyons and black holes in CHL orbifolds using Borcherds lift,''
  JHEP {\bf 0711} (2007) 077
 {\tt  [hep-th/0603066]}.
  %%CITATION = HEP-TH/0603066;%%

\bibitem{GritsenkoNikulinSimplest}
V.~A.~Gritsenko and V.~V.~Nikulin, ``The Igusa modular form and the ''simplest'' Lorentzian Kac-Moody algebras'', Sb. Math. {\bf 187} (11), 1601--1641 (1996) {\tt
[arXiv:alg-geom/9603010]}.



\bibitem{GritsenkoNikulin}
V.~A.~Gritenko and V.~V.~Nikulin, ``Siegel automorphic form corrections of some Lorentzian Kac--Moody Lie algebras'', Amer. J. Math. {\bf 119} (1), 181--224 (1997)
{\tt [arXiv:alg-geom/9504006]}.





\bibitem{Eguchi:2011aj} 
  T.~Eguchi and K.~Hikami,
  ``Twisted Elliptic Genus for K3 and Borcherds Product,''
  Lett.\ Math.\ Phys.\  {\bf 102}, 203 (2012)
  {\tt [arXiv:1112.5928 [hep-th]]}.

\bibitem{GritsenkoClery}
 F.~Clery and V.~A.~Gritsenko, ``The Siegel modular forms of genus 2 with the simplest divisor'',
Proc. Lond. Math. Soc. {\bf 102} (6), 1024--1052 (2011) {\tt [arXiv:0812.3962 [math.NT]]}.


\bibitem{David:2006ji} 
J.R.~David, D.P.~Jatkar and A.~Sen,
``Product representation of Dyon partition function in CHL models,''
JHEP {\bf 0606}, 064 (2006)
{\tt [arXiv:hep-th/0602254]}.
 %%CITATION = HEP-TH/0602254;%%

\bibitem{David:2006ud} 
J.R.~David, D.P.~Jatkar and A.~Sen,
``Dyon spectrum in generic N=4 supersymmetric Z(N) orbifolds,''
JHEP {\bf 0701}, 016 (2007)
{\tt [arXiv:hep-th/0609109]}.
%%CITATION = HEP-TH/0609109;%%


\bibitem{Bass}
H.~Bass, J.~Milnor and J.~P.~Serre,
     ``Solution of the congruence subgroup problem for {${\rm SL}_{n}\,(n\geq 3)$} and {${\rm Sp}_{2n}\,(n\geq 2)$}'',
  Inst. Hautes \'Etudes Sci. Publ. Math. {\bf 33} (1967).

%\cite{Cheng:2013wca}
\bibitem{Cheng:2013wca}
  M.~C.~N.~Cheng, J.~F.~R.~Duncan and J.~A.~Harvey,
  ``Umbral Moonshine and the Niemeier Lattices,'' Res Math Sci 2014, {\bf 1}:3;
[arXiv:1307.5793 [math.RT]].
  %%CITATION = ARXIV:1307.5793;%%


\bibitem{Cheng:2008fc}
M.C.N.~Cheng and E.P.~Verlinde,
``Wall crossing, discrete attractor flow, and Borcherds algebra,''
SIGMA {\bf 4}, 068 (2008) 
{\tt [arXiv:0806.2337 [hep-th]]}.
%%CITATION = ARXIV:0806.2337;%%


%\cite{Cheng:2008kt}
\bibitem{Cheng:2008kt}
  M.~C.~N.~Cheng and A.~Dabholkar,
  ``Borcherds-Kac-Moody Symmetry of N=4 Dyons,''
  Commun.\ Num.\ Theor.\ Phys.\  {\bf 3} (2009) 59
  {\tt [arXiv:0809.4258 [hep-th]]}.
  %%CITATION = ARXIV:0809.4258;%%

%\cite{Govindarajan:2008vi}
\bibitem{Govindarajan:2008vi}
  S.~Govindarajan and K.~Gopala Krishna,
  ``Generalized Kac-Moody Algebras from CHL dyons,''
  JHEP {\bf 0904} (2009) 032
  {\tt [arXiv:0807.4451 [hep-th]]}.
  %%CITATION = ARXIV:0807.4451;%%
  
  %\cite{Govindarajan:2009qt}
\bibitem{Govindarajan:2009qt}
  S.~Govindarajan and K.~Gopala Krishna,
  ``BKM Lie superalgebras from dyon spectra in Z(N) CHL orbifolds for composite N,''
  JHEP {\bf 1005} (2010) 014
 {\tt [arXiv:0907.1410 [hep-th]]}.
  %%CITATION = ARXIV:0907.1410;%%

  


\bibitem{Karpilovsky1993}
G.~Karpilovsky,
``Group Representations: Volume 2," 
North-Holland Mathematical Studies {\bf 177}, Elsevier (1993).


\bibitem{deWildPropitius:1995cf}
M.D.F.~de Wild Propitius,
``Topological interactions in broken gauge theories,''
{\tt arXiv:hep-th/9511195}.



\end{thebibliography}
\end{document}